%% file: carbon_fractionation-paper-I.tex
\documentclass[oldversion]{aa}
\usepackage{natbib}
\usepackage[T1]{fontenc}
\usepackage[version=3]{mhchem}
\usepackage{times}

\usepackage{amsmath}
\usepackage{amssymb}
\usepackage{wasysym}
\usepackage{epstopdf}

\usepackage{ulem}

\usepackage{graphicx}
\usepackage{url}

\newcommand{\CII}{[C\,{\sc ii}]}
\newcommand{\CI}{[C\,{\sc i}]}  
\newcommand{\thirteenCII}{[$^{13}$C\,{\sc ii}]}

\newcommand{\msol}{M$_{\odot}$}   

\newcommand{\ktau}{\textrm{KOSMA-}$\mathrm{\tau}$}

\newcommand{\emm}[1]{\ensuremath{#1}}   
\newcommand{\emr}[1]{\emm{\mathrm{#1}}} 

\newcommand{\unit}[1]{\emm{\, \emr{#1}}}

\newcommand{\Kkms}{\unit{K\,km\,s^{-1}}}

\newcommand{\thCO} {\mbox{$^{13}$CO}}   
\newcommand{\thC} {\mbox{$^{13}$C}}     
\newcommand{\eiO} {\mbox{$^{18}$O}}     
\newcommand{\ER}{\mbox{\it ER}}
\newcommand{\FR}{\mbox{\it FR}}
\newcommand{\IR}{\mbox{\it IR}}

\makeatletter
\newcounter{reaction}
\renewcommand\thereaction{C\,\arabic{reaction}}
\newcommand\reactiontag{\refstepcounter{reaction}\tag{\thereaction}}
\newcommand\reaction@[2][]{\begin{equation}\ce{#2}%
\ifx\@empty#1\@empty\else\label{#1}\fi%
\reactiontag\end{equation}}
\newcommand\reaction@nonumber[1]{\begin{equation*}\ce{#1}%
\end{equation*}}
\newcommand\reaction{\@ifstar{\reaction@nonumber}{\reaction@}}
\makeatother

\begin{document}
\input{journals.tex}

\title{Carbon Fractionation in PDRs}
\titlerunning{Carbon Fractionation in PDRs}
\author{M. R\"ollig \and V. Ossenkopf}
\institute{I. Physikalisches Institut, Universit\"at zu K\"oln, Z\"ulpicher Str. 77, D-50937 K\"oln,
Germany}
\offprints{M. R\"ollig,\\
 \email{roellig@ph1.uni-koeln.de}}

\abstract{
We upgraded the chemical network from the UMIST Database for Astrochemistry 2006  to include isotopes such as \thC\  and \eiO. This includes all corresponding isotopologues, their chemical reactions and the properly scaled reaction rate coefficients. We study the fractionation behavior of astrochemically relevant species over a wide range of model parameters, relevant for modelling of photo-dissociation regions (PDRs). We separately analyze the fractionation of the local abundances, fractionation of the total column densities, and fractionation visible in the emission line ratios.
We find that strong \ce{C+} fractionation is possible in cool \ce{C+} gas. Optical thickness as well as excitation effects produce intensity ratios between 40 and 400. The fractionation of \ce{CO} in PDRs is significantly different from the diffuse interstellar medium. 
PDR model results never show 
 a fractionation ratio of the \ce{CO} column density larger than the elemental ratio. Isotope-selective photo-dissociation is always dominated by the isotope-selective chemistry in dense PDR gas. The fractionation of \ce{C}, \ce{CH}, \ce{CH+} and \ce{HCO+} is studied in detail, showing that the fractionation of \ce{C}, \ce{CH} and \ce{CH+} is dominated by the fractionation of their parental species. The light hydrides chemically derive from \ce{C+}, and, consequently, their fractionation state is coupled to that of \ce{C+}. The fractionation of \ce{C} is a mixed case depending on whether formation from \ce{CO} or \ce{HCO+} dominates.
Ratios of the emission lines of \CII, \CI, \thCO, and \ce{H^{13}CO+} provide individual diagnostics to the fractionation status of \ce{C+}, \ce{C}, and \ce{CO}. 
}

\maketitle

\section{Introduction}
Astronomical observations of molecules and their respective isotopologues reveal, that abundance ratios of the main species to their respective isotopologues may differ significantly from e.g. solar system isotope ratios. While isotopic fractionation in the interstellar medium is widely discussed in the framework of deuterium chemistry, its relevance for the $^{12}$C/$^{13}$C ratio in various species usually gains much less attention. In this paper we investigate chemical fractionation in the context of models of photo-dissociation regions (PDR) with the focus on effects that result from introducing \ce{^{13}C} into the applied chemical network.
 
The most important fractionation reaction is
\reaction{^{13}C+ + CO <=> C+ + ^{13}CO + \Delta E=35\,K\label{13eq1}}
\citep[see][ and references therein]{woods09}.
At high temperature back and forth reaction are equally probable and no apparent deviation from the elemental isotope ratio takes place. The lower the temperature gets, the less probable the back reaction becomes, resulting in a one-way channel shifting \ce{^{13}C} into \ce{^{13}CO} and decreasing the abundance ratio of \ce{^{12}CO}/\ce{^{13}CO}. At the same time the abundance ratio of \ce{^{12}C+}/\ce{^{13}C+} is shifted oppositely. 
\citet{langer84} performed numerical calculations of a time-dependant chemical network for a variety of physical parameter and given kinetic temperatures concluding that chemical fractionation of carbon bearing species is of increasing significance the lower the temperate is, confirming the relevance of zero-point energy differences of a few ten K  at low temperatures. In dark cloud models, where radiation is usually neglected, the kinetic temperature is the major parameter in opening and closing reaction channels.
For a given density, the chemical network only depends on the temperature and the cosmic ray ionization rate $\zeta_\mathrm{CR}$ (, and history in case of time-dependant calculations).

Carbon fractionation in molecular and diffuse clouds has been 
discussed systematically by \citet{Keene1998} and \citet{liszt07}.
Following \citet{wakelam08} we assume a standard elemental
abundance of $^{13}$C/$^{12}$C of 67 in the solar neighbourhood.
Observing $^{13}$\CI{} and $^{13}$C$^{18}$O in the Orion Bar
\citet{Keene1998} found little evidence for chemical fractionation.
Their observations showed a slight enhancement of 
C$^{18}$O$/^{13}$C$^{18}$O=75 and no enhancement of 
[$^{13}$C{\sc I}]/[C{\sc I}] relative to the standard elemental 
abundance while the chemical models predicted the opposite. The systematic
study of the C$^{18}$O$/^{13}$C$^{18}$O by \citet{LangerPenzias1990,
LangerPenzias1993} showed a systematic gradient with Galactocentric
radius, i.e. significantly higher $^{13}$C abundances in the inner
Milky Way, but also variations between 57 and 78 at the solar circle.
\citet{WouterlootBrand1996} showed that the trend continues to the
outer Galaxy with ratios above 100 in WB89-437. Optical spectroscopy
of $^{13}$CH$^+$ in diffuse clouds has shown that the $^{13}$CH$^+$/CH$^+$
ratio matches the elemental abundance ratio in the solar vicinity
very closely \citep[see e.g.][]{Centurion1995}. 
\citet{liszt07} showed that the fractionation reaction (\ref{13eq1})
is even the dominating CO destruction process for high and moderate
densities. They also report fractionation ratios of the \ce{CO} column density between 15 and 170 with a tendency for the ratio to drop with increasing column density. Their analysis is based on observation of diffuse clouds and covers a \ce{CO} column density range up to $N(\ce{^{12}CO})\approx 2\times 10^{16}$~cm$^{-2}$ and densities $\apprle 100$~cm$^{-3}$.
Here we complement this study by concentrating on PDRs with higher densities ($n \ge 10^3$~cm$^{-3}$).

This paper is organized as follows: in Sect. \ref{sec-ktau} we will briefly overview the \ktau\ PDR model which has been used to perform the model computations. The updated isotope chemistry is described in detail in Sect. \ref{sec-chemistry}. In Sect. \ref{sec-modelresults} we will present results from our model calculations. In a second paper \citep[][ Paper~II]{paper2}  we present observations of the \CII/\thirteenCII\  ratio in various PDRs and investigate the fractionation ratio of \ce{C+} in more detail. 

\section{The \ktau\ PDR model}\label{sec-ktau}
A large number of numerical PDR codes is presently in use and an 
overview of many established PDR models is presented in \citet{comparison07}\footnote{The comparison results and code descriptions are available at \url{http://www.astro.uni-koeln.de/pdr-comparison}}. 
We use the \ktau\ PDR code \citep{stoerzer1996,roellig06}\footnote{We recently updated \ktau\  to self-consistently account for various dust compositions and dust size distributions including wavelength-dependent continuum radiative transfer, dust temperature computation and photo-electric heating. For details see \citet{roellig11dust}. However, the update was not  fully available when we started this study, so we do not fully use the updated code capabilities here.}  to numerically solve the coupled equations of energy balance
(heating and cooling), chemical equilibrium, and radiative transfer.
The main features of the \ktau\ PDR model are: a) spherical model symmetry, i.e finite model clouds, b) modular chemistry, which means, that chemical species can easily be added or removed from the network and the network will rebuild dynamically, c) isotope chemistry including \ce{^{13}C} and \ce{^{18}O}, and  d) optimization toward large model grids in parameter space, allowing for example to build up any composition of individual clouds in order to simulate clumpy material \citep[for details see][]{cubick08}. The \ktau\ results can be accessed on-line at: \url{http://www.astro.uni-koeln.de/~pdr}.

\subsection{Model physics}\label{sec-physics}
Individual PDR-clumps are characterized by 
the total gas density $n$ at the cloud surface,
the clump mass $M$ in units of the solar mass,
the incident, isotropic 
far ultraviolet (FUV: 6 eV $\le$ E $\le$ 13.6 eV)
intensity $\chi$, given in units of the mean interstellar radiation
field of Draine (1978), and the metallicity $Z$. 
We assume a density power-law profile $n(r)=n_0(r/R_\mathrm{tot})^{-\alpha}$ for $R_\mathrm{core}\le r \le R_\mathrm{tot}$, and $n(r)=const.$ for $0\le r \le R_\mathrm{core}$. The standard parameters are: $\alpha=1.5$, $R_\mathrm{core}=0.2\, R_\mathrm{tot}$, roughly approximating the structure of Bonnor-Ebert spheres. Figure~\ref{denprof} shows the applied density structure. In Appendix~\ref{colstruct} we describe how to compute mean column densities for spherical model results.
\begin{figure}
\resizebox{\hsize}{!}{\includegraphics{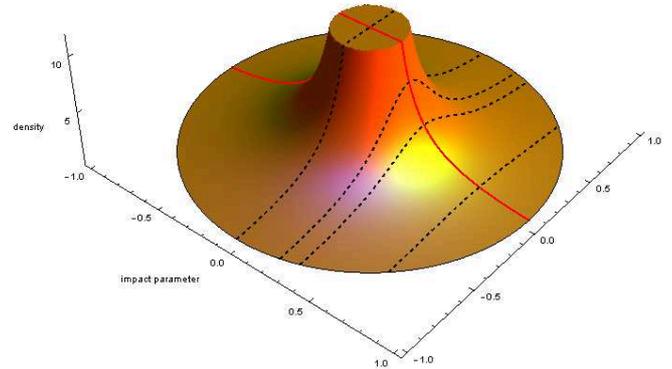}}
\caption{Radial density profile of a model clump. The dashed lines show the density profile along various lines of sight (for different impact parameters $p$).  }
\label{denprof}
\end{figure}

Excitation of the \ce{H2} molecule is computed by collapsing all rotational levels with the same vibrational quantum number into a corresponding, virtual $v$ level. We then solve the detailed population problem accounting for 15 ground-state levels ($v=0-14$) and 24 levels from the Lyman band as well as 10 Werner band levels. We assume that chemical reactions with the  population of vibrationally excited \ce{H2} have no activation energy barrier to overcome. This is especially important for species such as \ce{CH} \citep{comparison07} and \ce{CH+} \citep{agundez2010}. The heating by collisional de-excitation of vibrationally excited \ce{H2} is  calculated from the detailed level population. Photo-electric heating is calculated according to \citet{bt94}. Overall, we account for 20 heating and cooling processes. For a detailed description see \citet{roellig06} and \citet{comparison07}. 
All model results in this paper are for single-clump models without contribution from interclump gas and without clump superposition.

\subsection{Model chemistry} \label{sec-chemistry}
In \ktau\ we solve pure gas-phase steady-state chemistry with the exception of \ce{H2} forming on grains \citep{SD95}. 
It relies on the availability of comprehensive databases of chemical reaction rate coefficients. Today a few databases are publicly available, such as UDfA\footnote{\url{http://www.udfa.net}} \citep{udfa06}, the Ohio database OSU\footnote{\url{http://www.physics.ohio-state.edu/~eric/research.html}}, and the NIST Chemistry Webbook\footnote{\url{http://webbook.nist.gov/chemistry/}}. There are also efforts to pool all available reaction data into a unified database (KIDA: KInetic Database for Astrochemistry)\footnote{\url{http://kida.obs.u-bordeaux1.fr/}}. 

In the following we use UDfA06.
Retrieved at the 09/29/2009 the database consists of 4556 reaction rates, involving 421 chemical species (420 species + electrons). 34 of these reactions are present with 2 or 3 entries, valid in different temperature ranges \citep{roellig2011refit}. 
All species in UDfA06 are  composed of the main isotopes only.
\subsubsection{Isotopization}
In order to calculate chemical reactions involving different isotopologues, i.e. molecules that differ only by their isotopic composition, it is necessary to extend the chemical reaction set. 
For example, reaction 1 in UDfA06:
\reaction*{H + CH -> C + H2}
becomes
\begin{align*}
\cee{H + ^{12}CH  &-> ^{12}C + H2 \\
H + ^{13}CH  &-> ^{13}C + H2}
\end{align*}
We developed a software routine to automatically implement isotopic reactions into a given reaction set\footnote{We realized the isotopization routine in Mathematica\copyright  by Wolfram Research.}. A similar automatic procedure was  used by \citet{leBurlot93,lepetit2006}. We applied our routine to the UDfA06 reaction set, but it can be applied to any given set of chemical reactions. The routine features are:
\begin{itemize}
\item inclusion of a single \ce{^{13}C} and a single \ce{^{18}O} isotope (multiple isotopizations are neglected in this study)
\item UDfA often does not give structural information, for instance \ce{C2H3} does not distinguish between linear and circular configurations (\ce{l-C2H3} and  \ce{c-C2H3}). In such cases we consider all carbon atoms (denoted by \ce{C_{n}}) as indistinguishable. However, if structure information is provided we account for each possible isotopologue individually:
\item molecular symmetries are preserved, i.e. \ce{NC^{13}CN} = \ce{N^{13}CCN}, but \ce{HC^{18}OOH} $\ne$ \ce{HCO^{18}OH} 
\item functional groups like \ce{CH_{n}} are preserved \citep[see also][]{woods09}
\item when the above assumptions are in conflict to each other we assume {\it minimal scrambling}, i.e. we choose reactions such, that the fewest possible number of particles switch partners.
\item we favor proton/H transfer over transfer of heavier atoms
\item we favor destruction of weaker bonds
\end{itemize} 
In  Appendix~\ref{appendix:isotopization} we describe in detail how isotopologues were introduced into the chemical network. The rescaling of the newly introduced reaction rates is described in Appendix~\ref{appendix:rescaling}.\footnote{ The isotopized chemical data set as well as the isotopization routine are available online: \url{http://www.astro.uni-koeln.de/kosma-tau}.}

\subsubsection{Choice of the chemical data set}
Some data sets, e.g. OSU, are more relevant for the cold ISM while others, for instance by including reactions with higher activation energy barriers, are better suited to describe the warm ISM. Consequently, the results of chemical model calculations may differ significantly depending on the applied chemical data set. Reactions that can be found in different chemical sets can have significantly different rate coefficients among the various sets. Even very prominent reactions, such as the photo-dissociation of CO (from now on we will omit the isotopic superscript when denoting the main isotope) are listed with very different rate coefficients: UDfA06 gives an unshielded rate coefficient of $2\times 10^{-10}$~s$^{-1}$, OSU and KIDA give $3.1\times 10^{-11}$~s$^{-1}$. This is a huge difference and will lead to a significantly different chemical structure of a model PDR.
It is not the purpose of this paper to perform a detailed analysis of how the choice of a chemical data set affects PDR model results. However, we show in Appendix~\ref{chemsets} how the (isotope-free) chemistry of the main species discussed later in this paper changes for three different chemical sets. For a similar discussion see also \citet{kida2012}.  

\subsubsection{Isotope exchange reactions}
In this frame reaction (\ref{13eq1}) turns into two reactions, one for the forward, one for the back reaction, where both have the value of $\alpha$ (the rate coefficient for reaction (\ref{13eq1}) is $k_{(\ref{13eq1}\rightarrow)}=4.42\times 10^{-10}(T/300 K)^{-0.29}$), but where the back reaction is suppressed by the factor $\exp(-\gamma/T)$ with $\gamma=35$~K.
\citet{watson76} proposed that carbon isotope transfer between interstellar species can occur as a result of reaction (\ref{13eq1}). At low temperatures, reaction (\ref{13eq1}) transfers \ce{^{13}C} isotopes from \ce{^{13}C+ -> ^{13}CO}, enhancing the abundances of \ce{^{13}CO} and \ce{^{12}C+}. This reaction needs three main ingredients: sufficient amounts of \ce{^{13}C+} and \ce{CO} and temperatures well below 100~K. In the PDR context, reaction (\ref{13eq1}) is especially interesting, because in the outer transition regions, where \ce{CO} is still strongly dissociated, the \ce{^{13}C+} abundance is very large while \ce{^{13}CO} is very rare. 
Even small numbers of \ce{^{13}CO} products from reaction (\ref{13eq1}) will have a significant influence on the total \ce{^{13}CO} abundance and thus on the [\ce{CO}]/[\ce{^{13}CO}] ratio. Vice versa,  \ce{^{13}C+} will be depleted strongly, increasing  [\ce{C+}]/[\ce{^{13}C+}].
\citet{smith80} measured another isotope-exchange reaction:
\reaction[13eq2]{HCO+ + ^{13}CO <=> H ^{13}CO+ + CO + \Delta\, E = 9\,K }
with less effect on the hotter parts of the PDR, due to the low differences in back and forward reaction rates at higher temperatures.
\citet{langer84} tabulated reaction rates and reaction enthalpies for the various isotopic variants of reactions (\ref{13eq1}) and  (\ref{13eq2}) and we use their values in our calculations. Slightly different reaction rate coefficients are also given by \citet{liszt07} and \citet{woods09} but the differences are small.

\section{Application}\label{sec-modelresults}
\subsection{Model parameter grid}\label{sec-grid}
We test the outcome of the isotopic network under various conditions by computing a large grid of models spanning the possible parameter space.
Our chemical network consists of 198 species, 
involved in a total of 3250 reactions. We did not include \ce{^{18}O} into the chemistry here.

We separate the model parameters into two sets: fixed and variable. The fixed parameters determine the fundamental physical and chemical conditions for the model clouds, e.g. gas density profile parameters $\alpha$ and $R_\mathrm{core}/R_\mathrm{tot}$, cosmic ray ionization rate $\zeta_\mathrm{CR}$ of molecular hydrogen elemental abundances $X_i$, metallicity, and dust composition. The variable parameters compose the final model parameter grid. A common set of variable parameters is: total surface gas density $n_0=n(R_\mathrm{tot})=n_\ce{H} + 2 n_\ce{H2}$, cloud mass $M$, and ambient FUV field strength $\chi$ in units of the Draine field \citep{draine78}. 
For a given density law, $\alpha$ and $f_c=R_\mathrm{core}/R_\mathrm{tot}$, the total cloud radius $R_\mathrm{tot}=5.3\times 10^{18} \sqrt[3]{M/n}$~cm and  the maximum (radial) column density $N_\mathrm{max}=4.7 n R$~cm$^{-2}$ (see also App.~\ref{colstruct}).

\begin{table}[htb]
\begin{center}
\caption{Overview of the most important model parameter. All abundances
are given with respect to the total H abundance. The numbers in parentheses indicate powers of ten.}\label{parameter}
\begin{tabular}{lll}
\hline\hline
\multicolumn{3}{c}{\rule[-3mm]{0mm}{8mm}\bf Model Parameters}\\ \hline
\rule[2mm]{0mm}{2mm}He/H&0.0851&\citet{asplund05}\\
O/H&$2.56(-4)$&\citet{wakelam08}\\
C/H&$1.2(-4)$&\citet{wakelam08}\\
\ce{^{13}C}/H&$1.8(-6)$\\
N/H&$6.03(-5)$&\citet{asplund05}\\
S/H&$3.5(-6)$&\citet{goicoechea06}\\
$Z$&1&solar metallicity\\
$\zeta_{CR}$&$5(-17)$~s$^{-1}$&CR ionization rate\\
$R_\mathrm{V}$&3.1& \citet{wd01} \\
$\sigma_\mathrm{D}$&$1.75(-21)$~cm$^2$& UV dust cross section per \ce{H}\\
$<A(\lambda)/A_\mathrm{V}>$&$3.339$&mean FUV extinction\\
$\tau_\mathrm{UV}$&$3.074 A_V$&FUV dust attenuation\\
$v_b$ & 1~km~s$^{-1}$&Doppler width\\
$n_0$&$10^{3,\ldots,6}$~cm$^{-3}$&total surface gas density\\
$M$&$10^{-2,\ldots,3}$~\msol&cloud mass\\
$\chi$ & $10^{0\ldots,6}$&FUV intensity w.r.t.\\
& & \cite{draine78} field\footnote{i.e. $\chi =1.71~G_0$}\\
$\alpha$&1.5&density power law index\\
$R_\mathrm{core}$&$0.2 R_\mathrm{tot}$&size of const. density core\\
$N_\mathrm{tot}/A_\mathrm{V}$&$1.62(21)$~cm$^{-2}$\\
\hline\\
\end{tabular}
\end{center}
\end{table}
For the present study we varied the clump parameters $n_0$, $M$, $\chi$ and kept all other parameters constant. 
Table \ref{parameter} gives an overview over the used parameters.
We assume a dust composition according to \citet{wd01} (entry 7 in their Tab.~1, which is equivalent to $R_\mathrm{V}=A_\mathrm{V}/E_\mathrm{B-V}=3.1$). From the extinction cross section of each dust component we compute an average, effective FUV dust cross section per \ce{H} $\sigma_\mathrm{D}$. For a given total gas column density $N_\mathrm{tot}$, then follows: $\tau_\mathrm{FUV}=N_\mathrm{tot} \sigma_\mathrm{D}$ , and  $A_\mathrm{V}=\sigma_D N_\mathrm{tot} 1.086/3.08$. The term in the denominator corrects from visual to FUV extinction. 
 Note, that the elemental abundances of carbon show an elemental ratio (\ER{}) of 67, close to the average ratio in the local ISM \citep{sheffer2007}. We computed 168 models. The computation times per model range from 36 to 930 minutes with a median of 100 minutes.\footnote{On an Intel Xeon, 2.5 GHz CPU}

\begin{figure}[htb]
\resizebox{\hsize}{!}{\includegraphics{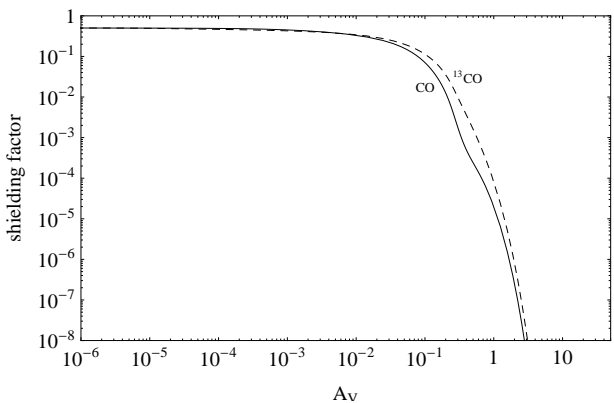}}
\caption{\ce{CO} shielding factors\protect\footnotemark of a model clump with the following model parameters: $n_0=10^5\,\mathrm{cm}^{-3}$, $M=100\,M_\odot$, $\chi=10$ (\ce{CO}: solid line, \ce{$^{13}$CO}: dashed line).}
\label{selfshielding}
\end{figure}
\begin{figure}[htb]
\resizebox{\hsize}{!}{\includegraphics{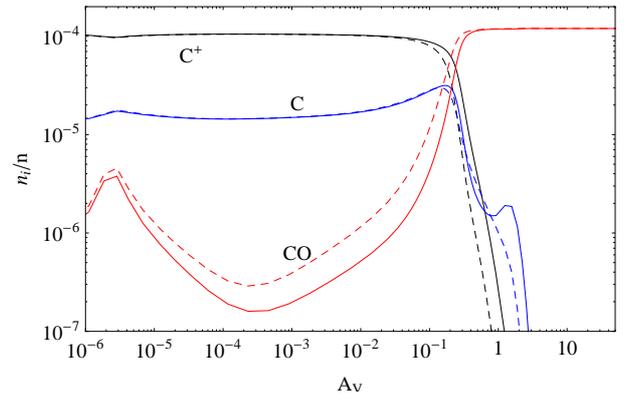}}
\caption{Chemical structure of a model clump with the following model parameters: $n_0=10^5\,\mathrm{cm}^{-3}$, $M=100\,M_\odot$, $\chi=10$. (main isotopologue: solid line, \ce{^{13}C} isotopologue multiplied by \ER{}=67: dashed line).}
\label{carbon-structure-1}
\end{figure} 
\subsection{Structure of the \ce{C+}/\ce{C}/\ce{CO} transition}
The chemical stratification of  \ce{C+}\,--\,\ce{C}\,--\,\ce{CO}, the carbon transition (CT), is a well known signature of PDR chemistry \citep[e.g. ][]{ht99}. The stratification is the result of a number of competing formation and destruction processes like photo-dissociation, dissociative recombination, and others. Photo-dissociation of CO by FUV photons is a line absorption process, and as such subject to shielding effects \citep{vDB88}. Numerically, this can be described by shielding factors that depend on the column densities of dust and of all species that absorb at the frequency of the line.
The shielding factor multiplicatively enters the photo-dissociation rate, i.e. describes the reduction of the photo-dissociation through the line absorption.\footnotetext{ Because of the isotropic illumination one has to compute the average shielding over $4\pi$ at each $A_\mathrm{V}$. Optical thickness and scattering then leads to a mean shielding factor at the cloud surface of approximately 0.5.}
 In the case of carbon monoxide the shielding depends on the columns of \ce{H2} and \ce{CO} \citep[e.g.][]{vDB88,lee96,warin1996,visser2009}\footnote{The most recent set of shielding rates is predicted by \citet{visser2009}. In this work we still used the rates from \citet{vDB88}. Under most parameter conditions the model results differ only marginally ($\sim 10\%$) when switching to the new shielding rates.}. The self-shielding of the \ce{CO} leads to stronger photo-dissociation of the rarer isotopologues at a given $A_V$\footnote{\ce{^{12}CO} could also shield its less abundant relatives, if their absorption lines are sufficiently close together. However, only very few of the efficiently dissociating lines overlap, too few for mutual shielding to be important \citep{warin1996}.}. This is also shown in Fig.~\ref{selfshielding}. Because of the lower column density of \ce{^{13}CO}, with respect to the main isotopologue, it takes a larger cloud depth for \ce{^{13}CO} to become optically thick than it takes for \ce{CO}. From Fig.~\ref{selfshielding} it can be seen  that photo-dissociation of \ce{^{13}CO} is still strong at $A_\mathrm{V}\approx0.3$  where \ce{CO} is already optically thick.

The physical conditions, such as density structure and FUV illumination, determine where the CT is situated. For the purpose of this paper we define CT as the position in a cloud where $n(\ce{C+})=n(\ce{CO})$.
The details of the \ce{C+}\,--\,\ce{C}\,--\,\ce{CO} structure are changed by various effects. \ce{C+} remains the least affected species. At the outside of the cloud photo-ionization turns basically all carbon into \ce{C+}. The strength of the FUV field and the attenuation by dust determine the depth where recombination dominates over ionization and the \ce{C+} abundances decreases. Carbon now becomes distributed between numerous species, but once shielding of \ce{CO} becomes effective, usually at $A_\mathrm{V} \gtrsim 1$, the large majority of all carbon atoms is bound into carbon monoxide. Both species, \ce{C+} and \ce{CO}, are quite insensitive to changes in the chemistry or the temperature structure, at least in cloud regions where they dominate the carbon population. However, in regions where they represent only a minor fraction of all carbon species, their chemical structure may depend sensitively on details of the cloud chemistry and physics. For example, increasing the \ce{H2} formation efficiency on hot dust grains, i.e. at low $A_\mathrm{V}$, increases the corresponding \ce{H2} formation heating efficiency, leading to higher gas temperatures in these cloud parts. This can produce an increase of the \ce{CO} population in the hot gas and consequently produce strong emission of high-$J$ emission \citep{lebourlot2012,roellig11dust}. The same effect can also positively affect the abundance of light hydrates, such as \ce{CH+} and \ce{CH}, which are primarily formed in these regions.

Atomic carbon is the species that is probably most affected  by changes in the chemistry and physics, because it is involved in the chemistry of \ce{C+} as well as \ce{CO} and chemically constitutes a transitional species. It is the major carbon species, that is least understood. Observations and model predictions of the spatial distribution of \ce{C} show big differences. The classical \ce{C+}-\ce{C}-\ce{CO} stratification, with \ce{C} being sandwiched between its two big brothers is hardly observed at all. Instead, atomic carbon shows a widespread distribution that remains to be understood \citep[e.g.][]{kramer04, mookerjea06a, kramer08, roellig2011, mookerjea2012}.

\subsection{Cloud fractionation structure}\label{fractstruct}
Unfortunately it is impossible to discuss the fractionation for all species from our chemical dataset so that we focus on a 
few molecules of particular astronomical interest.
A complete coverage of the fractionation ratio (\FR{}) for the selected species in our model grid is presented in Appendix~\ref{FRplots}.
\subsubsection{\ce{C+} and \ce{CO}}

\begin{figure}[htb]
\resizebox{\hsize}{!}{\includegraphics{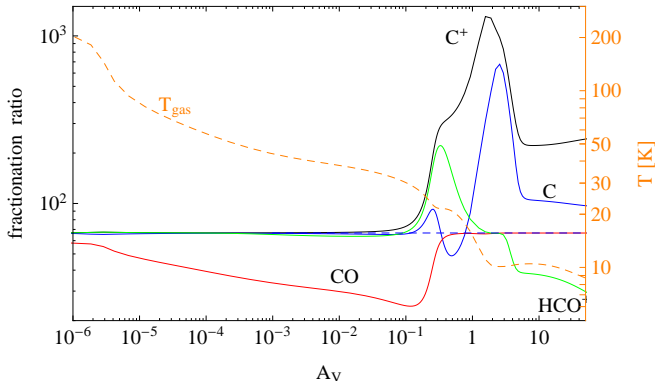}}
\caption{Fractionation structure of the same model clump as shown in Fig.~\ref{carbon-structure-1} (solid lines: fractionation ratios $n(\ce{X})/n(\ce{^{13}X})$, dashed line: kinetic gas temperature (right axis).}
\label{carbon-fractionation-1}
\end{figure}

Figure~\ref{carbon-structure-1} shows the chemical structure of the main carbon species in the model clump. Solid lines show the main isotopologues, dashed lines the \ce{^{13}C} variants. In the outer parts of the clump, most of the carbon is in the form of \ce{C+}. The CT for the model shown in Fig.~\ref{carbon-structure-1} is at $A_\mathrm{V}\approx 0.2$ where the \ce{CO} photo-dissociation rate has dropped sufficiently in order to build large quantities of \ce{CO}, which leads to a steep decline in $n(\cee{C+})$ and $n(\cee{C})$. 
In Fig.~\ref{carbon-fractionation-1} we show the corresponding \FR{} of the species from Fig.~\ref{carbon-structure-1}.
We note a number of features:
\begin{enumerate}
\item the \FR{} of \ce{C+} is always larger than or equal to the \ER{}., i.e. \ce{^{13}C+} is always under-abundant with respect to \ce{C+}
\item the \FR{} of \ce{C+} equals the \ER{} at low $A_V$
\item the \FR{} of \ce{C+} increases significantly at large $A_V$
\item the \FR{} of \ce{CO} is always smaller or equal to the \ER{} except for conditions described in Sect.~\ref{highdensity}.
\item the \FR{} of \ce{CO} deviates the strongest from the \ER{} at low $A_V$ and equals the \ER{} at large $A_V$
\item the \FR{} of \ce{C} and \ce{HCO+} show mixed behavior
\end{enumerate}

Points 1-3 are a direct consequence of reaction~(\ref{13eq1}). Any fractionation of \ce{C+} has to be a direct result of this reaction. \ce{C+} stands at the beginning of the chemical chains and there is no other direct mechanism acting in the opposite direction, such as isotope-selective photo-destruction. At low $A_V$ temperatures are sufficiently large for the reaction to work equally well in both directions. Once the temperature drops below 50~K the back reaction becomes less probable and the \FR{} increases rapidly and can be kept at large values as long as enough \ce{^{13}C+} ions are available to feed the reaction. Deep inside the clump, the \FR{} is a result of the balance between \ce{^{13}C+} formation via \ce{He+ + ^{13}CO} and destruction via the fractionation reaction. Dominance of reaction~(\ref{13eq1}) automatically leads to fractionation, in this case to \ce{C+} enrichment relative to \ce{^{13}C+}. The absolute magnitude of the \FR{} is controlled by the \ce{He+} abundance which is a direct result of the cosmic ray ionization rate. In Fig. \ref{carbon-fractionation-1} this can be seen by the roughly constant \FR{} of \ce{C+} deep inside the cloud. The same qualitative behavior is visible for all other model parameters in our model grid. Only the particular position, width and height of the \FR{} peak of \ce{C+} varies with density and FUV field strength. 
\citet{woods09} find the same fractionation behavior in their protoplanetary disk model calculations. At the surface \ce{C+} is not fractionated while at large depths they find \FR{}>\ER{}. 

The fractionation of \ce{CO} is different because a second isotope-selective process is at work, the shielding of \ce{CO} and \ce{^{13}CO} from FUV photons.
We would expect that if photo-dissociation was the dominant process, i.e. if \ce{^{13}CO} photo-dissociation was relatively stronger than the photo-dissociation of the main isotopologue, then this would result in \FR{}(\ce{CO}) > \ER{}. This is not the case for any model clump in our calculations.
It happens in thinner clouds with $n<10^2$~cm$^{-3}$ as discussed by \citet{liszt07}.
All models show a \FR{}(\ce{CO}) < \ER{} indicating that the \FR{} is dominated by the chemistry. i.e. by reaction~(\ref{13eq1}) which can only produce \FR{} < \ER{}. Exceptions to this behavior are discussed in Sect.~\ref{highdensity}.

\begin{figure}[htb]
\resizebox{\hsize}{!}{\includegraphics{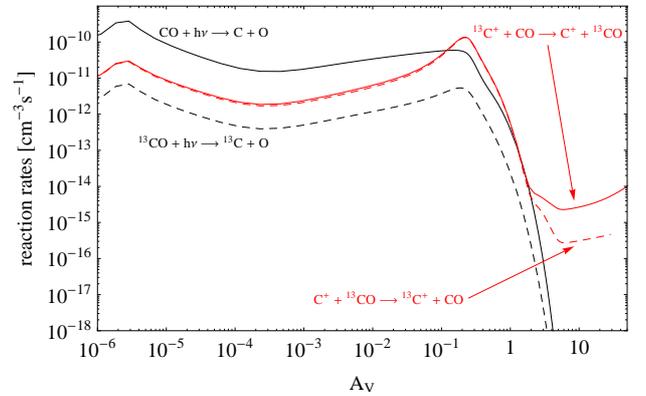}}
\caption{Reaction rates for the same model clump as shown in Figure \ref{carbon-structure-1} (photo-dissociation of \ce{CO}: black, solid;  \ce{^{13}CO}: black, dashed; back and forth reaction rates in reaction~(\ref{13eq1}): red). }
\label{co-reaction-rates}
\end{figure}
This is also visible from Fig. \ref{co-reaction-rates}, where the destruction of \ce{CO} (black, solid line) and \ce{^{13}CO} (black, dashed line) via photo-dissociation is compared to the respective formation via reaction~(\ref{13eq1}) (red lines). For \ce{CO} photo-dissociation is the major destruction process until $A_V>0.1$. Then chemical destruction by the fractionation reaction takes over.
Formation via the fractionation reaction is not the dominant formation channel of \ce{CO} for most of the clump. Only for the small $A_V$ range of $0.1\le A_V \le 0.6$ \ce{CO} formation is dominated by the fractionation reaction.  
 This is different for \ce{^{13}CO} where destruction via photo-dissociation is weaker than formation by reaction~(\ref{13eq1}) throughout the clump.
Both, formation and destruction of \ce{^{13}CO} is governed by reaction~(\ref{13eq1}) (both red lines in Fig. \ref{co-reaction-rates}). At $A_V>2$ electron recombination with \ce{H^{13}CO+} becomes the main formation channel. Hence, for the whole low $A_V$ part of the clump, the \ce{^{13}CO} abundance is controlled by the chemical fractionation reaction and accordingly, \ce{^{13}CO} is  significantly enriched relative to \ce{CO}.

\begin{figure*}[htb]
 \centering
 \includegraphics[width=8.5cm]{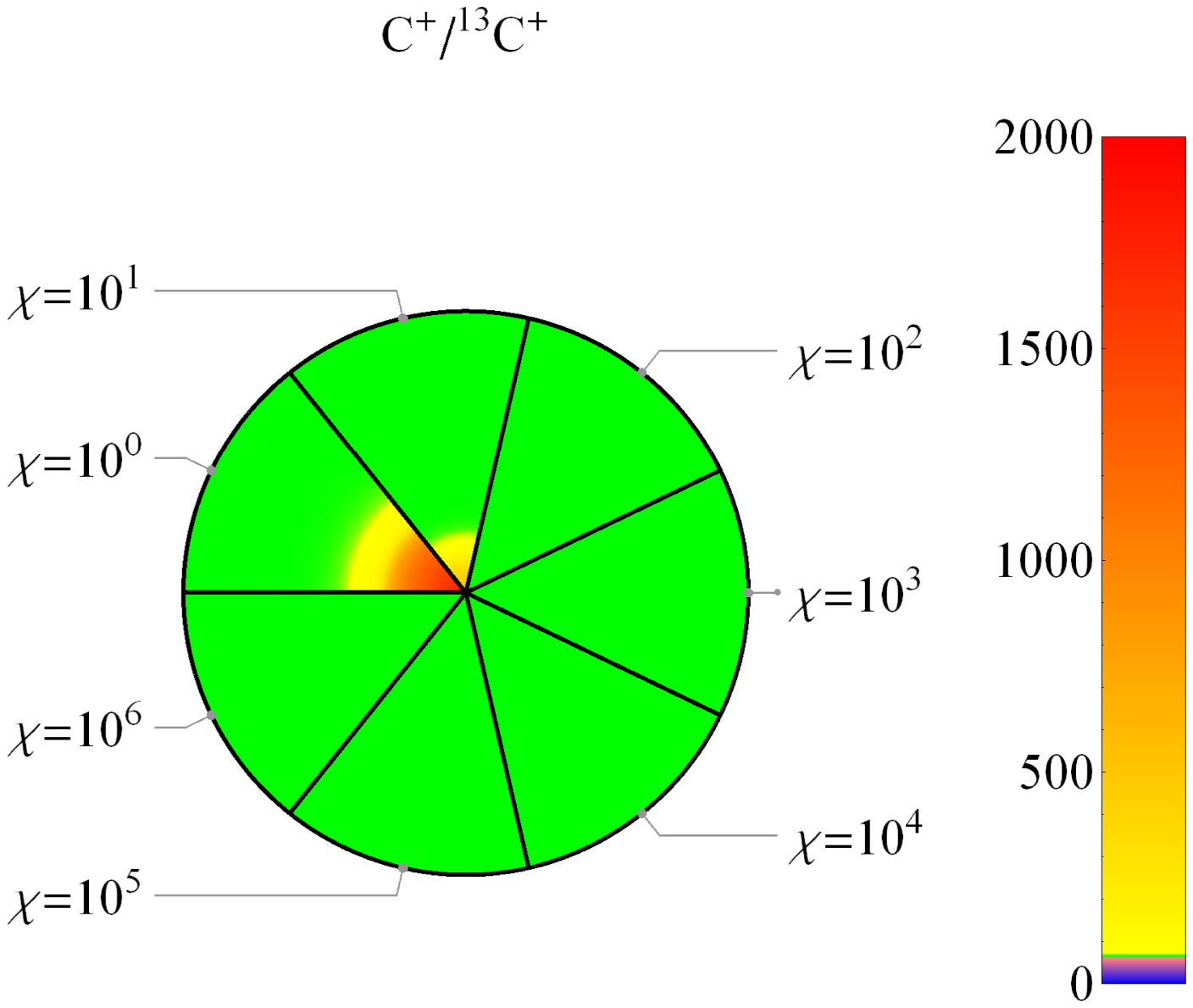}
\hfill
 \includegraphics[width=8.5cm]{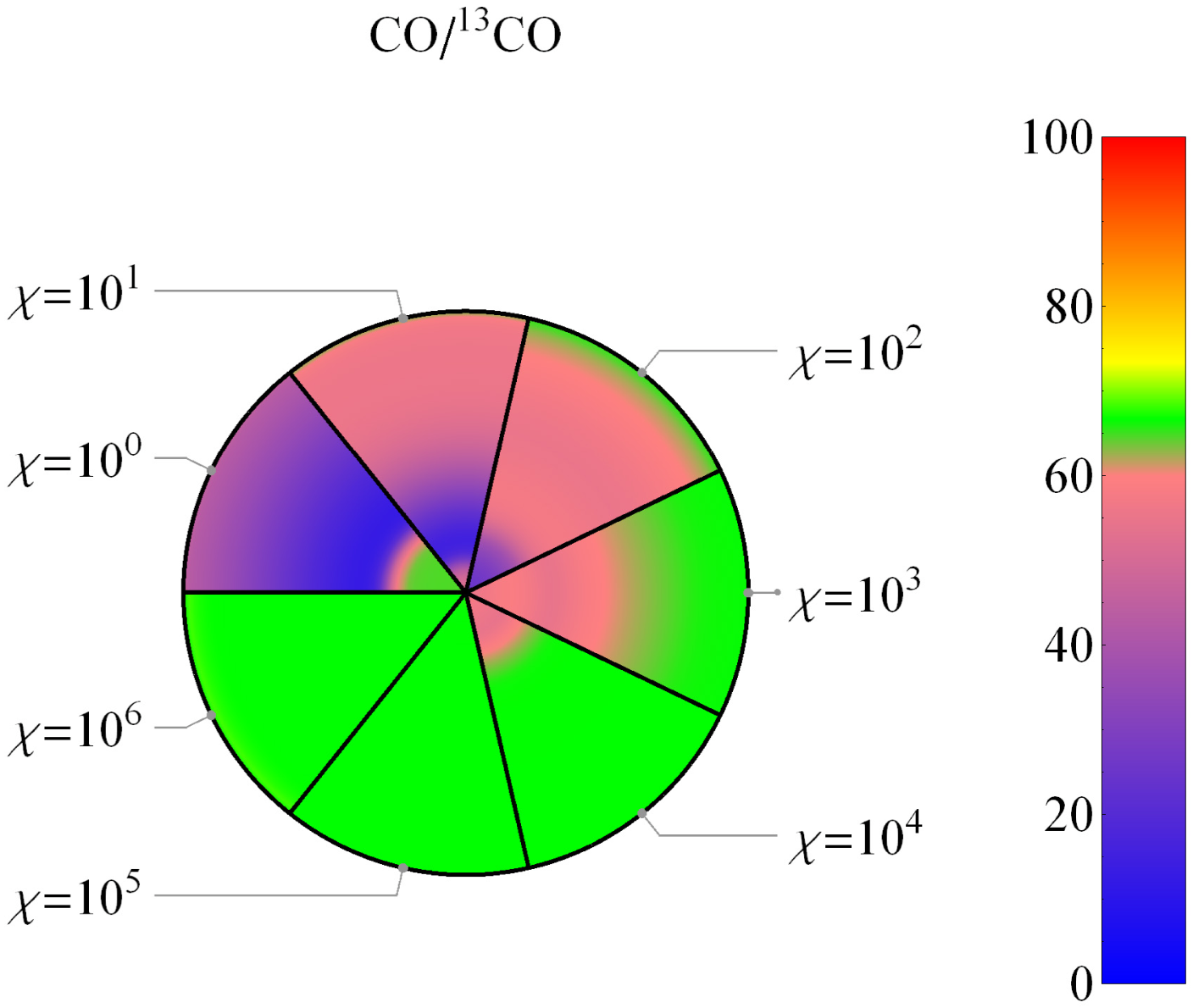}
\caption{Fractionation structure as function of relative clump radius $r/R_\mathrm{tot}$ for $n=10^3$~cm$^{-3}$ and $M=1$~M$_\odot$. Each sector corresponds to a different $\chi$ value. The \FR{} is color coded, ratios within $\pm 10$\% of the \ER{} are shown in green. Left panel: \FR{} of \ce{C+}, the color scale goes from 0 to 2000. Right panel: \FR{} of \ce{CO}, the color scale goes from 0 to 100.} \label{SP-n30m00}
 \end{figure*}

\begin{figure*}[htb]
 \centering
 \includegraphics[width=8.5cm]{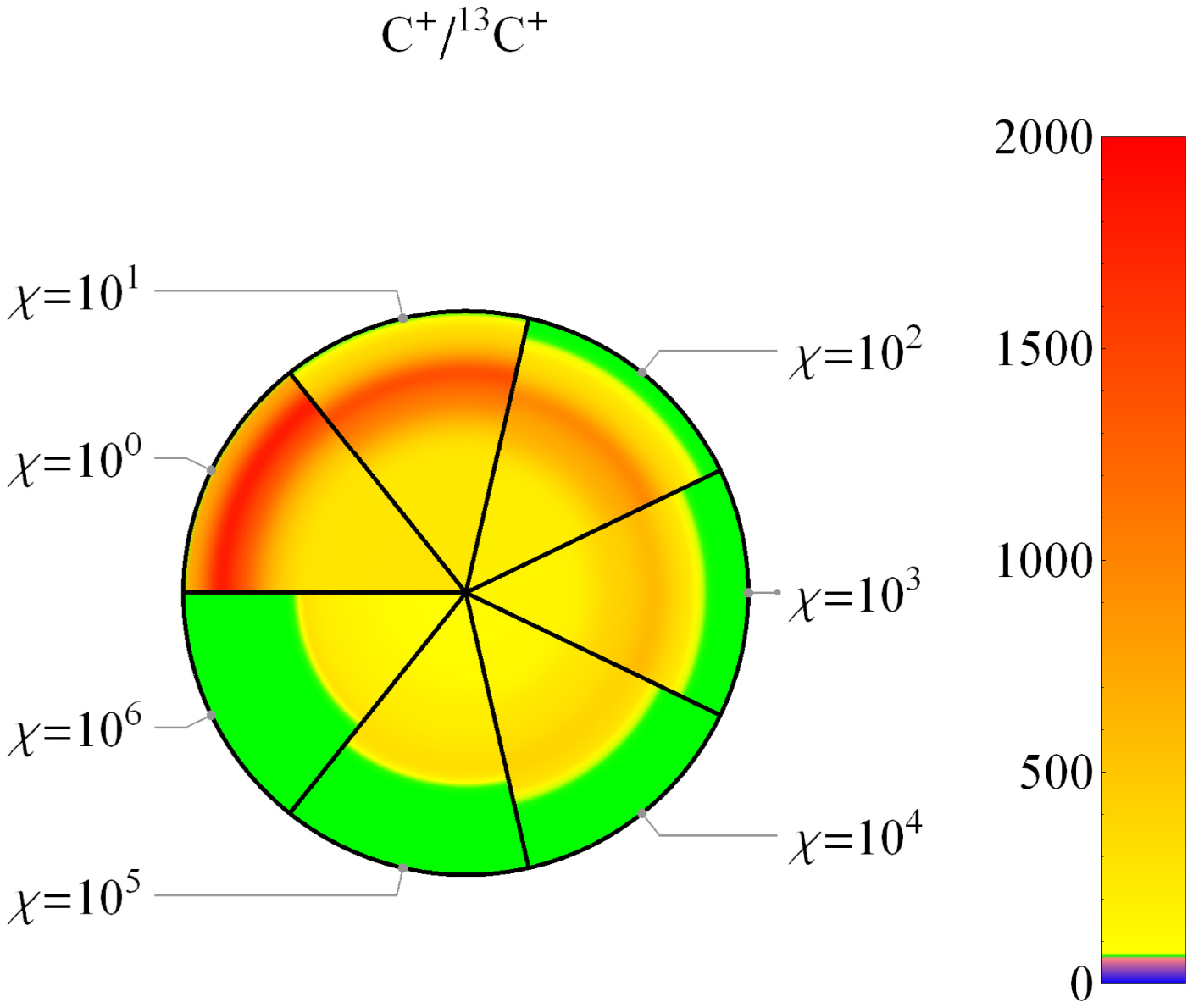}
\hfill
 \includegraphics[width=8.5cm]{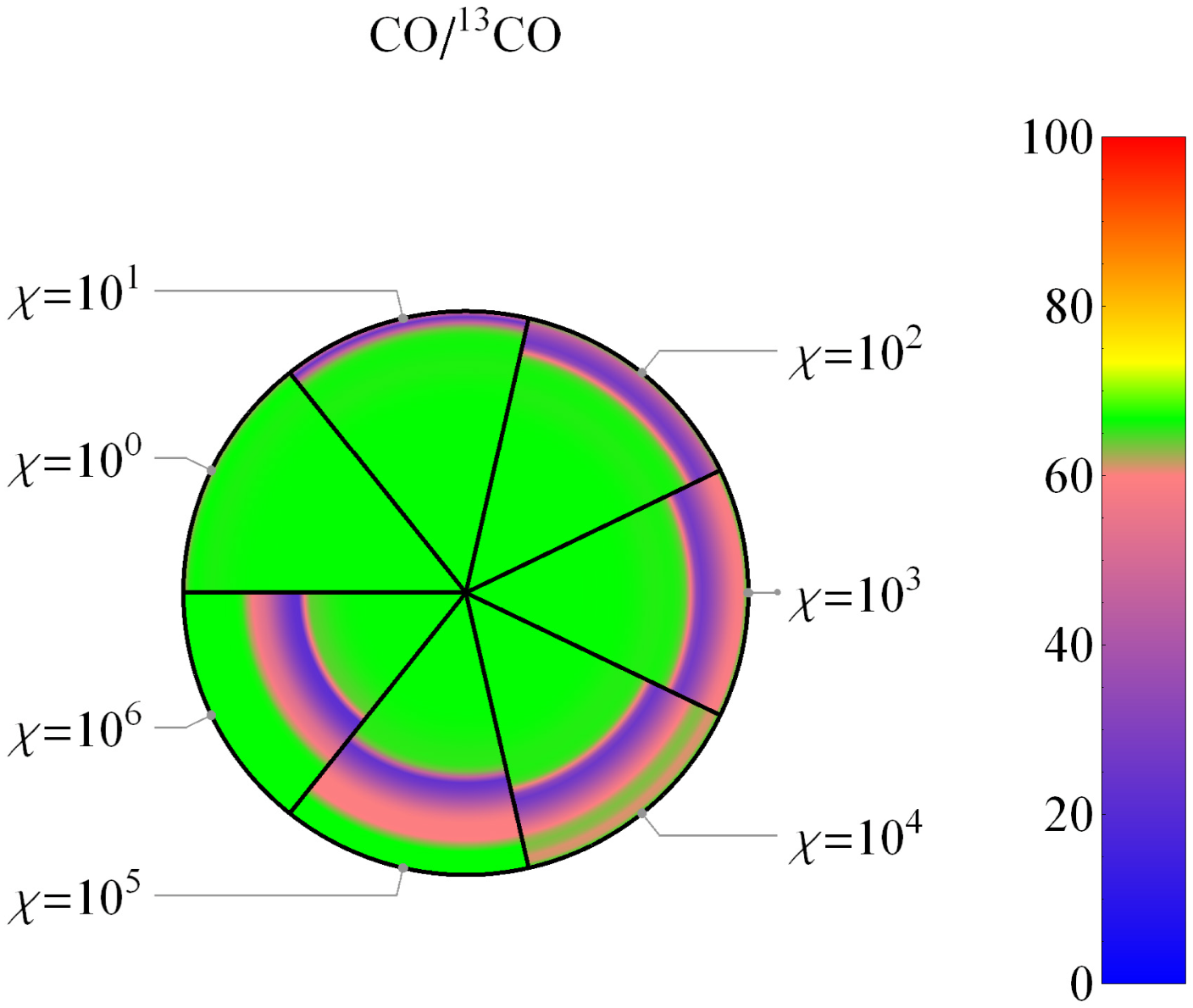}
\caption{Same as Fig. \ref{SP-n30m00} for $n=10^5$~cm$^{-3}$ and $M=1$~M$_\odot$.} \label{SP-n50m00}
 \end{figure*}

Figure \ref{SP-n30m00} compares the \FR{} of \ce{C+} (left panel) and \ce{CO} (right panel) for $n=10^3$~cm$^{-3}$ and $M=1$~M$_\odot$.  Each sector in the figure corresponds to a different FUV  intensity $\chi$. The \FR{}, as function of the relative clump radius $r/R_\mathrm{tot}$, is color coded, ratios within $\pm 10$\% of the \ER{} are shown in green. Figure~\ref{SP-n30m00} gives a visual summary of the analysis above. 
The pie chart representation visualizes the relative contribution of the abundance profile at different radii of our spherical clumps to the integrated clump ratio. At $n=10^3$~cm$^{-3}$, 
fractionation of \ce{C+} only occurs deep inside the clump, while \ce{CO}
shows most fractionation further out. However, we also note, that for higher values of $\chi$, 
the cloud is so hot that no fractionation occurs any more.
The stronger the FUV intensity the deeper the dominance of the photo-dissociation of CO. No shielding or selective photo-dissociation can yet take place and the \FR{} equals the \ER{}. However, in case of $n=10^3$~cm$^{-3}$ and $\chi\ge10^2$, CO can not be shielded efficiently  and most of the carbon is locked in its ionized form. This is different for higher densities.

The situation at higher gas density is presented in Fig.~\ref{SP-n50m00}. The fractionation of \ce{C+} is much more prominent and dominates a much larger clump volume compared to lower densities. As explained above, cold (T<100~K) \ce{C+} is always fractionated with \FR{}$\gg$\ER{}. This is true for the entire parameter grid. \ce{CO} on the other hand, requires \ce{CO} and sufficient \ce{^{13}C+} to become fractionated. These conditions are only met in a limited radius range, that is pushed to larger depths if $\chi$ increases. Deeper inside and further outside, the \FR{} of \ce{CO} equals the \ER{}. 

The effect of different clump masses is easier to understand. Adding mass is effectively equivalent to adding shielded, cold material to the clump, since it approximately requires a constant column of gas to attenuate the FUV radiation. Once this column is reached any additional material will be shielded and therefore located in the center of the clump. The appendix gives sector plots such as Figs.~\ref{SP-n30m00} and \ref{SP-n50m00} for the full parameter grid.

\begin{figure}[h]
\resizebox{\hsize}{!}{\includegraphics{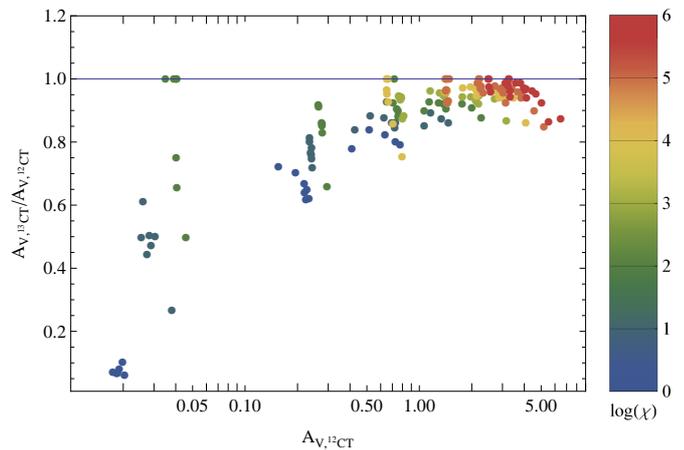}}
\caption{Comparison of the $A_\mathrm{V,^{12}CT}$ and $A_\mathrm{V,^{13}CT}$, i.e. the $A_\mathrm{V,}$ where $n(\ce{C+})=n(\ce{CO})$ and $n(\ce{^{13}C+})=n(\ce{^{13}CO})$ for all models in our parameter grid. Each data point is colored according to the ambient FUV field $\chi$. The solid line corresponds to $A_\mathrm{V,^{12}CT}=A_\mathrm{V,^{13}CT}$.}
\label{CTZ}
\end{figure}

As an additional effect, the \ce{^{13}CO} recombination occurs at
 lower values of $A_V$ than the recombination of the main isotopologues, despite the lower shielding capabilities of \ce{^{13}CO} compared to \ce{CO}. This is true for all models in our parameter grid. Across our parameter grid, the CT of the main isotopologue occurs at a $\log( N_\ce{CO})=15.8\pm0.4$, while for the \ce{^{13}C} variant ($^{13}$CT)  $\log( N_\ce{CO})=15.4\pm0.4$. The difference between both is smallest for models where photo-dissociation is more important, i.e. for larger values of $\chi$. The comparison of the \ce{CO} column density at the CTs of \ce{CO} and \ce{^{13}CO} for all models in our parameter grid is shown in Fig. \ref{CTZ}. The colors of the data points represent the FUV field strength $\chi$ of the respective model.

\begin{figure*}
\includegraphics[width=17cm]{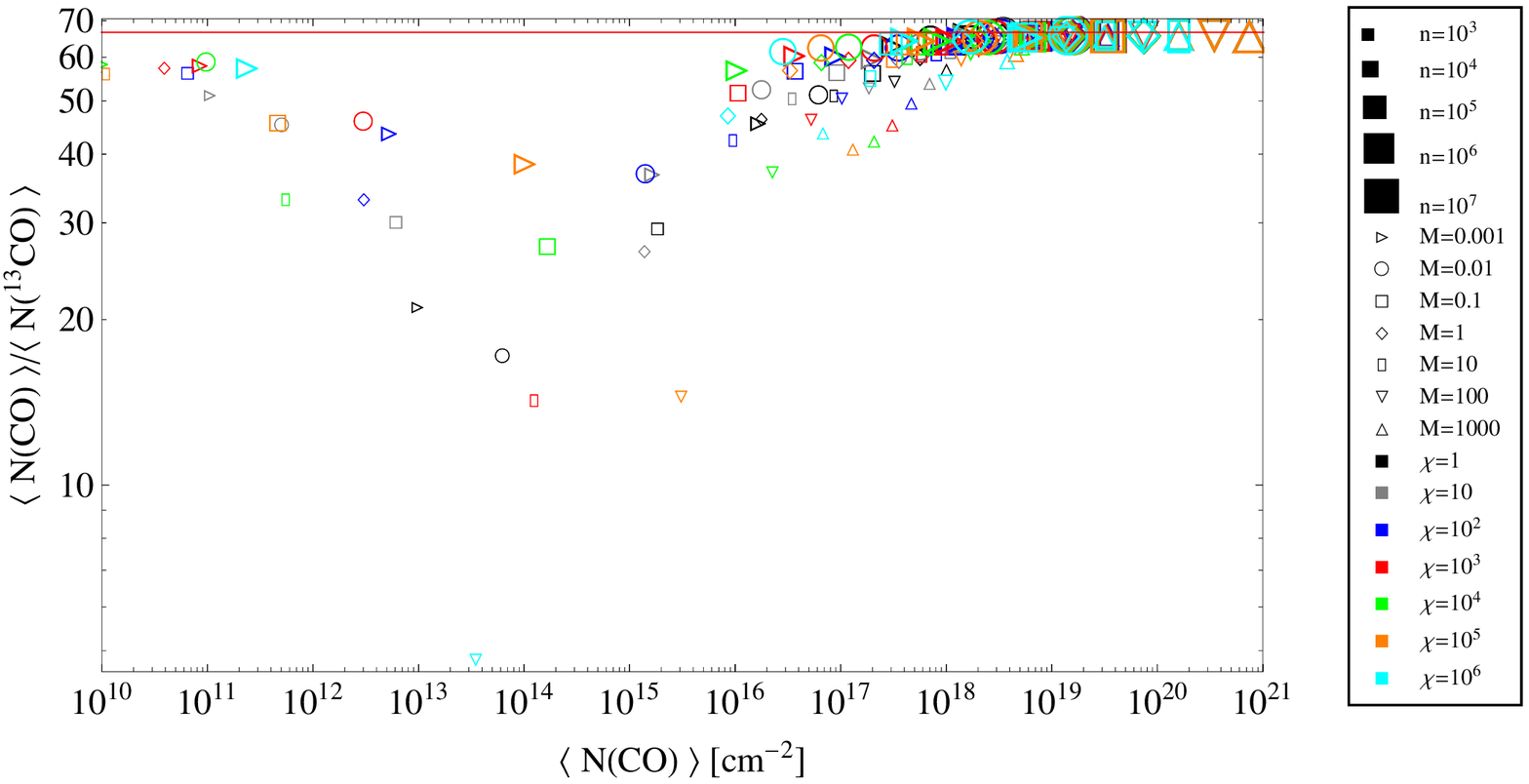}
\caption{Scatter plot of the mean column density fractionation ratio vs. the mean column density of the main isotopologue of \ce{CO}/\ce{^{13}CO} of the whole parameter space. The model parameters $n$, $M$, and $\chi$, are coded as size, shape, and color of the respective symbols. The red line denotes the model elemental abundance [\ce{^{12}C}]/\ce[\ce{^{13}C}]=67.}
\label{meancolco}
\end{figure*}

 The resulting impact on the column densities is shown in Fig.~\ref{meancolco}. Each symbol represents the column density  $\langle N(\ce{CO})\rangle/\langle N(\ce{^{13}CO})\rangle$ of a model clump with given density $n$, mass $M$, and FUV irradiation $\chi$ (clump column densities are defined in Appendix~\ref{colstruct}. Clumps with low \ce{CO} column densities deviate  most from the \ER{}. Measurements of the column density ratio $N(\ce{CO})/N(\ce{^{13}CO})$ are usually performed on diffuse or translucent clouds and thus naturally confined to a low $N(\ce{CO})$ regime. 

\begin{figure*}
\includegraphics[width=17cm]{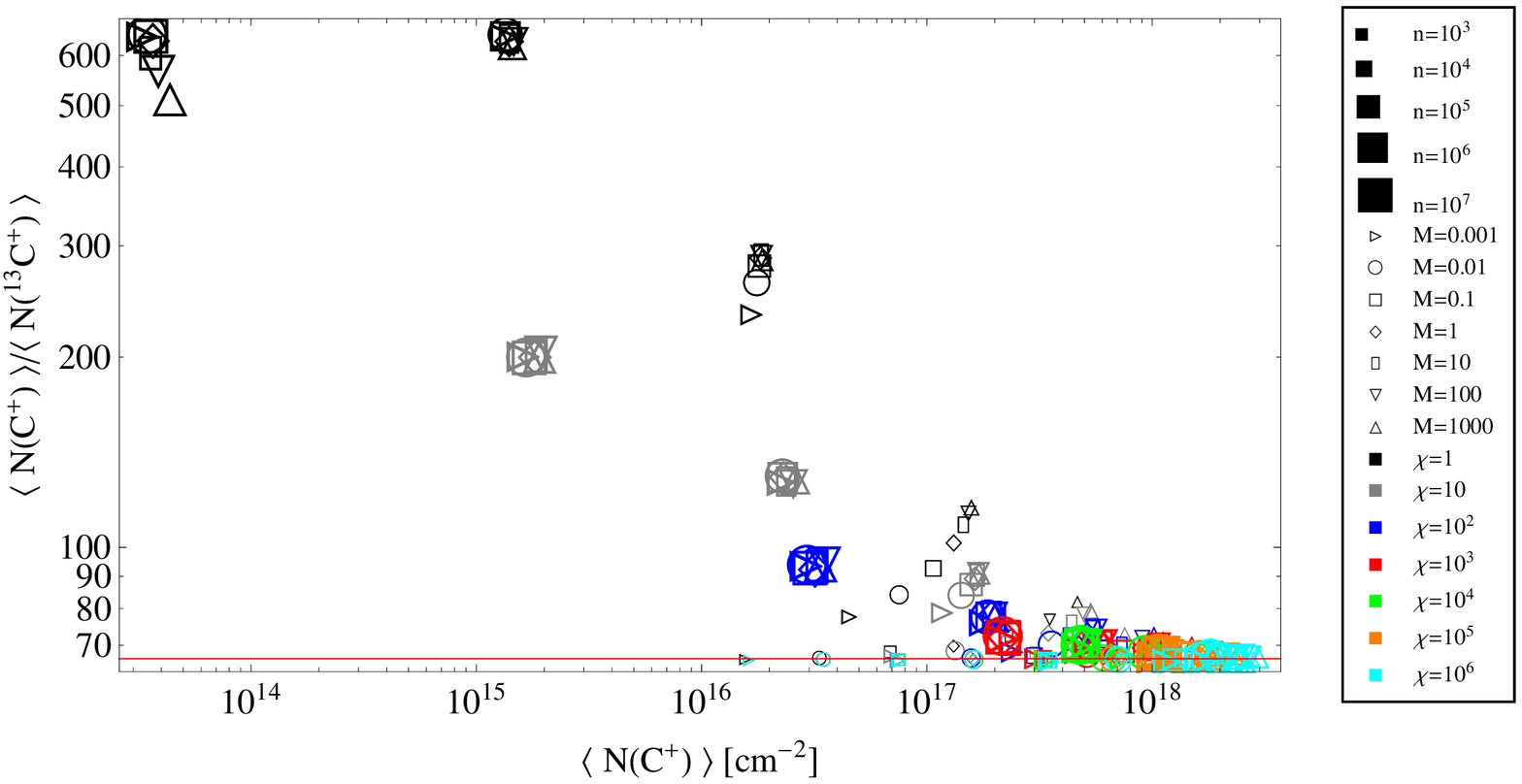}
\caption{Same as Fig.~\ref{meancolco} for \ce{C+}. 
}
\label{meancolc+}
\end{figure*}

The low column density region of Fig.~\ref{meancolco} is roughly consistent with UV absorption-line observations by \citet{sonnentrucker2007}. For translucent clouds they found an anti-correlation of $\langle N(\ce{CO})\rangle/\langle N(\ce{^{13}CO})\rangle$ with $\langle N(\ce{CO})\rangle$ in the range of $10^{14}\,\mathrm{cm}^{-2}\le \langle N(\ce{CO})\rangle \le 10^{16.5}\,\mathrm{cm}^{-2}$. \citet{liszt07} found $15<N(\ce{CO})/N(\ce{^{13}CO})<170$ with a tendency for the ratio to decline for higher column densities and a total CO column densities of a few $10^{16}$~cm$^{-2}$ from Galactic CO absorption and emission at 1.3 and 2.1~mm wavelengths for clouds with a total \ce{CO} column density $N(\ce{CO})\le 10^{16}$~cm$^{-2}$. 
\citet{sheffer2007} showed, that UV data toward diffuse/translucent lines of sight can give $0.5\le \FR{(\ce{CO})}/\ER{}\le 2$.

Clumps with stronger FUV fields show almost no fractionation, either because the molecular inner parts are so small that the gas temperatures throughout the clump are too high or because  fractionation only affects  the \ce{CO} at the outer clump regions but not the bulk of the \ce{CO} gas. For sufficiently large \ce{CO} column densities, the column density ratio of \ce{CO}/\ce{^{13}CO} turns out to be a relatively good tracer of the elemental abundance ratio of a given cloud.

The fractionation of the column density ratio of \ce{C+} is shown in Fig.~\ref{meancolc+}. The lower FUV models show the largest \FR{} while the models with very large FUV fields have a \FR{}=\ER{}. Models with low density show a much weaker fractionation (see also Fig.~\ref{SP-C+}) because the molecular part of the low density clouds contributes less to the total \ce{C+} column density. Increasing the model density (larger symbols in the figure) for given model mass and FUV illumination moves the models in Fig.~\ref{meancolc+} to the top left because larger densities lead to a stronger shielding of the gas from the FUV and therefore a CT closer to the clump surface. Consequently, the total \ce{C+} column density is decreased. On the other hand, the larger amount of cold \ce{CO} gas acts in favor of the stronger \ce{C+} fractionation. However, these strongly fractionated model clumps show the lowest \ce{C+} column density making observations difficult. Observability will be further discussed in Sect.~\ref{intensitystruct}.

As an additional consequence, when keeping density and FUV constant, the \FR{} of \ce{C+} is proportional to the clump mass. Increasing the clump mass leads to an increased \FR{}, moving model points, visible as different symbols in Fig. \ref{meancolc+}, to the top right.

\subsubsection{High density conditions with \FR{}(\ce{CO})>\ER{}\label{highdensity}}

\begin{figure}[htb]
\resizebox{\hsize}{!}{\includegraphics{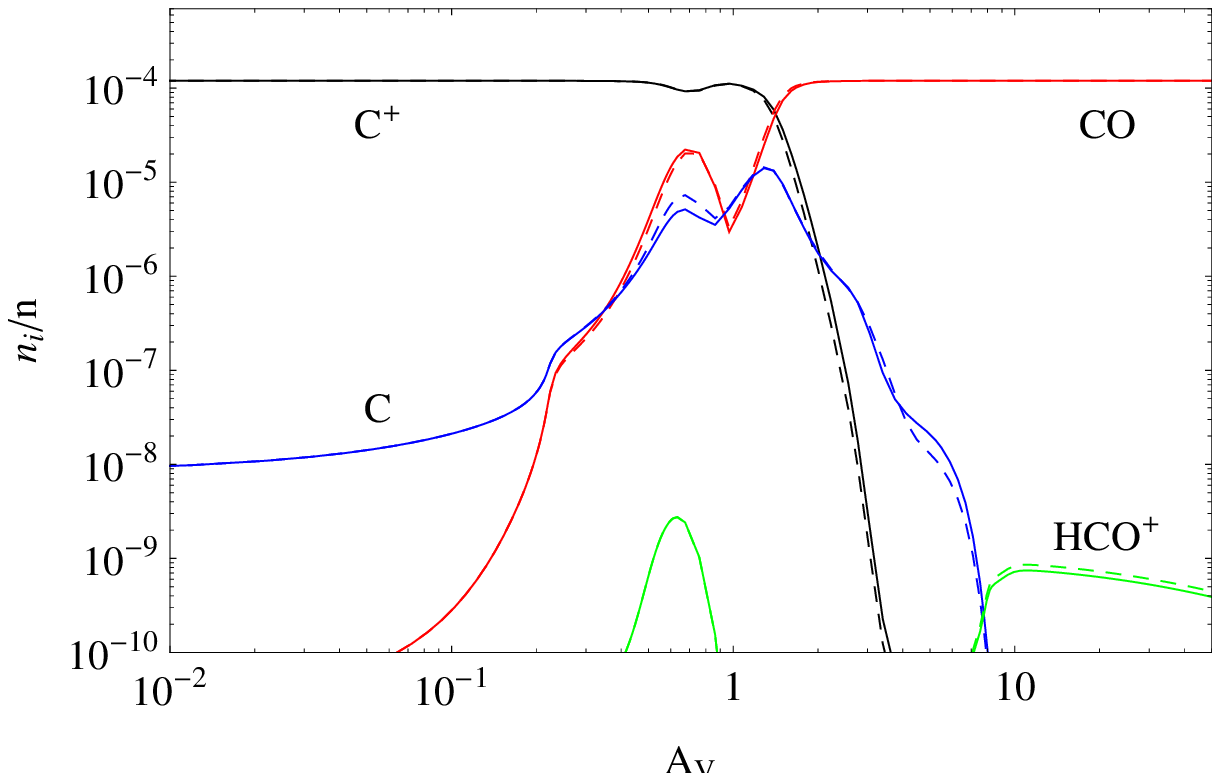}}
\caption{Fractionation structure of the very dense model clump. The model parameters are $n=10^7\,\mathrm{cm}^{-3}, M=0.01\,M_\odot, \chi=10^5$. (main isotopologue: solid lines, \ce{^{13}C} isotopologue multiplied by \ER{}=67: dashed lines)}
\label{densestruct}
\end{figure}

\begin{figure}[htb]
\resizebox{\hsize}{!}{\includegraphics{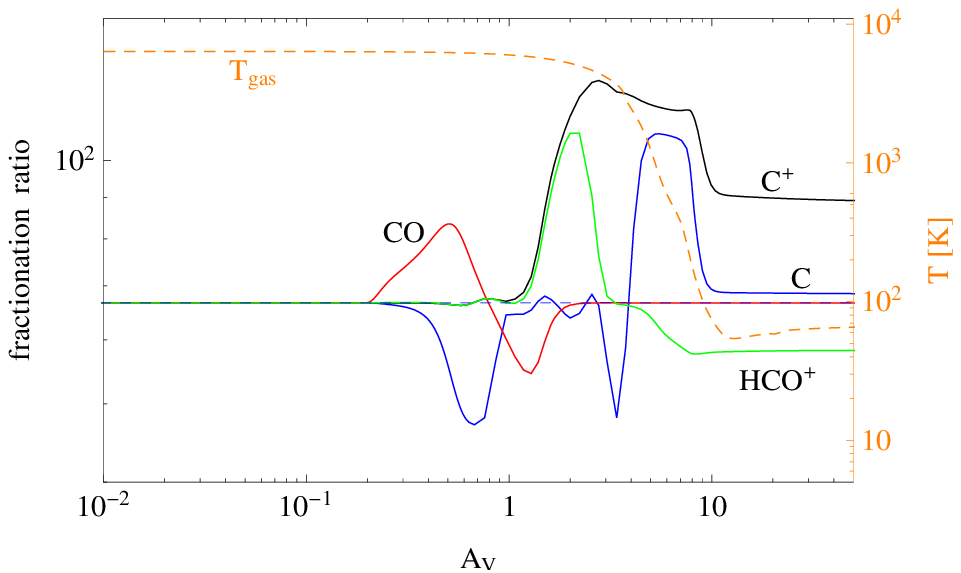}}
\caption{Fractionation structure of the same model clump as shown in Figure \ref{densestruct}.}
\label{densefract}
\end{figure}

For typical molecular cloud conditions isotope-selective photo-dissociation is never strong enough to keep \ce{^{13}CO} photo-dissociated while \ce{CO} is already recombined. If isotope-selective photo-dissociation was dominating the \ce{CO} chemistry one would obtain \FR{} > \ER{}, like it is found e.g. in diffuse clouds \cite{liszt07}.  For special conditions, particularly for densities of $n\ge10^6$ and sufficient FUV illumination, \FR{}>\ER{} is still possible across a limited $A_\mathrm{V}$ range. Figures~\ref{densestruct} and \ref{densefract} show the density and fractionation structure of such a clump. \ce{CO} shows a first, relatively strong abundance peak at $A_V=0.6$, before the CT. The gas temperature at this $A_V$ is around 600~K. At this part of the clump, the \FR{} of \ce{CO} is higher than the \ER{}.
Closer examination of the parameter reveals that similar behavior can be found for models with  $n\ge10^6$ and $\chi\ge10^4$ (compare Fig.~\ref{SP-CO}).

Detailed chemical analysis of these models shows that \FR{}>\ER{} occurs at significantly lower cloud depth than the CT (see Fig.~\ref{densestruct}).
The already mentioned \ce{CO} abundance peak results from dissociative recombination of \ce{HCO+}, which is effective in hot gas \citep{SD89} and gives rise to the \ce{CO} peak before the CT. At these cloud depths \ce{HCO+} itself is primarily formed by collisions of \ce{HOC+} and \ce{CO+} with \ce{H2} (\ce{CO+} forms through \ce{C+ + OH -> CO+ + H} and \ce{HOC+} has two main formation reactions: \ce{C+ + H2O -> HOC+ + H} and \ce{CO+ + H2 -> HOC+ + H}).

The maximum \ce{CO} abundance in this peak strongly increases with the total gas density. Even so, we have not yet reached the CT, the total \ce{CO} column density from the cloud's edge to that peak position can already reach values $\apprge 10^{15}$~cm$^{-2}$, which is sufficient for self-shielding \citep{vDB88}. Any \ce{CO} self-shielding will be stronger for the main isotopologue than for \ce{^{13}CO} and if photo-dissociation is the main destruction process for both isotopologues than the stronger shielding of \ce{CO} can give rise to \FR{}>\ER{}. If the destruction of \ce{^{13}CO} is controlled by reaction (\ref{13eq1}) than \FR{}$\le$\ER{}.

To understand which conditions can lead to \FR{}>\ER{} we balance the main formation and destruction processes of \ce{^{13}CO}. We already noted that formation via dissociative recombination of \ce{HCO+} is a general requirement, otherwise \ce{CO} self-shielding will not be effective. A second possible formation route is by the isotope exchange reaction (\ref{13eq1}). Destruction can be either via the back reaction  (\ref{13eq1}) or by photo-dissociation.  The gas temperatures are high enough to neglect the energy barrier of the back reaction and assume $k_{(\ref{13eq1}\rightarrow)}\approx k_{(\mathrm{\ref{13eq1}}\leftarrow)}$. The rate coefficient for dissociative recombination of \ce{HCO+} is $k_{(\mathrm{DR})}=2.4\times 10^{-7}(T/300 K)^{-0.29}$~s$^{-1}$~cm$^{3}$, and $\zeta_{13}$ is the photo-dissociation rate of \ce{^{13}CO}. Hence

\begin{eqnarray}
k_{(\mathrm{\ref{13eq1}}\rightarrow^)}n(\ce{C+})n(\ce{^{13}CO})+n(\ce{^{13}CO})\zeta_{13}  = \nonumber\\
 k_{(\mathrm{\ref{13eq1}}\leftarrow)}n(\ce{^{13}C+})n(\ce{CO})+n(\ce{H^{13}CO+})n(e^-)k_{(\mathrm{DR})}.
\end{eqnarray}

With $n(\ce{CO})=\FR{}\times n(\ce{^{13}CO})$, $n(e^-)\approx n(\ce{C^+})\approx 1.2\times 10^{-4} n_\mathrm{tot}$, and $n(\ce{C^+})=\ER{}\times n(\ce{^{13}C^+})$ follows

\begin{equation}
\FR{}\approx\ER{}\left(1-\frac{5317}{T^{0.4}}\frac{n(\ce{H^{13}CO+})}{ n(\ce{^{13}CO})}+3.6\times 10^{12}T^{0.29}\frac{\zeta_{13}}{n_\mathrm{tot}}\right)
\label{eqFR}
\end{equation}

The last two terms in parentheses compete in changing the \FR{} relative to \ER{}. In the relevant cloud regime, each term lies between $\approx 0.1-10$, depending on the detailed conditions, and \FR{}(\ce{CO}) can increase to  $\approx 80-100$.

We emphasize, that this behavior is not just a simple competition between chemistry and photo-dissociation as \citet{liszt07} described for diffuse clouds. A \FR{}>\ER{} in PDRs is the result of a local, chemically induced, dominance of the photo-dissociation of \ce{^{13}CO} over its chemical destruction. Only for $10^6\le n \le 10^7$~cm$^{-3}$ and  $10^4\le n\chi \le 10^6$ are the conditions such that in a narrow $A_\mathrm{V}$ range isotope-selective photo-dissociation leads to \FR{} >\ER{}. 

\subsubsection{\ce{C}}
\begin{figure}
\includegraphics[width=8.5cm]{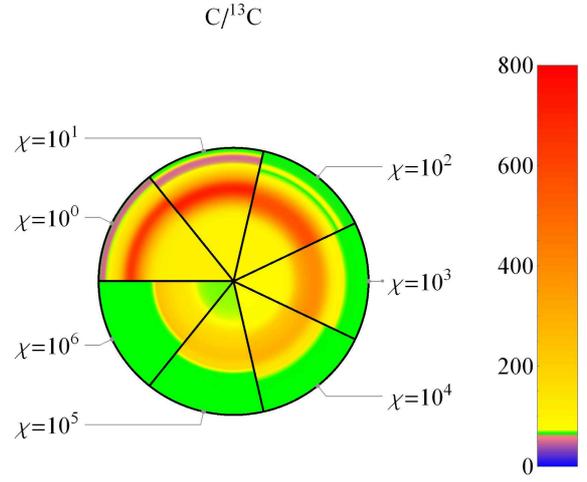}
\caption{Same as Fig.~\ref{SP-n30m00} for \ce{C} for $n=10^5$~cm$^{-3}$ and $M=1$~M$_\odot$.} \label{SP-Cn50}
 \end{figure}
From Figs.~\ref{carbon-fractionation-1} and \ref{SP-Cn50} we see that the \FR{} of \ce{C} has a small regime at low $A_\mathrm{V}$ and low FUV intensities where \FR{}<\ER{}, while under all other conditions the \FR{}$\ge$\ER{}. 
The \FR{}(\ce{C}) starts to peak at the rise of \FR{}(\ce{C+}) followed by a dip where \ce{CO} turns to the \ER{} and by a second peak at the declining flank of \FR{}(\ce{C+}).
 Until $A_\mathrm{V} < 5-10$ atomic carbon is dominantly formed through one of the following reactions:
\begin{align*}
\cee{
C^{+} + e- &-> C \\
C+ + S &-> C + S+ \\
CO + h\nu &-> C + O }
\end{align*}
while destruction occurs mainly via photo-ionization.
As discussed, the \FR{} of \ce{CO} and \ce{C+} behave oppositely  in distinct cloud depths; \FR{}(\ce{CO})$\le$\ER{}, mainly in outer layers, while \FR{}(\ce{C+})$\ge$\ER{}, somewhat deeper in. 
\begin{figure*}[ht]
 \centering
 \includegraphics[width=8.5cm]{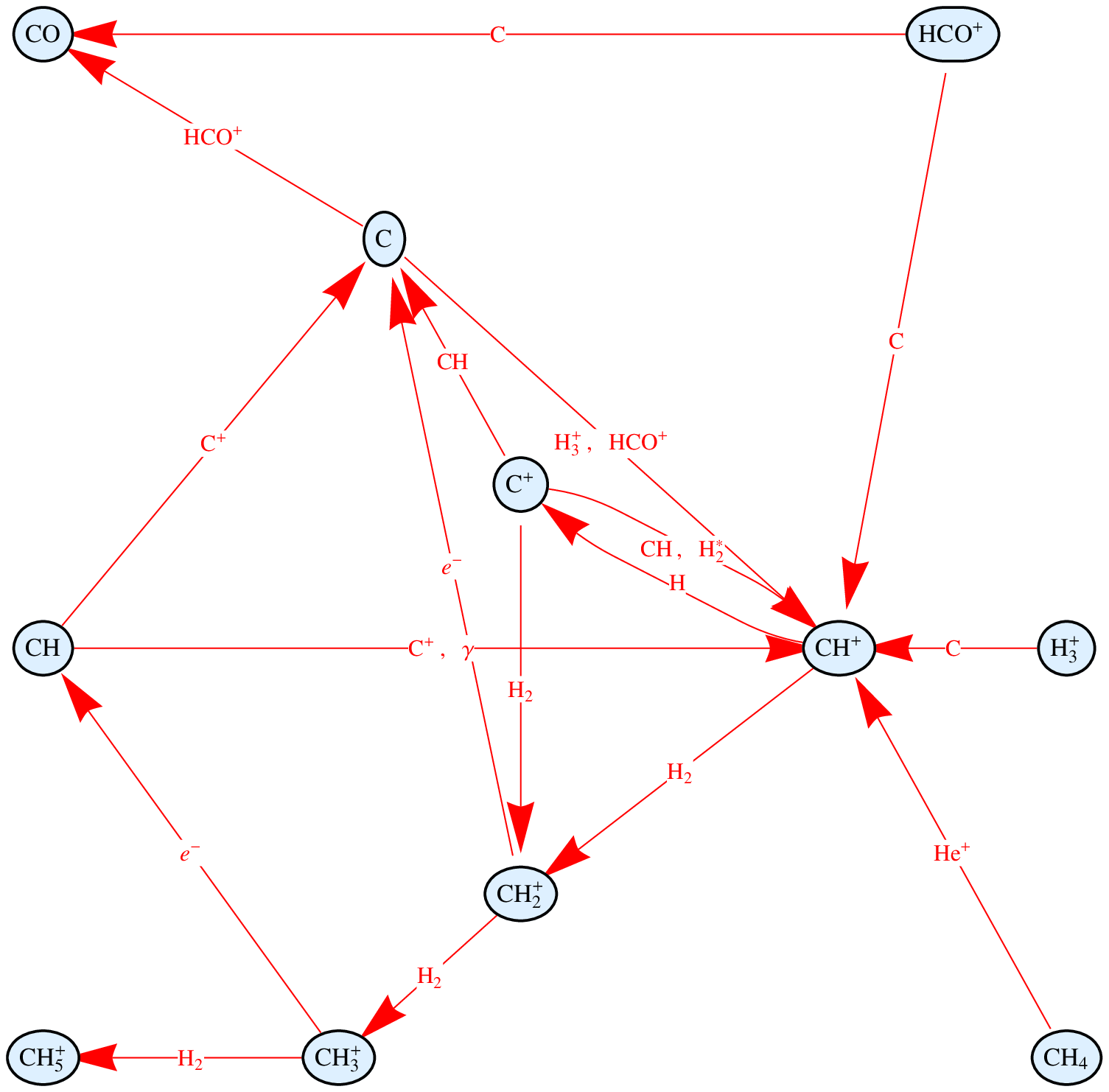}
\hfill
 \includegraphics[width=8.5cm]{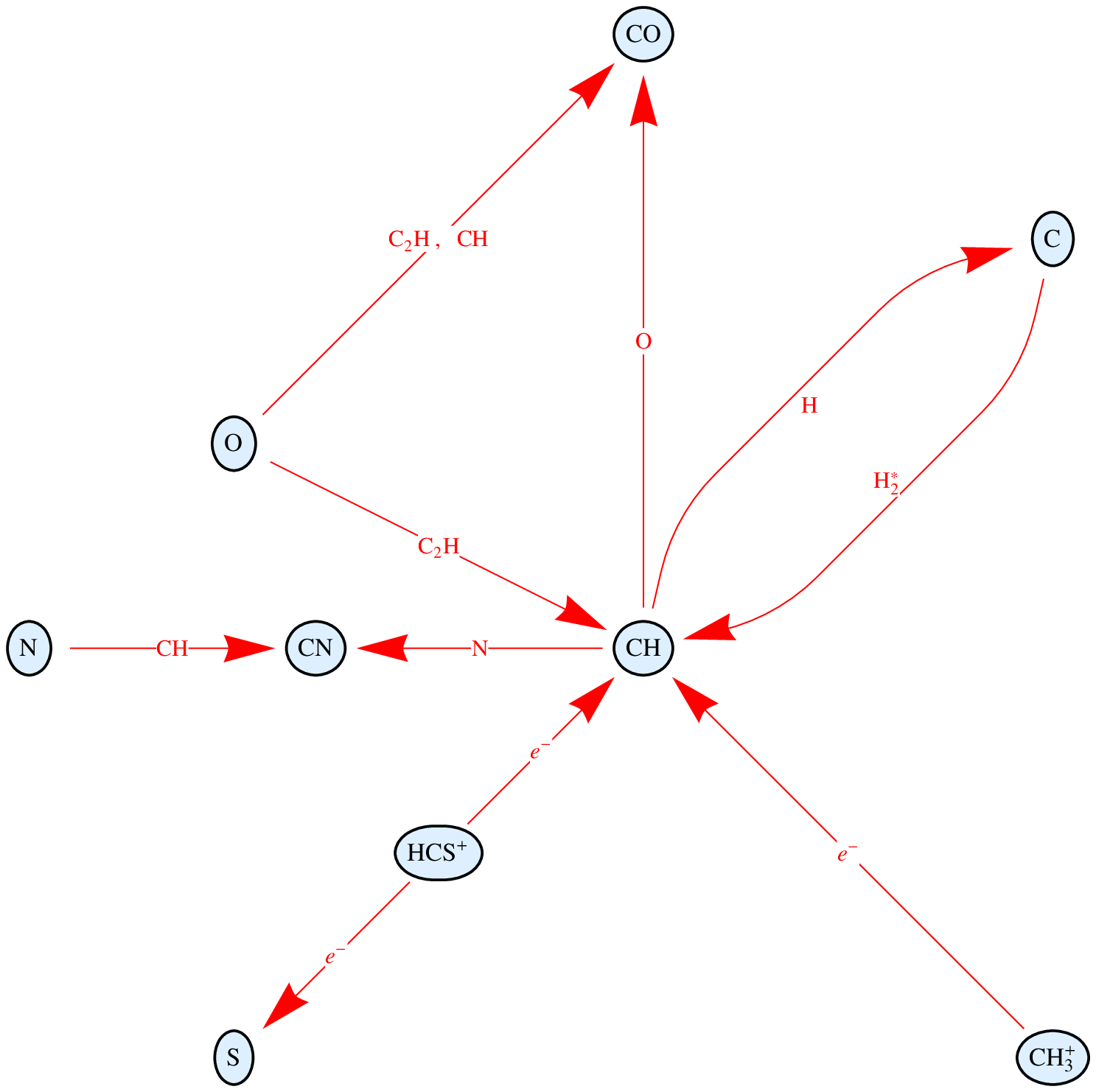}
\caption{Chemical network of the dominant formation and destruction channels for: {\bf left panel:} \ce{CH+ , CH2+}, and \ce{CH3+};  {\bf right panel:} \ce{CH} } \label{network}
 \end{figure*}

At low $A_\mathrm{V}$ both isotopic variants of atomic carbon will be dominantly formed from the recombination of their ionized forms and \ce{C} will share the \FR{} of \ce{C+}. This is visible as first peak in \FR{}(\ce{C}).The weaker shielding of \ce{^{13}CO} and the strong fractionation of \ce{C+} makes photo-dissociation of \ce{^{13}CO} the major formation reaction for \ce{^{13}C} for $A_\mathrm{V}>0.2$. The isotope-selective shielding drives the \FR{} of \ce{C} toward lower values and gives rise to the dip in the curve in Fig.~\ref{SP-Cn50}. The magnitude of this diminishment depends on the differences in the shielding of \ce{^{12}CO} and \ce{^{13}CO} and the cloud depth where the shielding is still weak. If this difference is still significant when \FR{}(\ce{C}) reaches its first peak it can push the \FR{} to values smaller than the \ER{}.

For $A_\mathrm{V}>0.2$, until FUV shielding becomes strong,  \ce{^{13}CO + h \nu} will be one order of magnitude faster than any of the other formation reactions while photo-dissociation of \ce{^{12}CO} will never be the dominant formation reaction. At $A_\mathrm{V}>1$, \FR{}(\ce{CO})=\ER{} and the \FR{}(\ce{C}) will increase again. Charge transfer between \ce{C+} and \ce{S} becomes the major formation reaction and consequently, \ce{C} will share the \FR{} of \ce{C+}. At very large cloud depths cosmic-ray-induced photo-dissociation of \ce{CO} becomes a main formation reaction together with charge transfer between \ce{C+} and \ce{SO} for \ce{C} and charge transfer between \ce{N+} and \ce{^{13}CO} for  \ce{^{13}C}. As a result, the \FR{}(\ce{C}) will slowly decrease with increasing $A_\mathrm{V}$.

\subsubsection{\ce{CH+} and \ce{CH}}
\begin{figure}[htb]
\resizebox{\hsize}{!}{\includegraphics{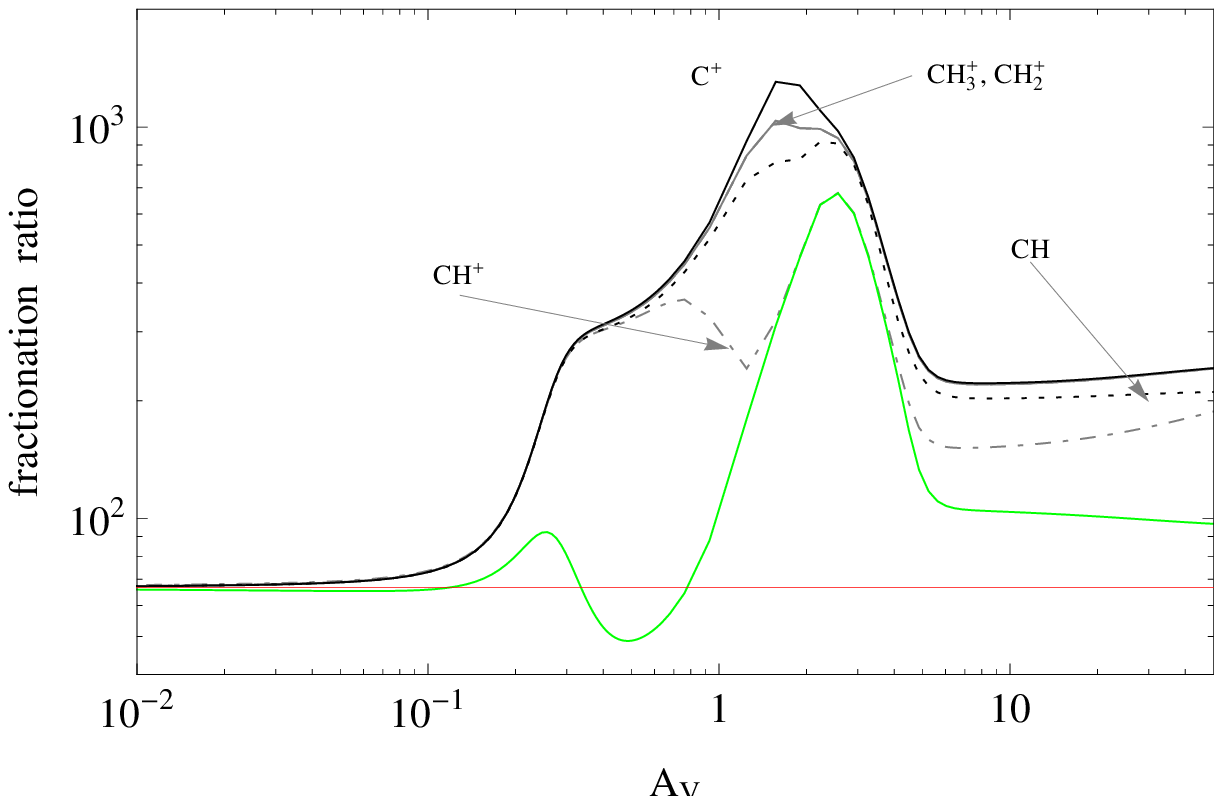}}
\caption{Fractionation structure of light hydrocarbons for $n_0=10^5\,\mathrm{cm}^{-3}$, $M=100\,M_\odot$, $\chi=10$ (green: \FR{}(\ce{C}), red \ER{}).}
\label{hydrocarbons}
\end{figure}

Figure~\ref{network} summarizes the dominant formation and destruction channels of the discussed hydrocarbons. The arrows denote the primary reaction channels for \ce{CH+ , CH2+}, and \ce{CH3+} (left panel) and for \ce{CH} (right panel) across the model clumps. Figure \ref{hydrocarbons} shows the \FR{} of \ce{C+ , CH+ , CH2+ , CH3+}, and \ce{CH} and Fig.~\ref{carbon-structure-2} shows the abundance profile of hydrocarbons and their respective isotopologues.

The reaction \ce{C+ + H2 +$\Delta$E -> CH+ + H} requires an activation energy of $\Delta$E=4600 K, but collisions with vibrationally excited \ce{H2^{*}} allow to overcome the barrier \citep{comparison07,agundez2010}. At the edge of the cloud, \ce{CH+} is primarily produced from \ce{C+} colliding with excited, molecular hydrogen. At $A_\mathrm{V}\sim 10^{-3}$, the proton exchange reaction \ce{C+ + CH -> CH+ + C} together with ionization of \ce{CH} become the main formation reactions. In those two regimes, the \FR{} is controlled by \ce{C+} (see also Fig. \ref{hydrocarbons}). At $A_\mathrm{V}\sim 1$ the main formation occurs via collisions of \ce{C} with \ce{HCO+} or \ce{H3+}.

Under very high density and very low $\chi$ conditions, \ce{C+} will be less abundant than \ce{C} throughout the clump. As a result the dip between the two peaks visible in the \FR{}(\ce{CH+}) in Fig.~\ref{hydrocarbons} becomes much more prominent and can reach values below \ER{}. This is visible in Fig.~\ref{SP-CH+}. 

The formation of \ce{CH} originates at \ce{C+}. Successive collisions with \ce{H2} form the chain: 
\begin{align*}
\cee{C^{+} &->[\ce{H2}] CH2+ ->[\ce{H2}] CH3+ ->[\ce{e-}] CH}
\end{align*}
 At the end of the reaction chain dissociative recombination leads to \ce{CH} and \ce{CH2}. At low $A_\mathrm{V}$, \ce{CH} can also be formed from \ce{C}, at high $A_\mathrm{V}$ it forms via dissociative recombination of \ce{HCS+}. The reaction paths to \ce{CH} (in order of cloud depths where they dominate) then are: 
\begin{align*}
\left.  
\begin{aligned}
\cee{
C^{+} ->[\ce{H_2^{*}}] &CH^{+} \\
C ->[\ce{H_2^{*}}] CH ->[\ce{C+}] &CH^{+} \\
C ->[\ce{H_2^{*}}] CH ->[h\nu] &CH^{+} \\
C ->[\ce{H_3^+}] &CH^{+}\\
C ->[\ce{HCO+}] &CH^{+} \\
}
\end{aligned} \right\}
\cee{
->[\ce{H_2}] & CH_2^{+}\\
C^{+} ->[\ce{H_2}] & CH2+ ->[\ce{H_2}] CH3+ ->[\ce{e-}] CH
}
\end{align*}

\begin{align*}
\cee{
C+ ->[\ce{HNC}] C2N+ ->[\ce{e-}] C2 ->[\ce{S+}] CS+ ->[\ce{H_2}] HCS+ ->[\ce{e-}] CH
}
\end{align*}
From Fig.~\ref{hydrocarbons} and from the above chain of reactions it is obvious, that the fractionation ratio of \ce{C+} will be handed down through the chain, unless other carbon species become involved. This is the case for \ce{CH}. At very low $A_\mathrm{V}$ where the formation via \ce{C + H2^{*}} is most important, and the \FR{} is closely related to \ce{C} which mostly equals the \ER{} under these conditions. Once the \FR{} of \ce{C+} starts increase, it will affect the fractionation of all the related \ce{CH_n^+} and consequently  that of \ce{CH}. 

Deeper in the cloud the same happens along a different chemical track. 
Recombination of \ce{CH3+} and of \ce{HCS+} (once $A_\mathrm{V}>3$) are the main formation reactions for \ce{CH}.
The chemical chain \ce{C+ , C2N+ , C2 , CS+ , HCS+ , CH} shares a common behavior of the \FR{} (see also Fig.~\ref{hydrocarbons}).  As a side remark we would like to emphasize this chain as a good example of how the chemical networks of different elements, sulphur and nitrogen in this case, are mixed. Consequently it is important to include both networks to correctly compute the carbon chemistry. 

\begin{figure*}
\includegraphics[width=17cm]{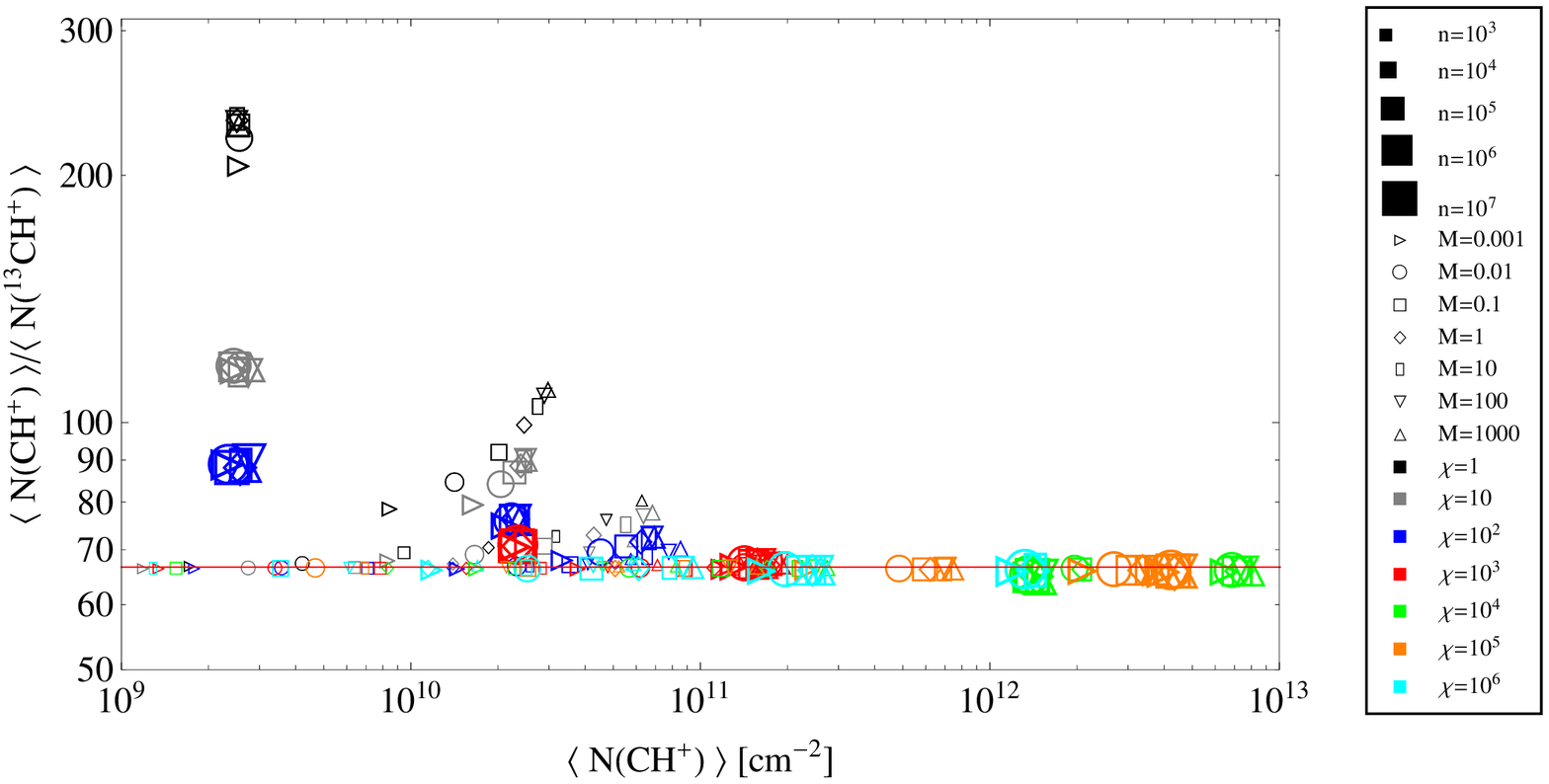}
\caption{Same as Fig. \ref{meancolco} for \ce{CH+}.}
\label{meancolch+}
\end{figure*}
\begin{figure*}
\includegraphics[width=17cm]{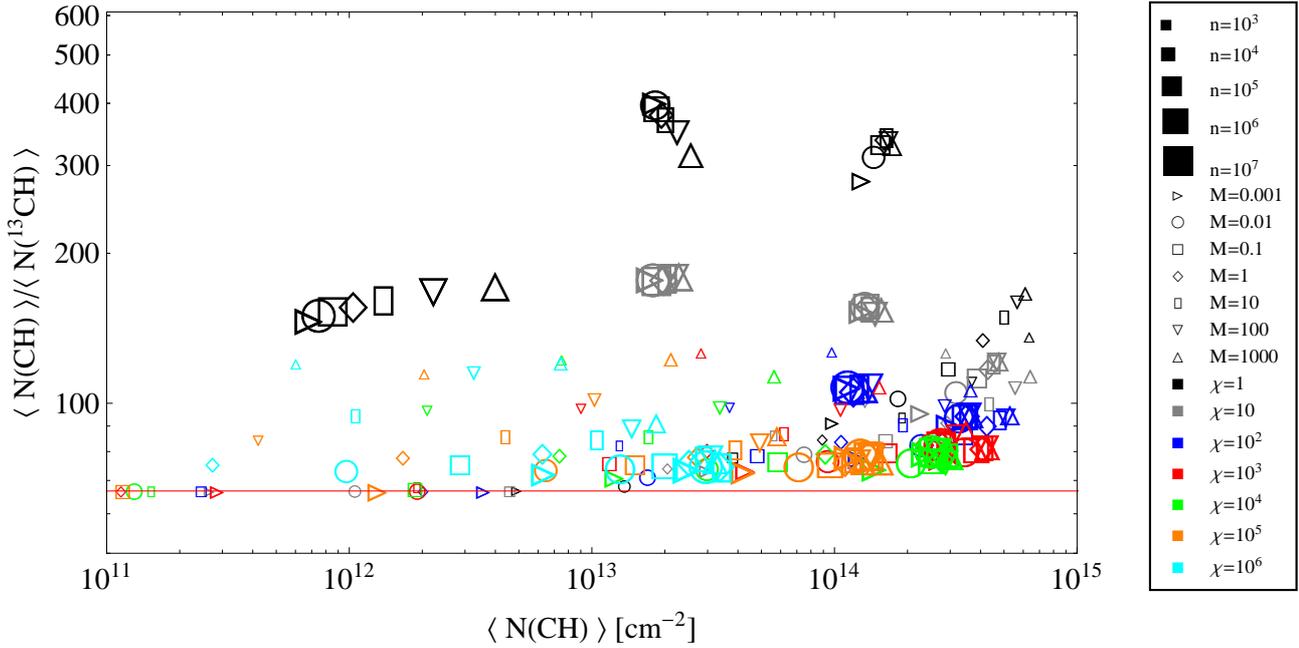}
\caption{Same as Fig. \ref{meancolco} for \ce{CH}.}
\label{meancolch}
\end{figure*}

In Figs.~\ref{meancolch+} and \ref{meancolch} we show the \FR{} of \ce{CH+} and \ce{CH} respectively.
The \FR{}(\ce{C+}) deviates from the \ER{} only at larger values of $A_\mathrm{V}$ and column densities of species whose \FR{} depends on \FR{}(\ce{C+}) will only be affected if these deeper cloud regions contribute significantly to their total column density. For \ce{CH+} this is the case for the same parameters where also \ce{C+} is fractionated and Fig.~\ref{meancolch+} shows their close chemical relationship. Deviations from the \ER{} occur only for the low $\chi$ models with sufficient total column densities, similar to \ce{C+} (see Fig.~\ref{meancolc+}).

The weak deviations of the \FR{}(\ce{CH+}) from the \ER{} is consistent with observations.  \citet{Centurion1995} found a mean value of the \ce{CH+}/\ce{^{13}CH+} column density ratio of 67$\pm$3 for five lines of sight, very close to the interstellar \ER{}. \citet{casassus2005} report an average ratio of 78$\pm$2 from  measurements along 9 lines of sight.  Recent absorption-line observations by \citet{ritchey2011} along 13 lines of sight through diffuse molecular clouds confirm a \FR{}(\ce{CH+}) close to the ambient \ER{}. They report total column densities of \ce{CH+} of a few $10^{13}$ cm$^{-2}$.

\ce{CH} remains abundant approximately until $A_\mathrm{V}$  approaches unity. As a consequence, a larger fraction of its total column density will be affected be the fractionation. This is visible in Fig.~\ref{meancolch}. Above a mean column density of $10^{13}$~cm$^{-2}$, all models show an enhanced \FR{}. Because of the strong coupling to \FR{}(\ce{C+}), weaker FUV models tend to have the strongest fractionation. Therefore, \ce{CH} promises to be a good observational fractionation tracer as it combines enhanced \FR{} with high column densities.

\subsubsection{\ce{HCO+}}
\begin{figure}[htb]
\includegraphics[width=8.5cm]{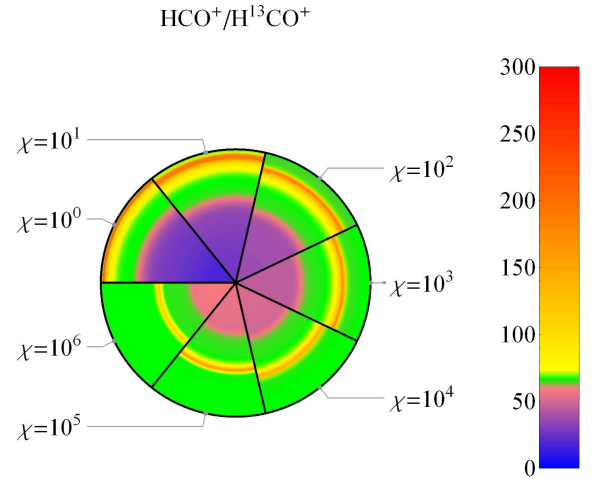}
\caption{Same as Fig.~\ref{SP-n30m00} for \ce{HCO+} for $n=10^5$~cm$^{-3}$ and $M=1$~M$_\odot$.} \label{SP-HCO+n50}
 \end{figure}

The fractionation of \ce{HCO+} is special because it is affected by two processes acting in opposite directions.
At very low $A_\mathrm{V}$ \ce{HCO+} is formed by \ce{H2} collision with \ce{HOC+} and \ce{CO+}. Both precursors are not fractionated, thus \FR{}(\ce{HCO+})$\approx$ \ER{} (see Sect.~\ref{highdensity}). 
A little deeper into the clump, the main formation reaction changes to \ce{CH + O -> HCO+ + e-}, thus its fractionation indirectly depends on reaction (\ref{13eq1}). \ce{CH} is strongly fractionated with \FR{} > \ER{} and passes down the fractionation to \ce{HCO+}. This fractionation peak is seen in Figs.~\ref{carbon-fractionation-1} and \ref{SP-HCO+n50}. With growing $\chi$ the peak is pushed to larger cloud depths. Once \ce{CO} is sufficiently abundant, the reaction \ce{H3+ + CO -> HCO+ + H2} takes over as dominant formation reaction and the \FR{} approaches that of \ce{CO} (compare with Fig.~\ref{SP-n50m00}, right panel). At even higher values of $A_\mathrm{V}$ the gas temperature becomes very low ($T\le 10$~K) and reaction (\ref{13eq2}) starts to dominate formation and destruction of \ce{HCO+} and pushes the \FR{} significantly below the \ER{}. In the appendix we show the \FR{} of \ce{HCO+} over a significant portion of our model grid. The central cloud regime with \FR{}<\ER{} is visible in all clumps that are sufficiently shielded from the external FUV radiation.
\begin{figure}[htb]
\resizebox{\hsize}{!}{\includegraphics{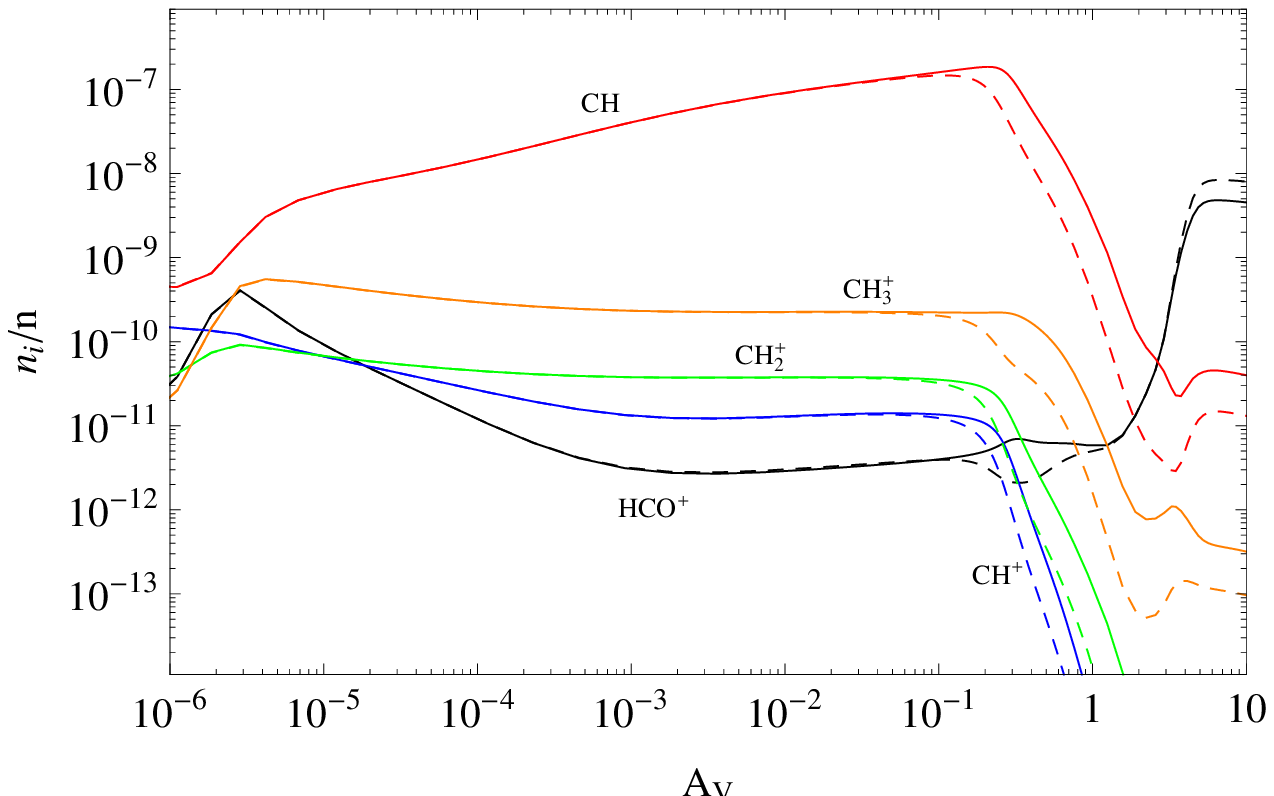}}
\caption{Same as Figure~\ref{carbon-structure-1} for species \ce{HCO+}, \ce{CH}, \ce{CH+}, \ce{CH2+}, and \ce{CH3+}.}
\label{carbon-structure-2}
\end{figure}

{In Fig.~\ref{meancolhco+} we show the column density fractionation ratio of \ce{HCO+}. Fractionation of \ce{HCO+} is strongest for clumps with large columns of cold gas, where reaction~(\ref{13eq2}) can contribute strongly to the total \ce{H^{13}CO+} abundance. Low mass and low density models have a \FR{} equal to the \ER{} or slightly higher.
For a given density, the \FR{} is largely independent of the model mass, which is consistent with the \ce{HCO+} and \ce{H^{13}CO+} density profiles shown in Fig.~\ref{carbon-structure-2} which show an increase for large $A_V$ and a roughly constant \FR{} (see also Fig.~\ref{carbon-fractionation-1}). Hence, the \FR{} of \ce{HCO+} is only marginally affected if the clump mass is increased.
The model results show a strong correlation of the column density ratio
with $\chi$.

\subsection{Emission line ratios}\label{intensitystruct}
Even though column densities are no direct observables,  they have to be derived from measured line strengths resulting from the full radiative transfer, including effects of a variable temperature of exciting collision partners and optical depths. 
However, the derivation of intensities requires numerous additional assumptions, such as collision rates, details of the geometry, assumptions on chemical and radiative pumping, and so on. Consequently, from a modellers perspective intensities  have larger uncertainties than column densities. Therefore we only compute intensities for a selected subset of commonly observed species and transitions.
Here,
we directly compare the isotopic ratio of clump averaged line intensities \citep[for definition see][]{roellig06} which can be compared directly to observations. We always assume unity beam filling factor. The large number of possible line combinations prohibits a complete presentation. We give just a few examples to demonstrate that line ratios between various isotopologues can differ from the corresponding column density ratios.  

\begin{figure*}
\includegraphics[width=17cm]{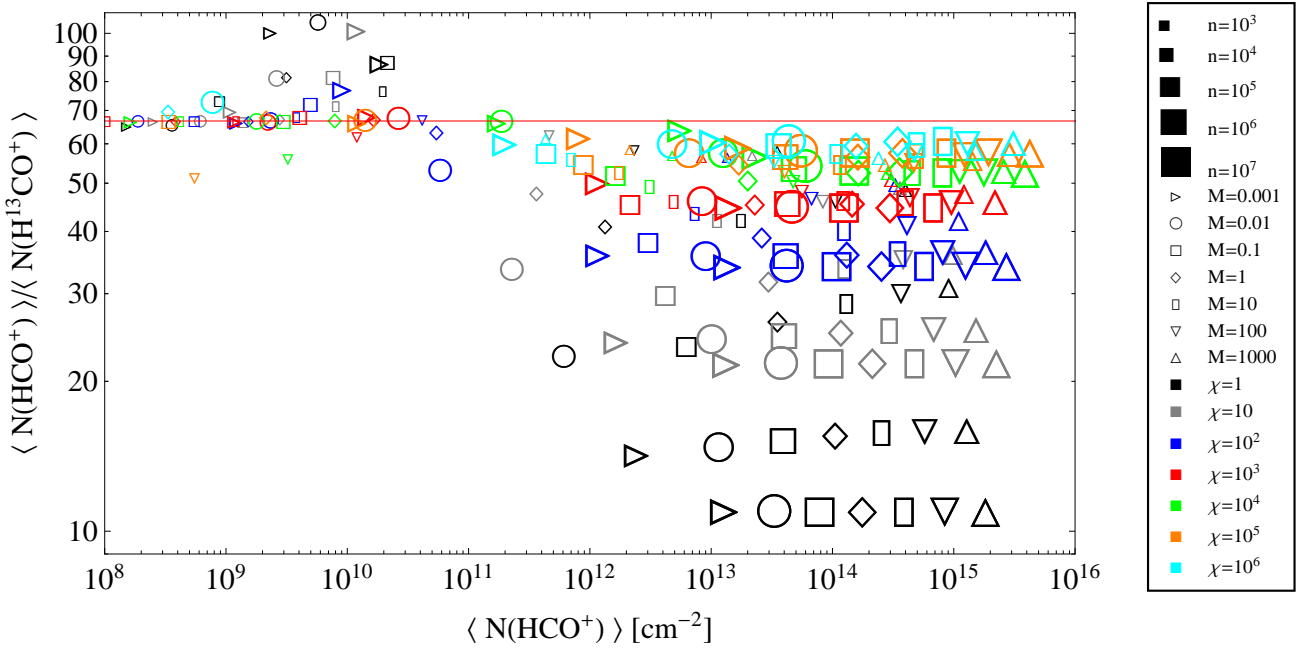}
\caption{Same as Fig. \ref{meancolco} for \ce{HCO+}.}
\label{meancolhco+}
\end{figure*}

\begin{figure*}
\includegraphics[width=17cm]{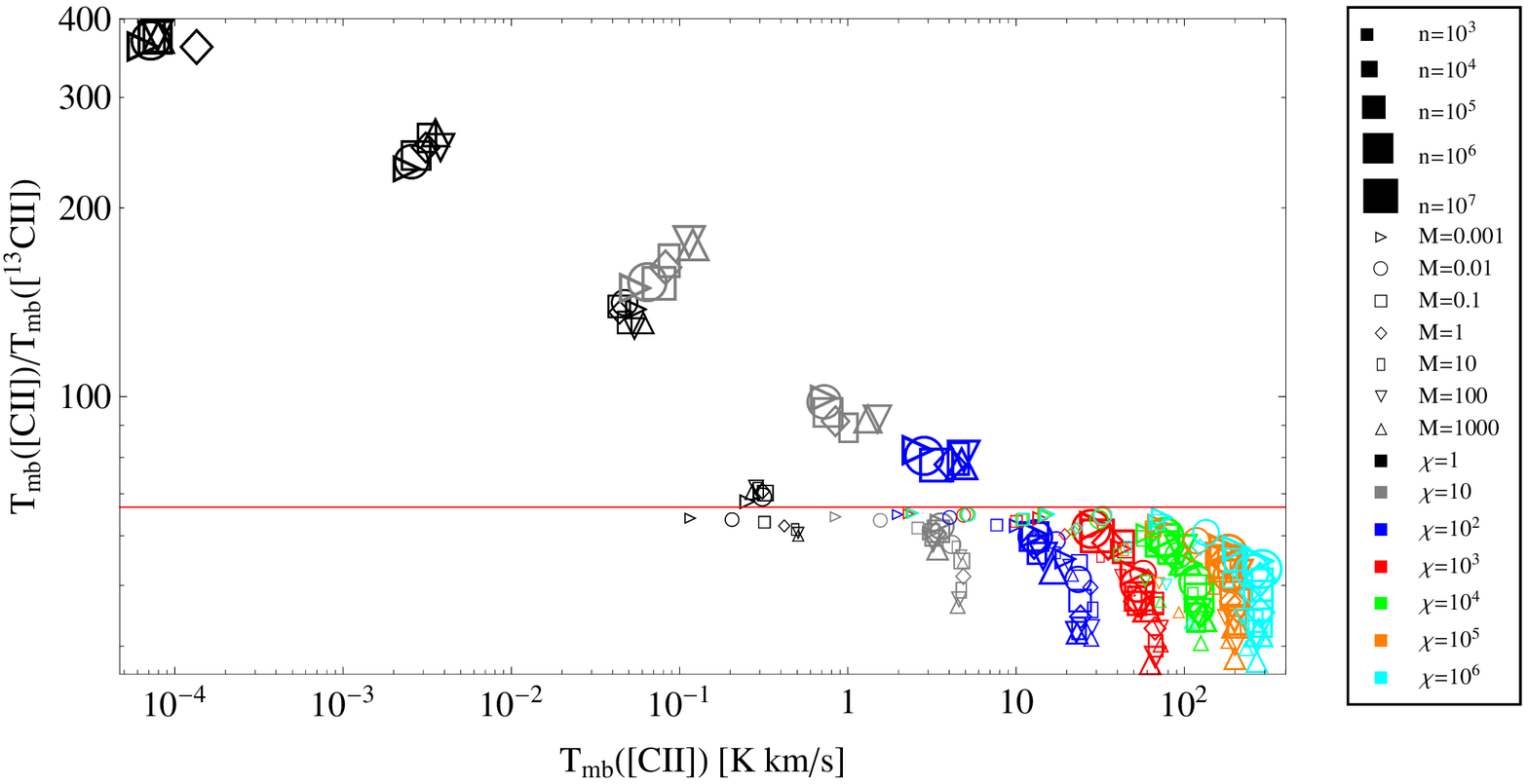}
\caption{Same as Fig. \ref{meancolco} for the intensity ratio  $T_\mathrm{mb}(\ce{[CII]})/T_\mathrm{mb}(\ce{[^{13}CII]})$ vs.$T_\mathrm{mb}(\ce{[CII]})$.}
\label{meanTmbc+}
\end{figure*}

\begin{figure*}
\includegraphics[width=17cm]{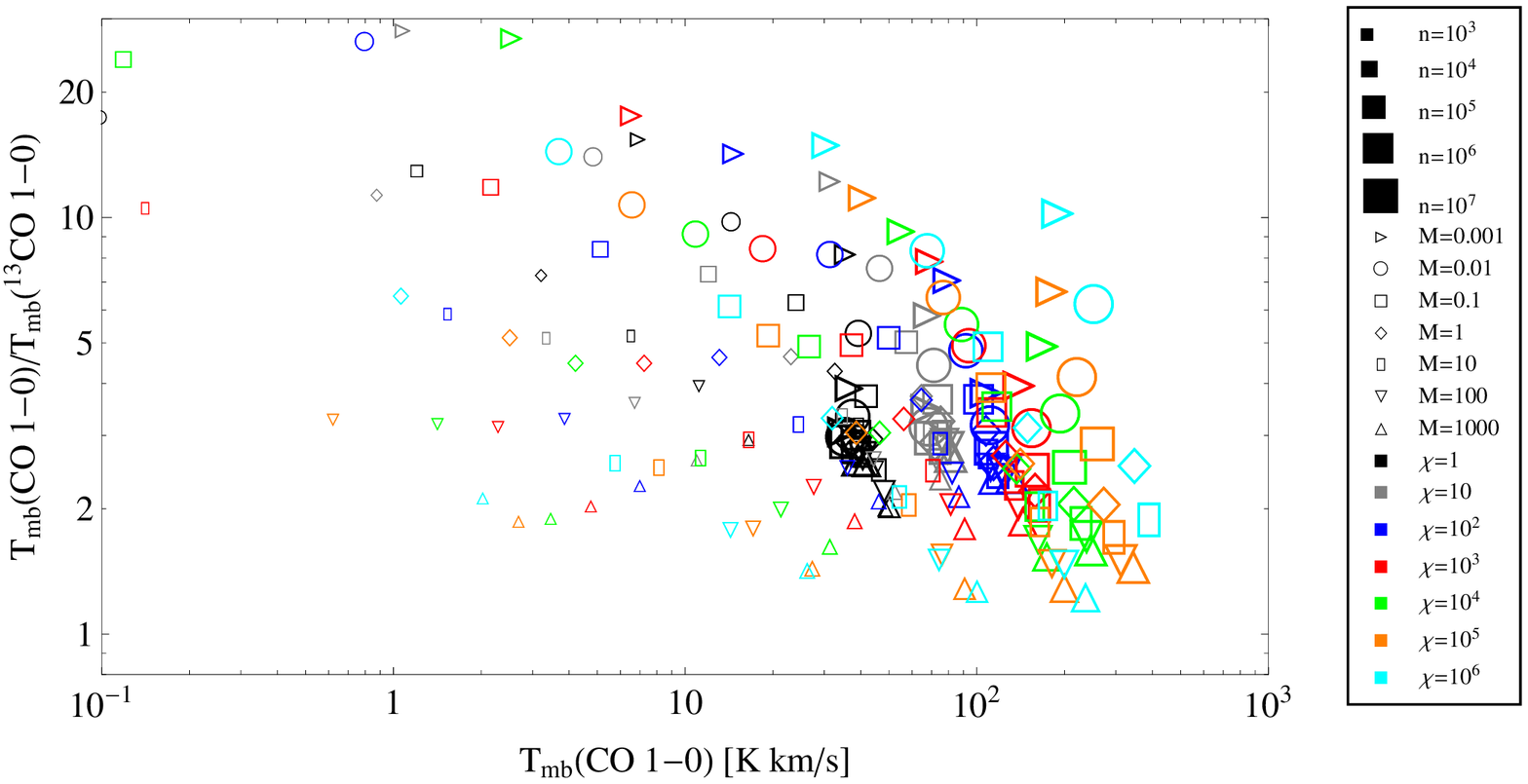}
\caption{Same as Fig. \ref{meanTmbc+} for $T_\mathrm{mb}(\ce{CO} (1-0))/T_\mathrm{mb}(\ce{^{13}CO} (1-0))$ vs. $T_\mathrm{mb}(\ce{CO} (1-0))$.}
\label{meantmbco}
\end{figure*}

Fig.~\ref{meanTmbc+} shows the ratio of $T_\mathrm{mb}(\ce{[CII]})/T_\mathrm{mb}(\ce{[^{13}CII]})$ with $T_\mathrm{mb}=\int T_\mathrm{mb}dv$ (For \ce{^{13}C+} we summed over all hyperfine components.). While the column density ratio is larger than the \ER{} for all model clumps the same is not true for the intensity ratio (\IR{}). All models with $\chi>100$ show a \IR{}<\ER{} down to values of 38. Only models with very low $\chi$ and large densities have a \IR{}>\ER{}. For a given density and mass, the \IR{} decreases with increasing $\chi$, because of the larger optical thickness of \ce{C+} relative to \ce{^{13}C+}. \ce{[^{13}CII]} observations of the Orion nebula by \citet{stacey1991:13CII} show a very similar behavior and are consistent with our model predictions. They report line ratios between 36 and 122 with the highest optical depths $\tau=3.3^{+1.1}_{-0.8}$  belonging to the lowest line ratio of $36\pm9$ and low optical depths where ratios are high.
\citet{BoreikoBetz1996} derive an intensity ratio of 46 in M42, consistent with an intrinsic \ER{} of $58^{+6}_{-5}$ and an optical depth of \ce{[CII]} of 1.3. 
 For a given $\chi$, the \IR{} drops with decreasing density and increasing mass. 
Increasing the mass for a given $\chi$ and $n$ will add cool, molecular mass. Provided that \ce{C+} is fractionated this will increase the \FR{} as long as the \ce{[^{12}CII]} line remains optically thin and decrease the \FR{} once it becomes optically thick (compare Fig.~\ref{meanTmbc+}).

In the previous section we showed, that the column density ratio of \ce{CO}/\ce{^{13}CO} is close to the \ER{} of the clump for most of the parameter space. \ce{CO} emission lines suffer much more from optical thickness effects than most other species. This is also shown in Fig.~\ref{meantmbco} where we plot the ratio of the integrated intensities of \ce{CO  (1-0)}/\ce{^{13}CO  (1-0)}. For most of the models, this ratio is between 1 and 10, much smaller than the \ER{}. Quite a few models have intensities of the rarer isotopologue comparable to those of the main species. 

The low mass, high density models show an increasing \IR{} with $\chi$,  opposite to the high mass, low density models. This is a result of the \ce{CO} abundance structure of these model (see Fig.~\ref{densestruct}). 
The lowest \IR{} is shown at the highest model mass model (\IR{} $\le 2$) while the highest \IR{} belongs to models with the lowest mass. Both can be explained as optical depth effect. Models with $n\ge 10^7$ cm$^{-3}$ have an increasing  $T_{mb}(\ce{CO  (1-0)})$ with $\chi$ and the \IR{} increases for the lower mass models and decreases for the higher mass models. The latter results from high optical thicknesses while the former results from  the emission of of very hot, strongly excited CO gas in the primarily ionized fraction of the clump.

In Fig.~\ref{emissratioj} we plot the intensity ratio of \ce{CO} lines  \mbox{$T_\mathrm{mb}(\ce{CO} (J\rightarrow J-1)) / T_\mathrm{mb}(\ce{^{13}CO} (J\rightarrow J-1))$} as a function of $J$ 
for models with $M=10_\odot$ and line intensities $>0.01$~\Kkms. For the lowest transition, the ratio lies between 1 and 10 and approaches \ER{} for high values of $J$, when both lines become optically thin. The $J$ where the ratio reaches the \ER{} increases with $n$ and $\chi$. For instance, models with $n=10^7$~cm$^{-3}$ and $\chi=10^6$ have a ratio of close to unity, until $J>10$ and reaches \ER{} at $J \apprge 25$, while models with the same density and $\chi=10^3$ reach \ER{} already at $J=15$.
We find that either the \ce{CO} lines are optically thick so that the \IR{} is lowered below the \ER{} or the \ce{^{13}CO} is too weak to be detectable. Only in dense clouds  the \ER{} is observable. The \ce{CO} \IR{} is therefore not a good diagnostics of carbon fractionation.

\begin{figure}[htb]
\resizebox{\hsize}{!}{\includegraphics{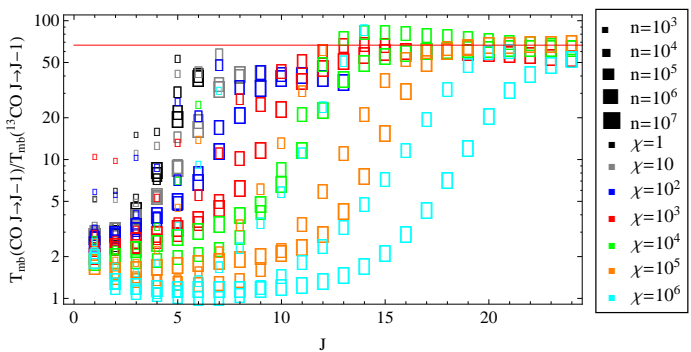}}
\caption{Emission line ratio $T_\mathrm{mb}(\ce{CO} (J\rightarrow J-1)) / T_\mathrm{mb}(\ce{^{13}CO} (J\rightarrow J-1))$ vs. $J$ for all models with $M=10\,\mathrm{M}_\odot$. Models with line intensities $<0.01$~\Kkms have been omitted. }
\label{emissratioj}
\end{figure}
\subsection{Diagnostics}\label{diagnostics}

Finally, we studied several line ratios with respect to their diagnostic value for the local \FR{} of \ce{CO} and \ce{C+} as well as to the local \ER{}. We already concluded 
 that \ce{CH} and \ce{HCO+} appear to be sensitive tracers of the \FR{}. Lacking collision rate coefficients for \ce{CH} we only calculated \ce{HCO+} (and \ce{H^{13}CO+}) intensities for our model clouds. 

We selected transitions that are observable from the ground or through SOFIA (Stratospheric Observatory For Infrared Astronomy) and show at most moderate optical depths of a few. Figure~\ref{diagfull} demonstrates how the emission line ratio  $T_\mathrm{mb}(\ce{[CII]})/T_\mathrm{mb}(\ce{H^{13}CO^+} (1-0))$ traces the column density \FR{} of \ce{C+}, \ce{C} and \ce{CO}.  
The figure reproduces the general trend of the \FR{} as discussed in the previous sections. 
We find a clear dependence of the emission line ratios on the column density \FR{} of \ce{C} and \ce{C+}, but not of \ce{CO}.
However, it is only partially applicable to observational data. 
All ratios above $\sim30000$ are practically not observable as they correspond to \ce{H^{13}CO+} intensities below 10 mK km/s.
If we exclude models with line intensities < 0.01 \Kkms  
the column density ratios of \ce{CO} and \ce{C+} show much weaker variations 
so that the emission line ratio only traces the fractionation of atomic carbon in the observable intensity range. We have repeated this exercise for many other line ratios
that are available through ground based and satellite observatories, and that are strong enough to be observable\footnote{Even though the PDR model takes optical thickness effects into account we tried, whenever possible, to find optically thin tracers in order to be as independent as possible from the local structure of the emitting source.}.
All three panels in Fig.~\ref{diagfull} show a transition around an emission ratio of $10^2-10^3$. This ratio corresponds to models with the highest $T_\mathrm{mb}(\ce{H^{13}CO^+} (1-0))\approx 1-2$~\Kkms and \FR{(\ce{C})} slightly below the \ER{}. It requires $\chi=10^4-10^5$
and densities of $n=10^7$ to reach these high values. Towards lower intensities, the models split into two branches: very high density models with $n=10^7$~cm$^{-3}$ and decreasing \FR{(\ce{C})}, and models with $n=10^6$~cm$^{-3}$, $\chi\ge100$ and increasing \FR{(\ce{C})}. This parameter regime corresponds to model clumps with the highest \ce{[CII]} emission, increasing with $n$ and $\chi$. Accordingly, the \ce{[CII]}/\ce{H^{13}CO+} intensity ratio can increase to very high values if the density and the FUV field are high enough, albeit with the visible transition. 

\begin{figure*}[htb]
 \centering
 \includegraphics[width=5.5cm]{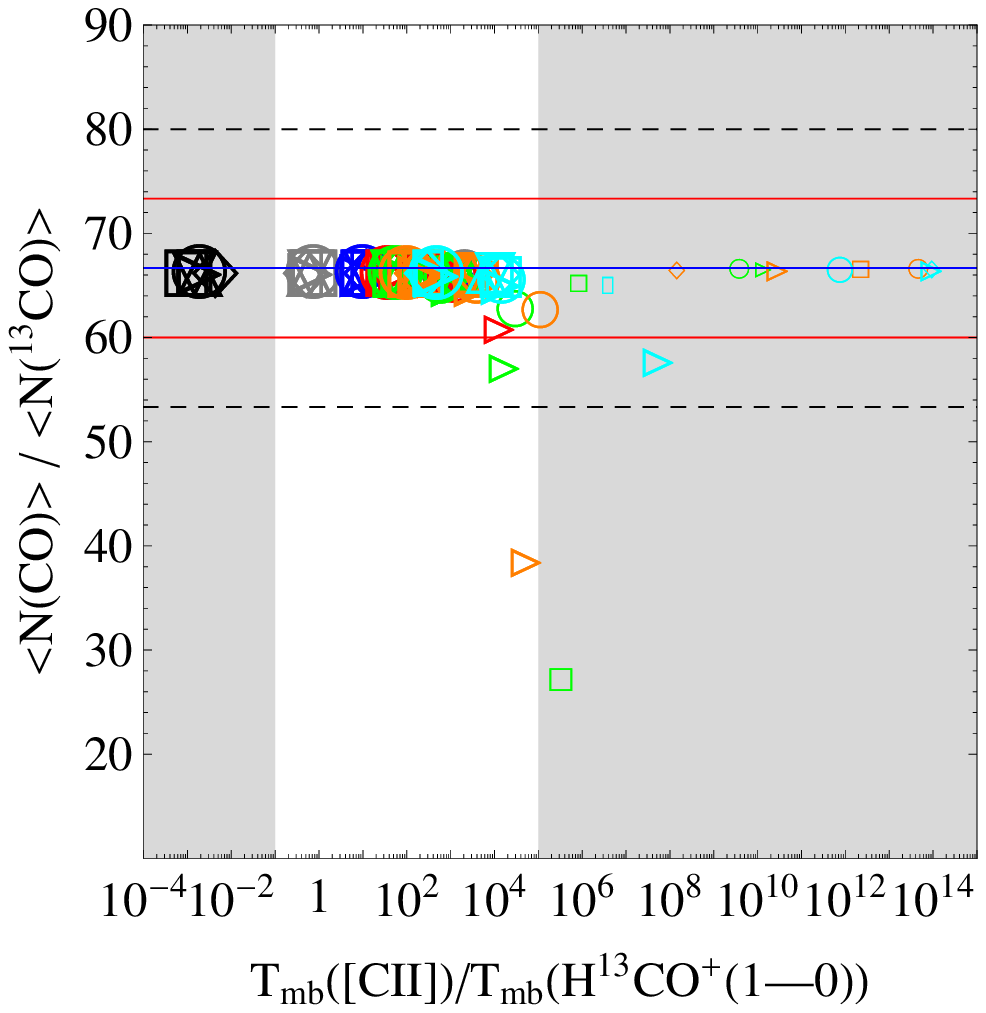}
\hfill
 \includegraphics[width=5.5cm]{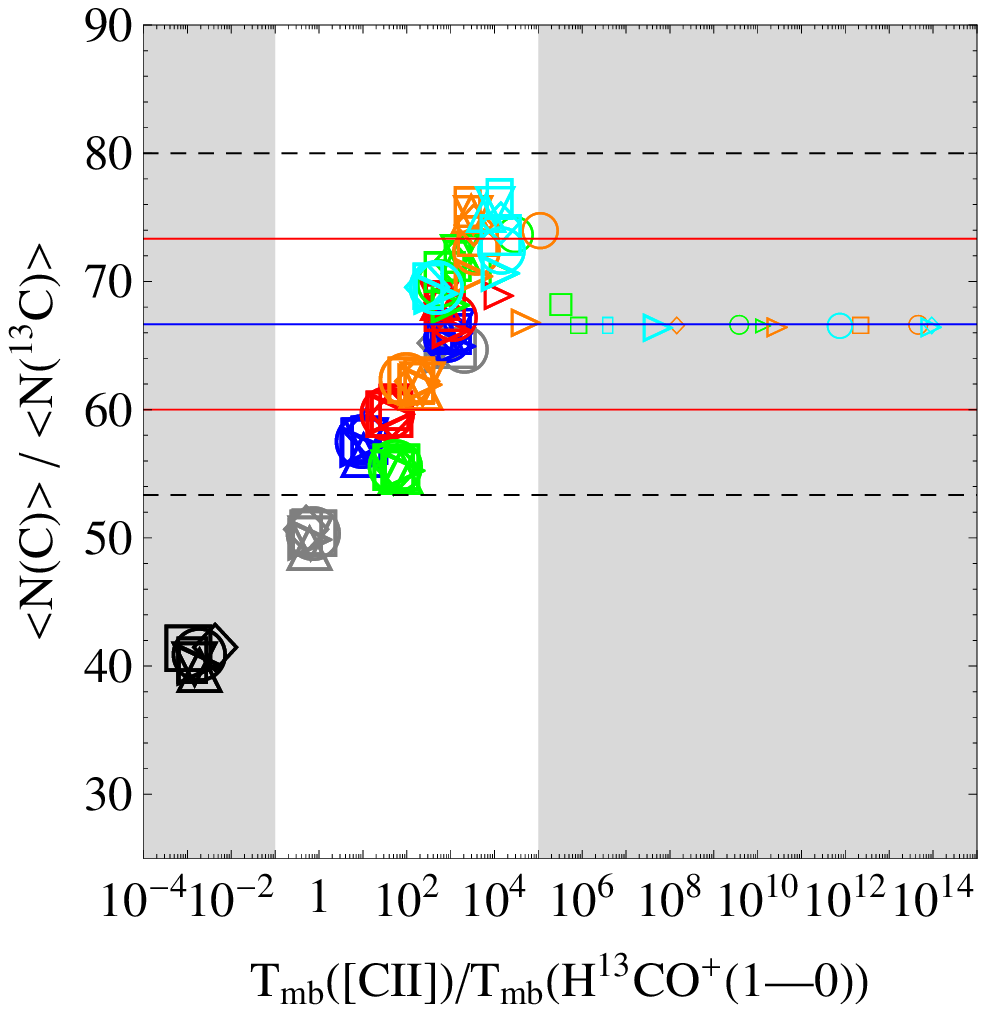}
\hfill
\includegraphics[width=5.5cm]{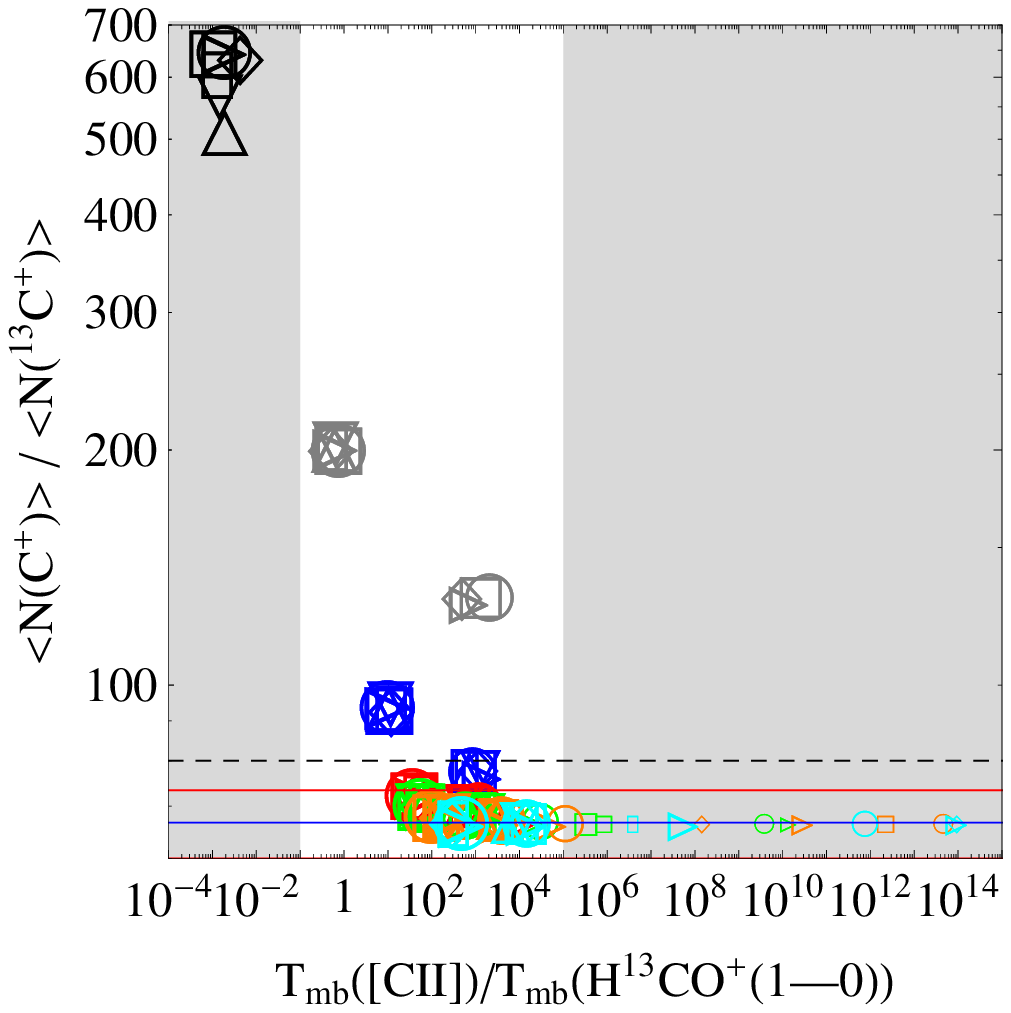}

\caption{Column density fractionation ratios plotted against the emission line ratio  $T_\mathrm{mb}(\ce{[CII]})/T_\mathrm{mb}(\ce{H^{13}CO^+} (1-0))$ for all models from our parameter space (blue: \ER{}, red: \ER{}$\pm 10\%$, dashed: \ER{}$\pm 20\%$). The symbols follow the coding from Fig.~\ref{meancolco}. The gray areas denote the regime below a sensitivity limit of $T_\mathrm{mb}<0.01$~\Kkms. {\bf Left panel:} $N(\ce{CO})/N(\ce{^{13}CO})$, {\bf middle panel:} $N(\ce{C})/N(\ce{^{13}C})$, 
 {\bf right panel:} $N(\ce{C+})/N(\ce{^{13}C+})$. 
  } \label{diagfull}
 \end{figure*}
 
\begin{figure*}[htb]
 \centering
 \includegraphics[width=5.5cm]{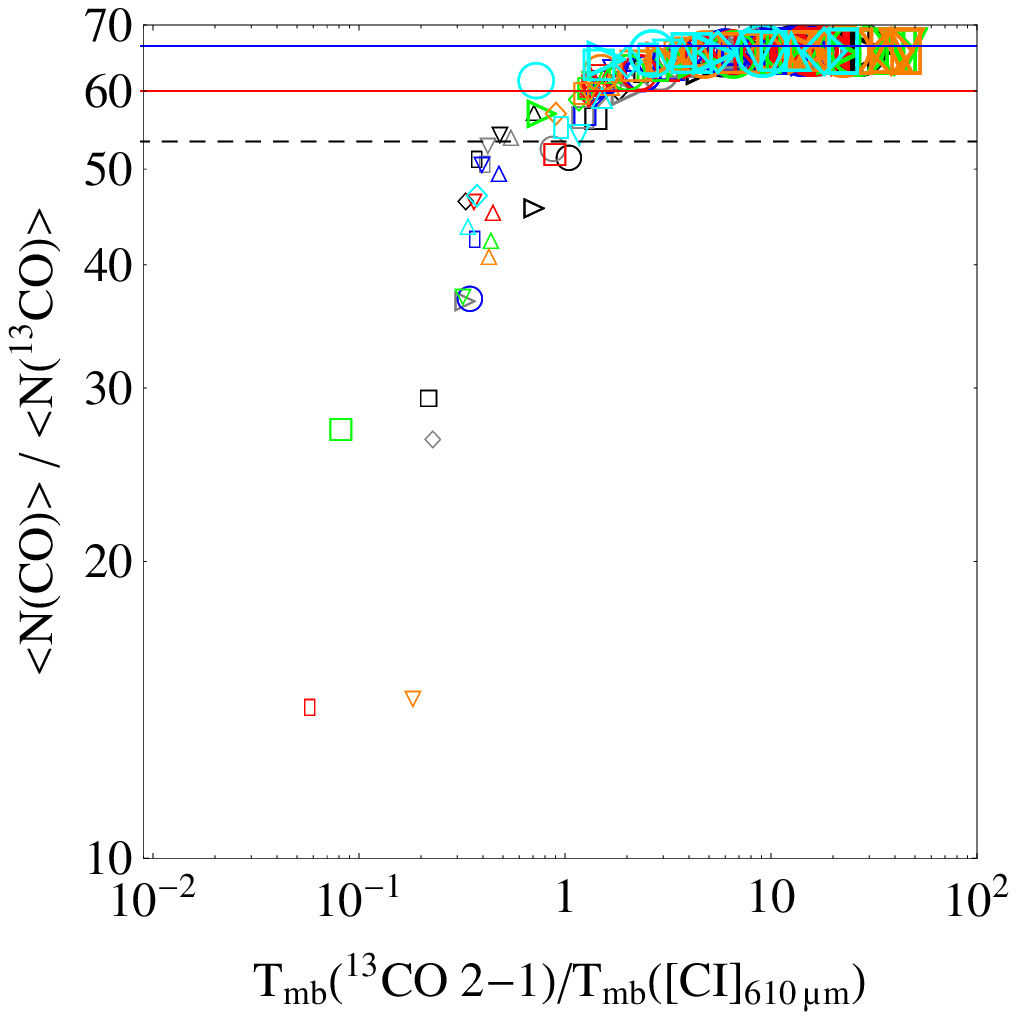}
\hfill
 \includegraphics[width=5.5cm]{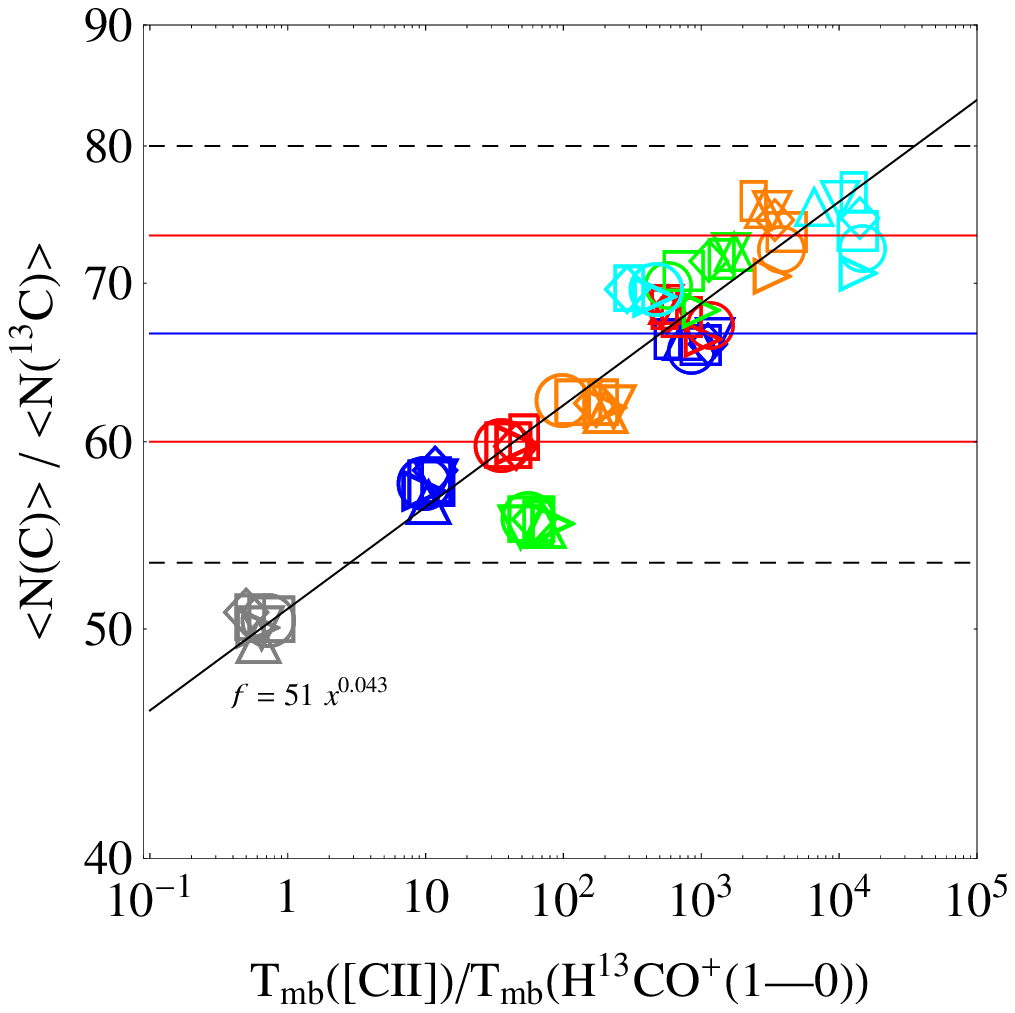}
\hfill
 \includegraphics[width=5.5cm]{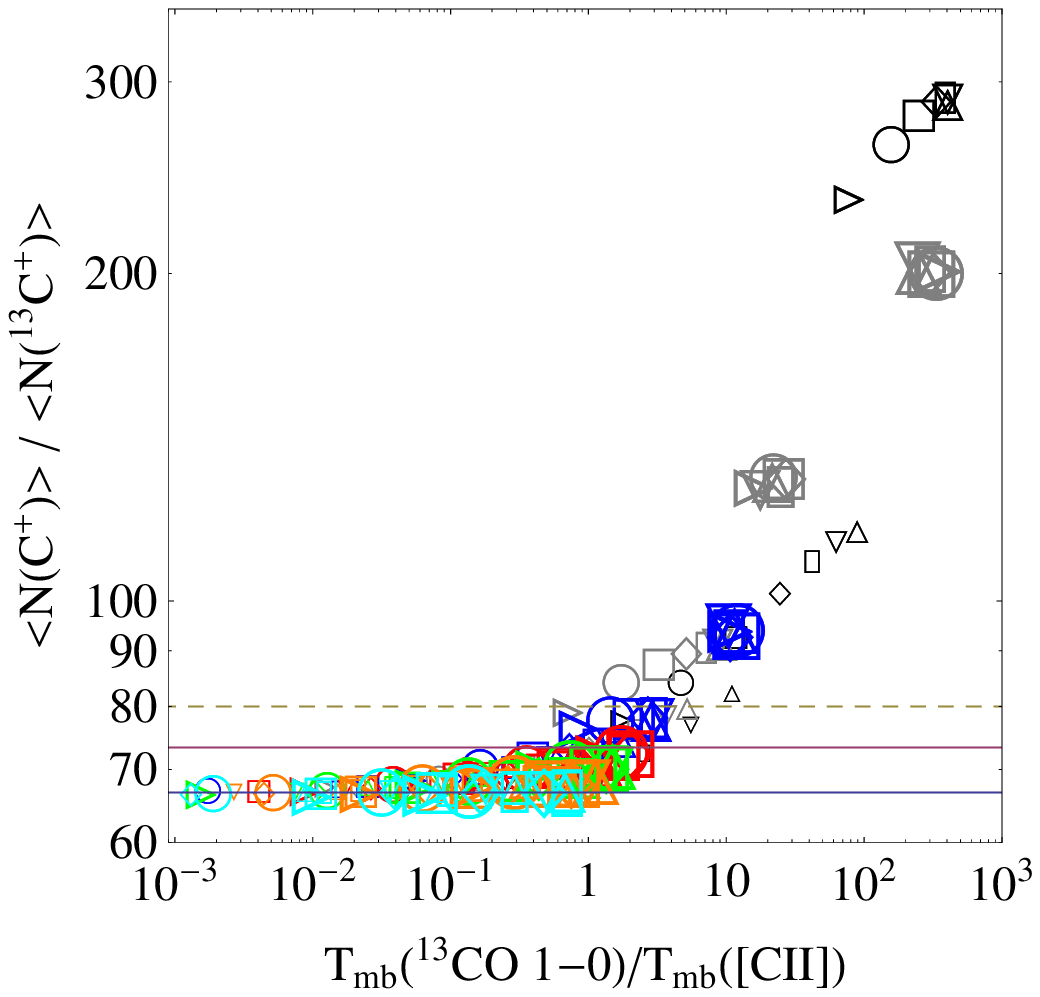}
\caption{Column density fractionation ratios plotted against emission line ratios for all models from our parameter space. (blue: \ER{}, red: \ER{}$\pm 10\%$, dashed: \ER{}$\pm 20\%$). Models with line intensities $<0.01$~\Kkms have been omitted. The symbols follow the coding from Fig.~\ref{meancolco}. {\bf Left panel:} $N(\ce{CO})/N(\ce{^{13}CO})$ vs. $T_\mathrm{mb}(\ce{^{13}CO} (2-1))/T_\mathrm{mb}(\ce{[CI] 610$µ$m})$, {\bf middle panel:} $N(\ce{C})/N(\ce{^{13}C})$ vs. $T_\mathrm{mb}(\ce{[CII]})/T_\mathrm{mb}(\ce{H^{13}CO^+} (1-0))$,
{\bf right panel:} $N(\ce{C+})/N(\ce{^{13}C+})$ vs. $T_\mathrm{mb}(\ce{^{13}CO} (3-2))/T_\mathrm{mb}(\ce{[CII]})$.
  } \label{diag1}
 \end{figure*}
 
Fig.~\ref{diag1} shows the result for the three ratios that seem to be suitable to trace the column density \FR{} of \ce{C+}, \ce{C}, and \ce{CO} individually.
The left panel shows $T_\mathrm{mb}(\ce{^{13}CO} (2-1))/T_\mathrm{mb}(\ce{[CI]_{610\mu m}})$ on the x-axis and the \ce{CO} column density ratio on the y-axis.  
The level energy of \ce{^{13}CO} (2-1) is 16~K, while the \ce{[CI]} line requires 24~K for excitation. 
Looking at the plot, we note, that any line ratio < 1 signals significantly fractionated \ce{CO} column densities. Any emission line ratio higher than a few reflects a normal \FR{} in the \ce{CO} gas. All models with an emission line ratio smaller than $\approx 1$ have $n\le 10^4$~cm$^{-3}$ and will consequently host a small \ce{^{13}CO} population but significant amounts of atomic carbon.

The middle panel  plots the  column density fractionation ratio of atomic carbon  versus $T_\mathrm{mb}(\ce{[CII]})/T_\mathrm{mb}(\ce{H^{13}CO^+} (1-0))$ (same as in Fig.~\ref{diagfull}, but excluding all model lines too weak to be detectable). The model results show a strong correlation between the \FR{} and the emission line ratio. The black line in the plot corresponds to a least-squares-fit $f=51 x^{0.043}$ with the emission line ratio $x$. 
 The \ce{[CII]} emission is strongest for low to intermediate densities while \ce{HCO+} is a typical density tracer. Both lines are typical PDR tracers, sensitive to significant FUV illumination.
Consequently, a line ratio higher than $10^3$ signals \FR{}>\ER{}. The corresponding points in the plot belong to models with $n\approx 10^6$~cm$^{-3}$ and $\chi\ge 10^4$. The points with a \FR{}<\ER{} correspond to models with even higher densities and somewhat lower $\chi$.

The panel on the right side shows the \ce{C+} column density ratio plotted against $T_\mathrm{mb}(\ce{^{13}CO} (1-0))/T_\mathrm{mb}(\ce{[CII]})$. All models with a line ratio higher than 2 contain fractionated \ce{C+}. We saw in Fig.~\ref{meancolc+} that the \FR{(\ce{C+})} is strongest for low UV models with high densities. These models have weak \ce{C+} emission, usually smaller than a few \Kkms, while these conditions favor \ce{^{13}CO} emission resulting in large line ratios.

\section{Summary}
We present an update of the isotope chemistry used in our PDR model code \ktau. An automated routine was created to allow for the inclusion of isotope reactions into the chemical database files that are used in numerical PDR computations. This is combined with a proper rescaling of the new isotope reactions. We computed  a large parameter grid of spherical PDR model clumps and investigated the effect of the isotope chemistry, particularly that of the isotope exchange reaction~(\ref{13eq1}), on the chemical structure of the model clumps as well as on their emission characteristics.

In the transition from ionized carbon to carbon monoxide the fractionation ratio of \ce{C+} is always larger than the elemental ratio in the gas. Strong \ce{C+} fractionation is possible in cool \ce{C+} gas. However, this is only partly visible in the corresponding intensity ratios of the model clumps. Optical thickness and excitation effects produce intensity ratios between 40 and 400, strongly dependant on the model parameters. 

In the dense ($n\ge10^3\,\mathrm{cm}^{-3}$) gas, \ce{CO} behaves differently and is never found with a fractionation ratio larger than the element ratio with the exception of a very limited $A_\mathrm{V}$ range under very special parameter conditions. 
In the diffuse gas \citet{liszt07} found qualitatively different behavior for $n<10^2$~cm$^{-3}$ and $M>10^3$~M$_\odot$.
It turns out that isotope-selective photo-dissociation, the major process able to produce a \FR{}>\ER{}, is always dominated by the chemistry in the denser PDR gas. This also affects the depth at which the transition from \ce{C+} to \ce{CO} occurs. The formation and destruction of \ce{^{13}CO} is much stronger controlled by reaction~(\ref{13eq1}). A direct consequence is that in all models in our grid \ce{^{13}CO} is formed at smaller $A_V$ than \ce{CO} despite the weaker shielding capabilities of the rarer isotopologue. 

The fractionation of other species can be understood in terms of their formation history. If their major formation channel originates from \ce{C+} their \FR{} is related to \FR{(\ce{C+})}. This is the case for many light hydrides, especially \ce{CH}, \ce{ CH2+}, and \ce{CH3+}. If the \FR{} of the parental species is controlled by other reactions than (\ref{13eq1}) the behavior might change. Atomic carbon is a mixed case with a regime at lower $A_\mathrm{V}$ where formation occurs mainly via recombination of \ce{C+} and a regime where it is chemically derived from \ce{CO} and \ce{HCO+}. Consequently, the \FR{(\ce{C})} exhibits a mixture of both cases. At particular depths, \ce{CH+} might be formed from \ce{C}. If that is the case the relation of its \FR{} to \FR{(\ce{C+})} breaks down until formation via a \ce{C+} route takes over again. 

Our computations have shown that \ce{CH} and \ce{HCO+} may be very sensitive tracers for chemical fractionation in PDRs. \ce{CH} amplifies the \FR{} of \ce{C+} and \ce{HCO+} amplifies the \FR{} from \ce{CO} in the CT. Both are abundant in the region where chemical fractionation plays a big role and suffer less from optical depth effects than their chemical progenitors.

We demonstrated that fractionation of the local densities not necessarily transforms into a fractionated column density. \ce{C+} is a prominent example. Fractionation of ionized carbon only takes part in cool \ce{C+} gas which in most clouds only makes up a small fraction of the total gas column. Only low FUV models are able to produce larger columns of cool \ce{C+} and will have a fractionated column density. \ce{CO} behaves oppositely in that it requires warm \ce{CO} in order to become fractionated. Again, this affects the column density only under certain conditions.

Finally, we provide  diagnostics for  the fractionation status of \ce{C+}, \ce{C} and \ce{CO} through suitable emission line ratios. 
We showed that a line ratio of $T_\mathrm{mb}(\ce{^{13}CO} (2-1))/T_\mathrm{mb}(\ce{[CI]_{610\mu m}})<1$ signals a significant fractionation of the \ce{CO} column density. 
The line ratio of $T_\mathrm{mb}(\ce{[CI]_{610\mu m}})/T_\mathrm{mb}(\ce{H^{13}CO^+} (1-0))$ has a power law dependence on the fractionation ratio of the column density of atomic carbon. The column density fractionation of ionized carbon is reflected in $T_\mathrm{mb}(\ce{^{13}CO} (1-0))/T_\mathrm{mb}(\ce{[CII]})>2$.

\begin{acknowledgements}
We acknowledge the
use of OSU (\url{http://www.physics.ohio-state.edu/~eric/research.html}),
KIDA (KInetic Database for Astrochemistry (\url{http://kida.obs.u-bordeaux1.fr}),
 and UDFA
(\url{http://www.udfa.net/}) chemical reaction data bases. This work was supported by the German \textit{Deut\-sche For\-schungs\-ge\-mein\-schaft, DFG\/} project number Os~177/1--1 as well as within the Collaborative Research Council 956, sub-project C1, funded by the \textit{DFG}. 

\end{acknowledgements}

\bibliographystyle{aa}
\bibliography{ref}
\Online
\begin{appendix}
\section{Isotopization rules\label{appendix:isotopization}}

Usually, only the main isotopes are considered in astrochemical databases, despite the fact that many isotopologues have been detected in astronomical observations so far. 
We will describe here our efforts to include  \ce{^{13}C} and \ce{^{18}O} isotopes into the chemical database. However,
in this paper we will only discuss the scientific implications of carbon fractionation not going into the details of the \ce{^{18}O} chemistry.

Usually, in a reaction formula like \ce{H + O2 -> OH + O} it is not possible to identify which oxygen atom binds with hydrogen. Both atoms in \ce{O2} are indistinguishable. This changes by including an isotope into a molecule. Now both atoms in \ce{O ^{18}O} are distinguishable and we get:
\begin{align*}
\cee{H + O ^{18}O &-> OH + ^{18}O\\
H + O ^{18}O &-> ^{18}OH +O}
\end{align*}

Unfortunately, this becomes more complicated if the isotope can be placed in more than one spot. The main isotope reaction
\begin{align*}  
\cee{ C       +        HCO+     &->       CO     +         CH+}
\end{align*}
splits into 5 isotopic reactions
\begin{align*}
\cee{ C       +        HC^{18}O+   &->       C^{18}O   +         CH+\\
      C       +        H ^{13}CO+   &->       ^{13}CO   +         CH+\\
      C       +        H^{13}C^{18}O+ &->       ^{13}C^{18}O +         CH+\\ 
      ^{13}C     +        HCO+     &->       CO     +         ^{13}CH+\\
     ^{13}C     +        HC^{18}O+   &->       C^{18}O   +        ^{13}CH+}
 \end{align*}
taking into account that the \ce{C=O} binding is preserved (again we omit the isotopic superscript when denoting the main isotope).
For more complex species the above scheme becomes much more complicated. Furthermore, the number of additional reactions is so large that it becomes difficult to perform the isotopization by hand, especially if one plans to update the chemical network regularly and manual isotopization is quite error-prone. We developed a software routine to automatically implement isotopic reactions into a given reaction set\footnote{We realized the isotopization routine in Mathematica\copyright  by Wolfram Research.}. 
The routine features are:
\begin{itemize}
\item inclusion of a single \ce{^{13}C} and a single \ce{^{18}O} isotope (multiple isotopizations are neglected in this study)
\item UDfA often does not give structural information, for instance \ce{C2H3} does not distinguish between linear and circular configurations (\ce{l-C2H3} and  \ce{c-C2H3}). In such cases we consider all carbon atoms (denoted by \ce{C_{n}}) as indistinguishable\footnote{Notation remark: isotopization of \ce{C_{n}} leads to \ce{C_{n-1} ^{13}C} , e.g. \ce{C2 -> C ^{13}C}. We always quote the full set of new reaction in the isotopic network derived from one UDfA reaction }, i.e.
\begin{align*}
\cee{C + C3H2 &-> C4H + H\\
C + C2 ^{13}CH2 &-> C3 ^{13}CH + H\\
^{13}C + C3H2 &-> C3 ^{13}CH + H  .}
\end{align*} 
However, if structure information is provided we account for each possible isotopologue individually:
\begin{align*}
\cee{C + H2CCC &-> C4H + H\\
C + H2CC ^{13}C &-> C3 ^{13}CH + H\\
C + H2C ^{13}CC &-> C3 ^{13}CH + H\\
C + H2 ^{13}CCC &-> C3 ^{13}CH + H\\
^{13}C + H2CCC &-> C3 ^{13}CH + H}
\end{align*}
\item molecular symmetries are preserved, i.e. \ce{NC^{13}CN} = \ce{N^{13}CCN}, but \ce{HC^{18}OOH} $\ne$ \ce{HCO^{18}OH} 
\item functional groups like \ce{CH_{n}} are preserved \citep[see also][]{woods09}, e.g.:
\begin{align*}
\cee{CH2CN+ + e- &-> CN + CH2\\
CH2 ^{13}CN+ + e- &-> ^{13}CN + CH2\\
^{13}CH2CN+ + e- &-> CN + ^{13}CH2}
\end{align*}
but not:
\begin{align*}
\cee{CH2 ^{13}CN+ + e- &-> CN + ^{13}CH2}
\end{align*} 
\item when the above assumptions are in conflict to each other we assume {\it minimal scrambling}, i.e. we choose reactions such, that the fewest possible number of particles switch partners. For example:
\begin{align*}
\cee{ CH3OH + C3H+ &-> HC3O+ + CH4 \\
 CH3OH + C2 ^{13}CH+  &-> HC2 ^{13}CO+ + CH4 \\
 CH3 ^{18}OH + C3H+  &-> HC3 ^{18}O+ + CH4 \\
 CH3 ^{18}OH + C2 ^{13}CH+  &-> HC2 ^{13}C ^{18}O+ + CH4\\ 
 ^{13}CH3OH + C3H+  &-> HC3O+ + ^{13}CH4 \\
 ^{13}CH3 ^{18}OH + C3H+  &->  HC3 ^{18}O+ + ^{13}CH4 }
\end{align*}  
but not 
\begin{align*}
\cee{ ^{13}CH3OH + C3H+ &-> HC2 ^{13}CO+ + CH4 }
\end{align*}
which would preserve the \ce{^{13}C=O} bond but would require 5 particles to switch partners. 
\item we favor proton/H transfer over transfer of heavier atoms
\item we favor destruction of weaker bonds, e.g.:
\begin{align*}
\cee{ C+ + OCS &->  CS+ + CO \\
 C+ + O ^{13}CS  &-> CS+ + ^{13}CO\\ 
 C+ + ^{18}OCS &->  CS+ + C ^{18}O \\
 C+ + ^{18}O ^{13}CS &->  CS+ + ^{13}C ^{18}O\\ 
 ^{13}C+ + OCS &->  ^{13}CS+ + CO \\
 ^{13}C+ + ^{18}OCS &->  ^{13}CS+ + C ^{18}O }
\end{align*}
but not:
\begin{align*}
\cee{ C+ + O ^{13}CS  &->  ^{13}CS+ +CO }
\end{align*}
\end{itemize}  
In the above example the binding enthalpies in \ce{O=C=S} are
745 kJ/mol for \ce{O=C} and 536kJ/mol \ce{C=S}. We only allow reactions, that break the weaker bond. In other words, we preserve \ce{O=C} bounds above others, since it is the strongest double bound. 
Please note that this is a strong assumption that can be sacrificed if more detailed knowledge on the reaction kinetics is at hand.

\section{Rescaling of reaction rates\label{appendix:rescaling}}
By introducing isotopologues into the chemistry we introduce many new reaction channels and we need to make sure that reaction rates are properly scaled. Unfortunately, not only the reaction rates for isotopologue reactions are unknown, we neither have information on branching ratios for reactions with several possible product channels. Thus, we assume equal probabilities for all branches, i.e. all isotopologue reactions possessing the same reactants but different products have their rate coefficient divided by the number of different product branches. For example:
\begin{align*}
\ce{C + O2} &\ce{-> CO + O} \;\;\;\; &\alpha&=4.7(-11)
\end{align*}
with $\alpha$ being the fit coefficient from UDfA06 at $T=300$~K and the number in parentheses indicates the decimal power. Introducing \ce{^{13}C} and \ce{^{18}O} into this reactions opens up additional channels:
\begin{align*}
\ce{C + O ^{18}O} &\ce{-> CO + ^{18}O}\;\;\;\; &\alpha&=2.35(-11)\\
\ce{C + O ^{18}O} &\ce{-> C ^{18}O + O}\;\;\;\; &\alpha&=2.35(-11)\\
\ce{^{13}C + O2}  &\ce{-> ^{13}CO + O}\;\;\;\; &\alpha&=4.7(-11)\\
\ce{^{13}C + O ^{18}O} &\ce{-> ^{13}CO + ^{18}O}\;\;\;\; &\alpha&=2.35(-11)\\
\ce{^{13}C + O ^{18}O} &\ce{-> ^{13}C ^{18}O + O}\;\;\;\; &\alpha&=2.35(-11)
\end{align*}
Assuming equal probabilities for different reaction branches is a strong assumption and the reader should keep in mind, that many of the introduced reactions might have  different reaction rate coefficients with potentially strong impact on the solution of the chemical network. The isotope exchange reactions discussed in the following are a notable exception from this.
\section{Influence of chemical data sets}\label{chemsets}
To illustrate how the choice of the chemical data set affects the outcome of our astrochemical calculations we calculate our reference model for three different chemical sets: UDfA06 \citep{udfa06}, OSU (version osu\_01\_2009), and KIDA \citep[version kida.uva.2011, ][]{kida2012}.
To allow a consistent computation it was necessary to slightly alter the original data sets:
\begin{itemize}
\item The formation of \ce{H2} on grain surfaces is calculated separately in \ktau\ and we removed the reaction \ce{H + H -> H2} from OSU.
\item The same holds for the photo-dissociation of \ce{H2},  we removed the reaction \ce{H2 + \text{h$\nu$} -> H + H}. \ktau\ explicitly calculates the \ce{H2} formation rate from the population of the (vibrational) $v$-levels (all rotational levels of one vibrational state summed, ground state $v=0-14$, Lyman band 24 level + Werner band 10 level)
\item The unshielded CO photo-dissociation rate coefficient differs significantly among the three sets. We recomputed the unshielded photo-dissociation rate for a standard Draine FUV field using absorption cross sections available from http://home.strw.leidenuniv.nl/$\sim$ewine/photo/ and can confirm the value of $2\times 10^{-10}$~s$^{-1}$ \citep{vdishoeck2006} from UDfA06. We replaced the corresponding $\alpha$ values in OSU and KIDA. For a discussion on differing photo-reaction rates see e.g. \citet{vhemert2008} and \citet{roellig11dust}.
\item We replaced reactions with very large negative rate coefficients $\gamma$ with a refitted expressions according to \citet{roellig2011refit}.
\item KIDA uses the Su-Chesnavich capture approach to compute rate coefficients for unmeasured reactions between ions and neutral species with a dipole moment \citep{woon2009}. Their formalism is incompatible with \ktau\ and we replace the corresponding 1877  reaction rate coefficients with a new set of rate coefficients $\alpha, \beta$, and $\gamma$, suitable for the Arrhenius-Kooij formula $\alpha (T/300 K)^\beta \exp(-\gamma T)$, which are fitted such that they approximate the original rates between 10 and 1000~K.
\item For species where \ktau\ does not distinguish between linear and cyclic isomeric forms we consider all species only in terms of their molecular formula, e.g. \ce{C3H} instead of $\mathrm{l-C}_3\mathrm{H}$ and $\mathrm{c-C}_3\mathrm{H}$.
\end{itemize} 

\begin{figure}[htb]
\resizebox{\hsize}{!}{\includegraphics{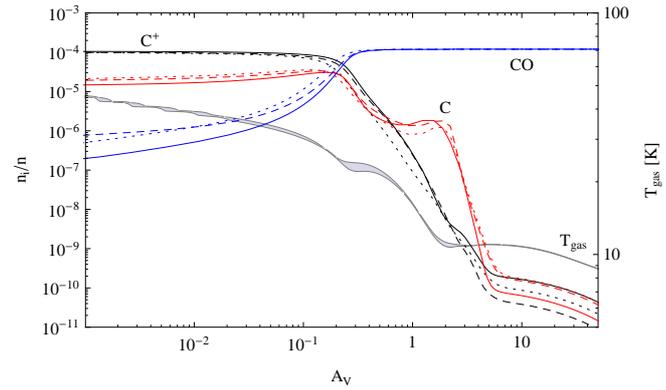}}
\caption{Chemical structure of a model clump with the following model parameters: $n_0=10^5\,\mathrm{cm}^{-3}$, $M=100\,M_\odot$, $\chi=10$. (UDfA06 (solid), OSU (dashed), KIDA (dotted). The gray shaded area shows the gas temperature spanned by the three models calculations. }
\label{databases1}

\end{figure}
\begin{figure}[htb]
\resizebox{\hsize}{!}{\includegraphics{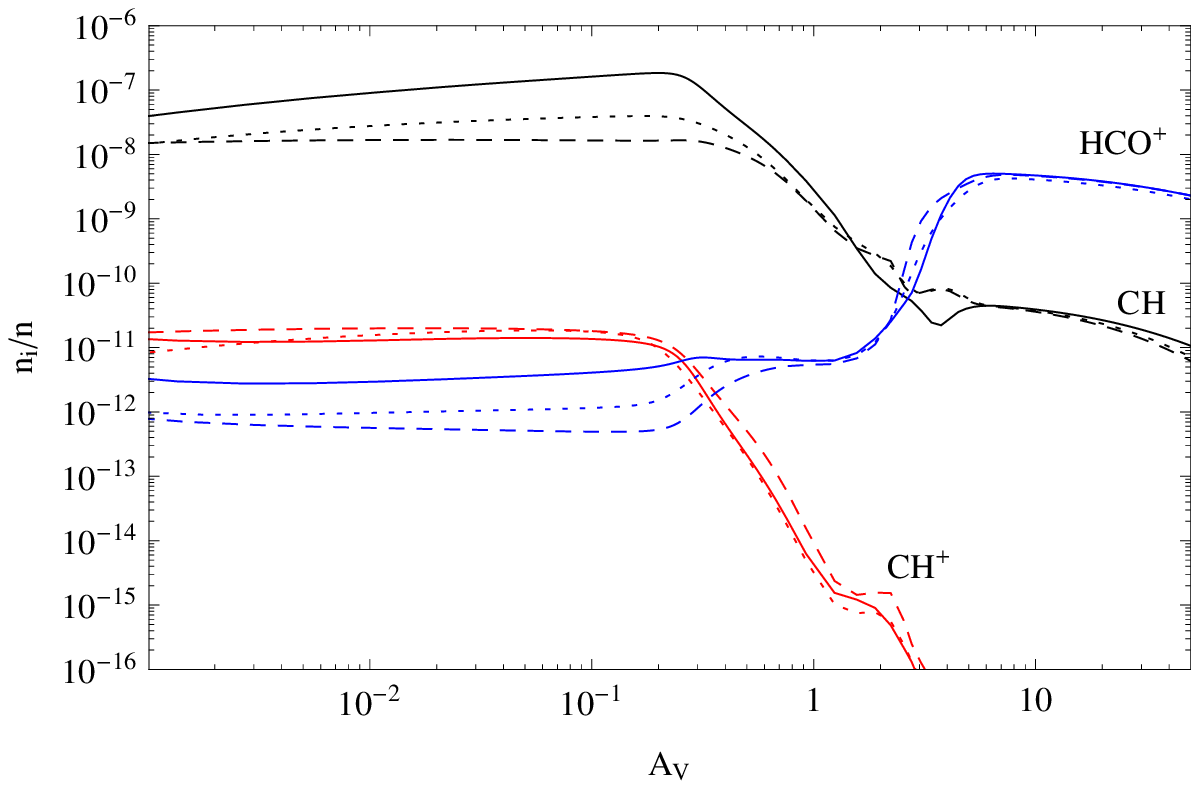}}
\caption{Same as Fig.~\ref{databases1} for \ce{HCO+}, \ce{CH}, and \ce{CH+}. }
\label{databases2}
\end{figure}
In Fig.~\ref{databases1} we show the relative abundances of \ce{C+}, \ce{C}, and \ce{CO} for model calculations with $n=10^5$~cm$^{-3}$, $M=1000$~M$_\odot$, and $\chi=10$ using different chemical data sets: UDfA06 (solid lines), OSU (dashed lines), KIDA (dotted lines) and the resulting range of gas temperatures. All three sets give the same chemical structure but with slight variations in the abundance profiles. The biggest difference is a higher  \ce{CO} abundance of OSU and KIDA at very low values of $A_\mathrm{V}$ (factor 2-3 at $A_\mathrm{V}=0.01$). This also leads to a carbon transition from \ce{C+} to \ce{CO} at slightly lower values of $A_\mathrm{V}$ compared to UDfA. Overall, the chemical structure  of the main carbon species is comparable, particularly the effect on the total column density is small. The similar chemical structure and the almost identical gas temperatures around the carbon transition will result in a consistent fractionation behavior of the three species across all three chemical sets.

Fig.~\ref{databases2} shows how the different chemical sets influence the chemistry of \ce{HCO+}, \ce{CH}, and \ce{CH+}. The differences are larger than those in Fig.~\ref{databases1}. UDfA06 produces a significantly higher CH abundance in the outer layer of the cloud, at some positions by a factor of 10 higher than OSU. However, we discussed earlier that fractionation of \ce{CH} is the result of \ce{C+} fractionation. We do not expect any significant deviations in the fractionation ratio (\FR{}) of \ce{C+})  for the three chemical sets, hence the effect on the \FR{(\ce{CH})} should be weak. The same should be the case for \ce{CH+} which shows even smaller differences for the three chemical sets. \ce{HCO+}shows the same abundance in the cloud center, but some significant differences for $A_\mathrm{V}<1$. This could lead to different fractionation behavior  at these parts of the cloud when using different chemical sets. It is unlikely that this  leads to observable effects because of the 100 times lower \ce{HCO+} abundance with respect to the center of the cloud and the respective weak influence on the total column density.
\begin{figure*}
\includegraphics[width=17cm]{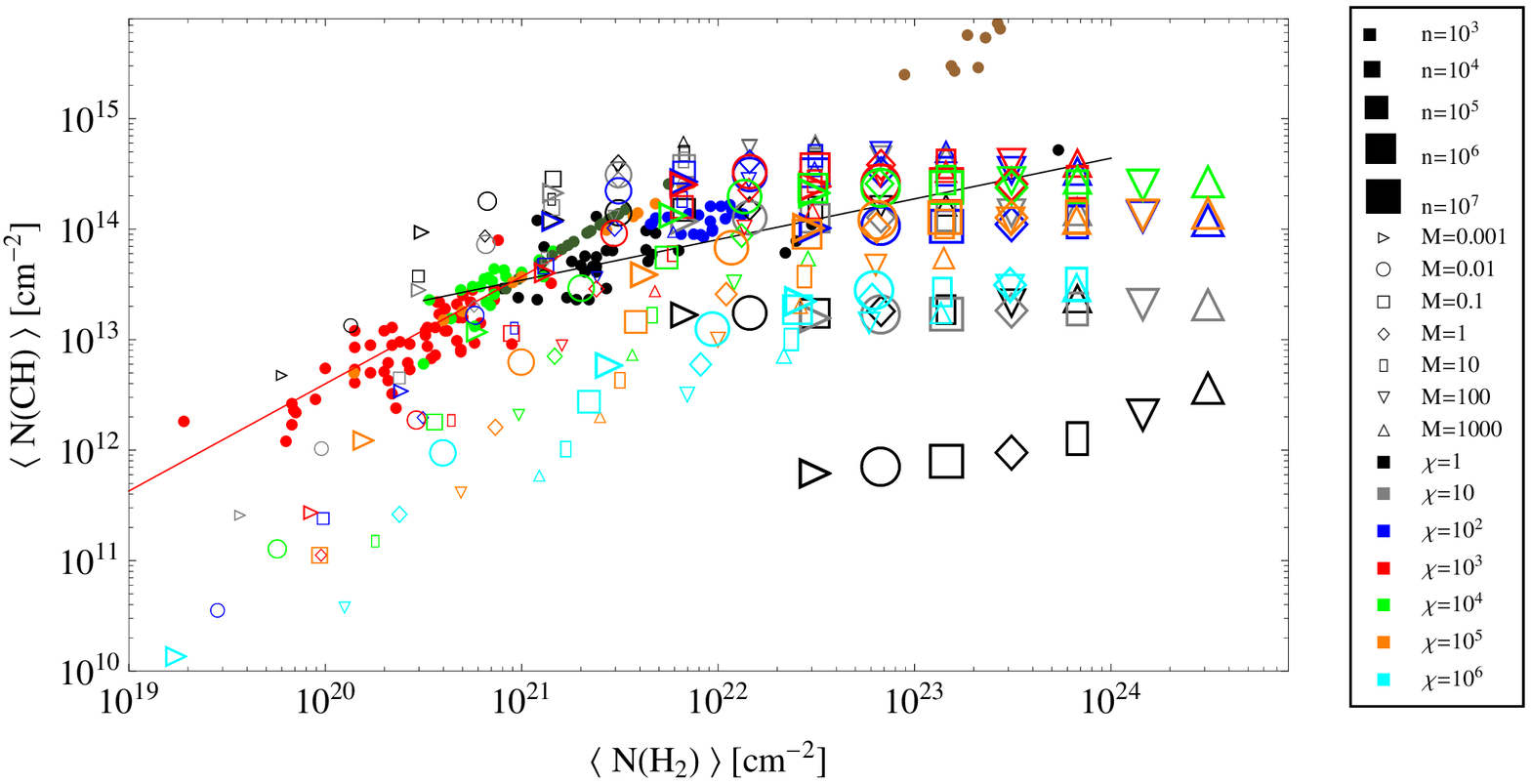}
\caption{Mean \ce{CH} column density  versus \ce{H2} column density (color/shape coding same as Fig. \ref{meancolco}). Observational results are plotted as colored points (red: \citet{sheffer2008};
green: \citet{rachford2002}; olive: \citet{magnani2005}; black: \citet{qin2010}; brown: \citet{magnani2006}; orange \citet{gerin2010}; blue \citet{suutarinen2011}). The two lines are the \ce{CH}-\ce{H2} relations from \citet{mattila1986} (red) and \citet{qin2010} (black).}
\label{obsch}
\end{figure*}
\section{Spherical model context}\label{colstruct}
\subsection{Cloud radius}
The model primarily computes radius dependent quantities. 
For a given density law, i.e., values $\alpha$ and $f_c=R_\mathrm{core}/R_\mathrm{tot}$, the total cloud radius $R_\mathrm{tot}$ (in cm) can be calculated from $n_0$ (in cm$^{-3}$) and $M$ (in M$_\odot$) using
\begin{equation}\label{rtot}
R_\mathrm{tot}=6.57\times 10^{18} \sqrt[3]{-\frac{(\alpha -3)
   M f_c^{\alpha }}{n_0 \left(3 f_c^{\alpha
   }-\alpha  f_c^3\right)}}\,\mathrm{cm}
\end{equation}
For $\alpha=3/2$ and $f_c=0.2$ this reduces to $R_\mathrm{tot}=5.3\times 10^{18} \sqrt[3]{M/n}$~cm.
\subsection{Column density}
The general expression for the maximum (radial) column density\footnote{$N_\mathrm{max}$ is the total gas column density from the cloud surface to the center along the cloud radius.} is
\begin{equation}
N_\mathrm{max}=\frac{n\, R_\mathrm{tot}\left(\alpha\, {f_c}^{1-\alpha
   }-1\right)}{\alpha -1}\,\mathrm{cm}^{-2}
\end{equation}  
that is $N_\mathrm{max}=4.7 n R$~cm$^{-2}$ for $\alpha=3/2$ and $f_c=0.2$.

However, they are not observables. 
Observations always yield an projected, beam-convolved figure. 
We describe them in terms of measurable column densities.

In the framework of spherical model clouds, 
column densities differ depending on where  we look at. 
To get a position-independent measure for the column density of a given species $i$, we calculate the average column density for the whole clump
\begin{equation}
<N_i>=\frac{4\pi}{\pi R^2}\int_0^R n_i(r)r^2dr \label{formula-meancol}
\end{equation} 
When referring to model column densities we always mean a clump averaged  column density according to Eq.~\ref{formula-meancol}.
This definition assumes that observations always cover whole clumps. This is equivalent to the traditional description of beam-filling factors for the observation of spatially unresolved clumps, i.e. we assume that we have (typically many) unresolved clumps within the telescope beam.
In that sense, the fractionation of the column density is always the result of a convolution of the fractionation structure with the absolute abundance profile.
\subsection{\ce{CH} column densities}
Modelling the formation of the light hydrides, such as \ce{CH+} and \ce{CH}, still poses a challenge to chemical models. In Fig.~\ref{obsch} we plot the mean \ce{CH} column density versus the total mean column density of the respective clump. The coloured dots are observations from absorption  and emission line measurements. The red and black lines are simple parametrized models to describe the column density \citep{mattila1986,qin2010}. We note, that the column density  splits into two distinct regimes with significantly different behavior. Diffuse clouds with total column densities below $10^{21}$~cm$^{-2}$ show a steeper slope than denser clouds, where the \ce{CH} column density appears to approach a limit of about $10^{15}$~cm$^{-2}$. This behavior is approximately reproduced by the model calculations. The column densities in our model are consistent with values for TMC-1 given by \citet{suutarinen2011} as well as with diffuse cloud observations presented by \citet{gerin2010} which give $N(\ce{CH})\approx 1-26\times 10^{13}$~cm$^{-2}$ for total \ce{H2} columns between $10^{21}-10^{22}$~cm$^{-2}$.

\citet{sheffer2008} present column densities  along 42 diffuse molecular Galactic sight lines. They find total columns between 10$^{12}$ and 10$^{14}$~cm$^{-2}$ and a very strong correlation between column densities of \ce{CH} and \ce{H2}. \citet{mattila1986} confirms this trend for dark clouds. \citet{magnani2005} derived \ce{CH} column densities from 3335 MHz observations in the Galactic plane and used the linear relation given by \citet{mattila1986} to derive corresponding \ce{H2} column densities. They find $10^{13}\le N(\ce{CH})<10^{15}$~cm$^{-2}$. 

\citet{magnani2006} observed the \ce{CH  ^2$\,\Pi$_{1/2}},$J=1/2,F=1-1$ transition toward the Galactic center. They find  $N(\ce{CH})\approx3-7\times 10^{15}$~cm$^{-2}$.  In addition to determining $N(\ce{H2})$ from its linear correlation to $N(\ce{CH})$ they also used a factor $1.8\times 10^{20}$ to derive  $N(\ce{H2})$ from integrated CO(1-0) line emission. 

\citet{qin2010} presented recent {\it Herschel/HIFI} observations against Sgr~B2(M) which revealed that the linear relationship between \ce{CH} and \ce{H2} flattens at higher visual extinctions. They give a log-log slope of 0.38$\pm$0.07 for $N(\ce{H2})\ge 10^{21}$~cm$^{-2}$. In Fig.~\ref{obsch} we show our model results for $N(\ce{CH})$ versus $N(\ce{H2})$ together with the observational data \citep{rachford2002,magnani2005,magnani2006,sheffer2008,qin2010,gerin2010,suutarinen2011}.

We cannot reproduce the large column densities derived by \citet{magnani2006}. Their data is derived from emission line measurements, while all other studies used absorption lines, and consequently suffers from higher uncertainties due to its dependence on the assumed excitation temperature.  To our knowledge  \ce{^{13}CH} has not yet been detected. The lack of \ce{^{13}CH} observations makes it difficult to assess the model results.
The total \ce{CH} column densities that we find in our model results stay below $10^{15}$~cm$^{-2}$. 
Nevertheless, the general behavior is well reproduced.
\section{Fractionation plots of selected species}\label{FRplots}
Here we show fractionation plots of selected species over a large portion of our model parameter space. We left out models with $M=10^3$~M$_\odot$ because they show very similar results as models with $M=10^2$~M$_\odot$.
The fractionation plots are also available for download: \url{http://www.astro.uni-koeln.de/kosma-tau}.
\begin{figure*}
\centering
\includegraphics[width=16cm]{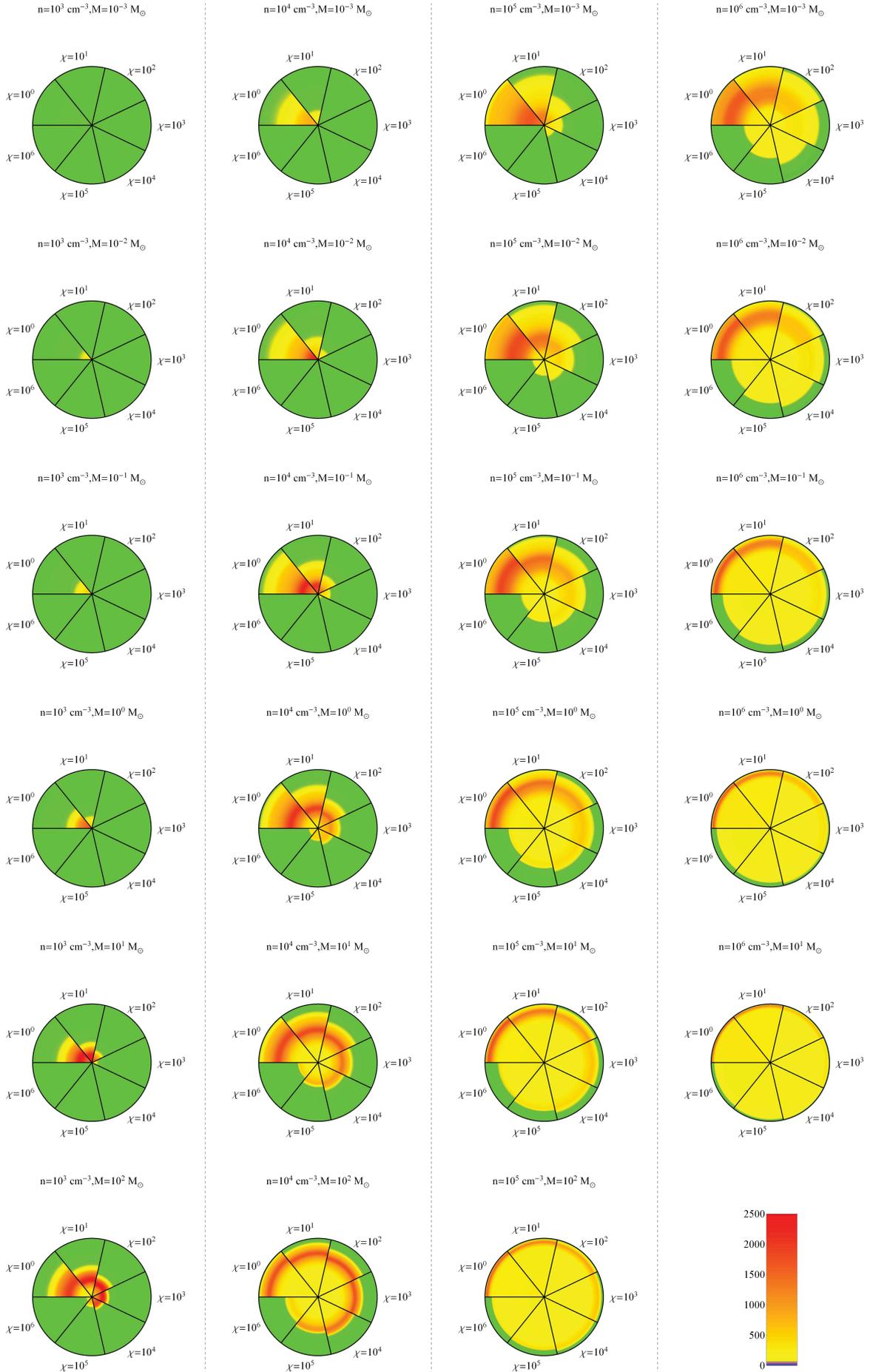}
\caption{\ce{C+} fractionation structure as function of relative clump radius $r/R_\mathrm{tot}$ for different values of $n$ and $M$. Each sector corresponds to a different $\chi$ value. The \FR{} is color coded, ratios within $\pm 10$\% of the \ER{} are shown in green. Blue and violet denotes \FR{}<\ER{}, yellow and red denotes \FR{}>\ER{}.}
\label{SP-C+}
\end{figure*}

\begin{figure*}
\centering
\includegraphics[width=16cm]{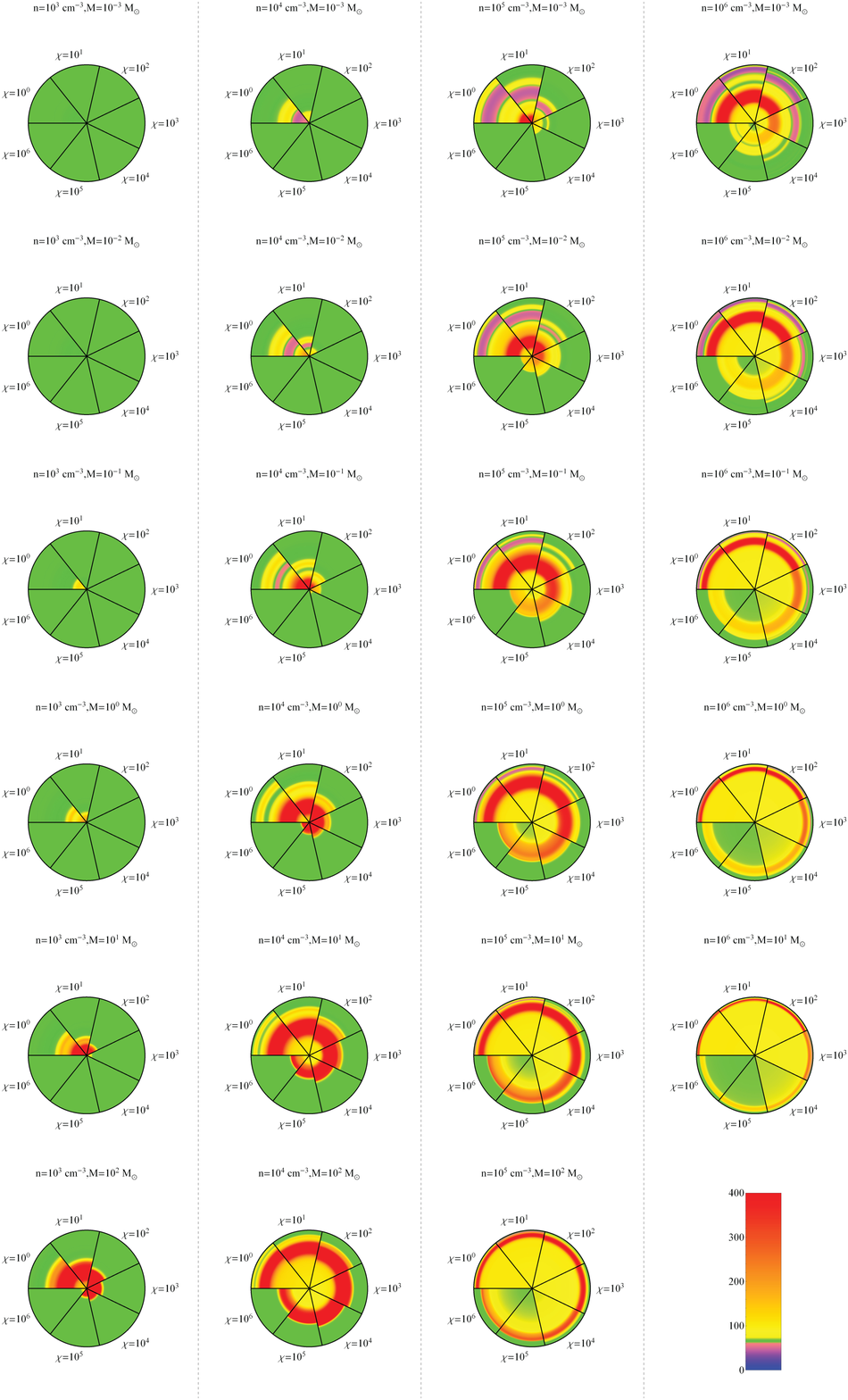}
\caption{Same as Fig.\ref{SP-C+} for \ce{C}.}
\label{SP-C}
\end{figure*}

\begin{figure*}
\centering
\includegraphics[width=16cm]{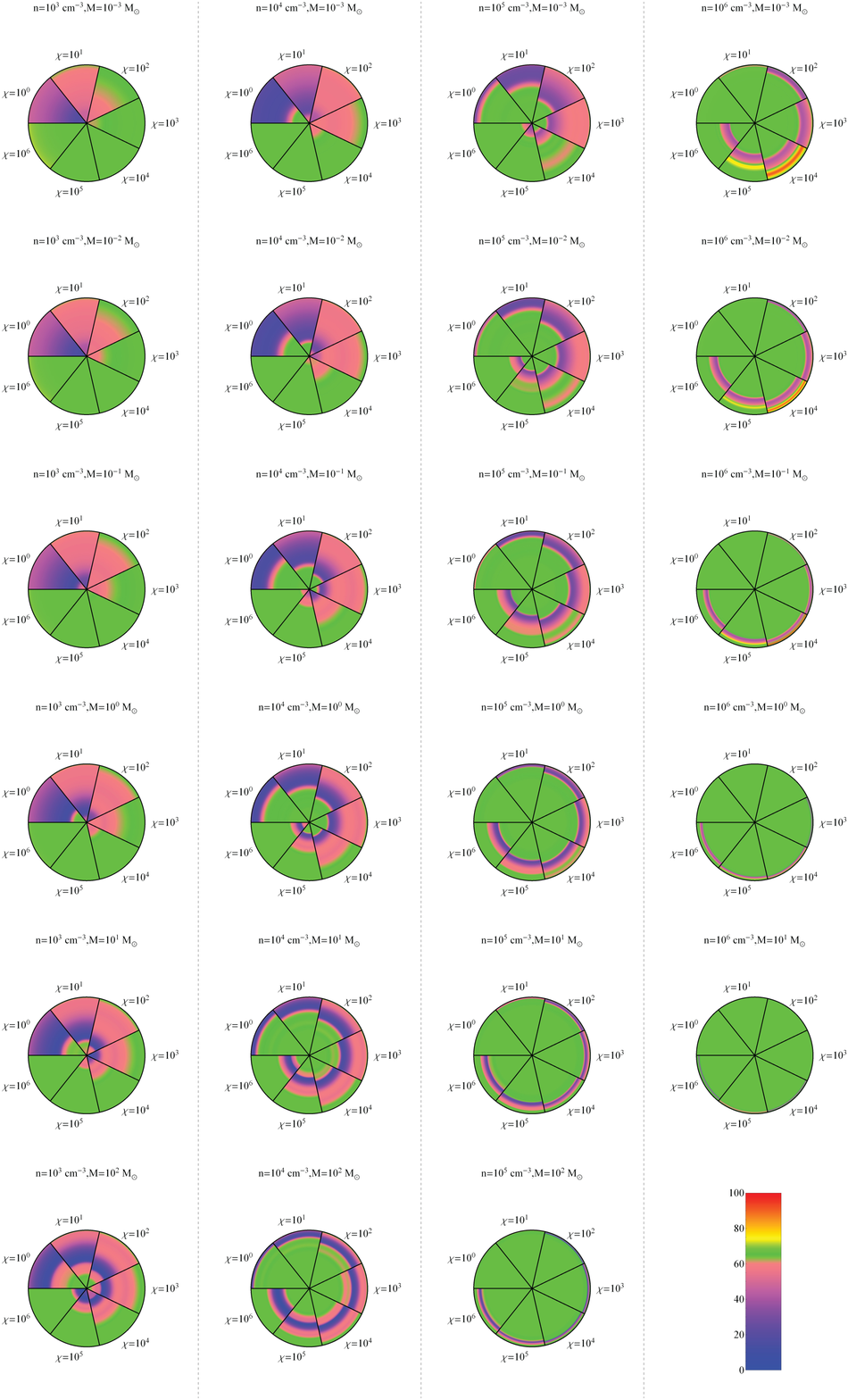}
\caption{Same as Fig.\ref{SP-C+} for \ce{CO}.}
\label{SP-CO}
\end{figure*}

\begin{figure*}
\centering
\includegraphics[width=16cm]{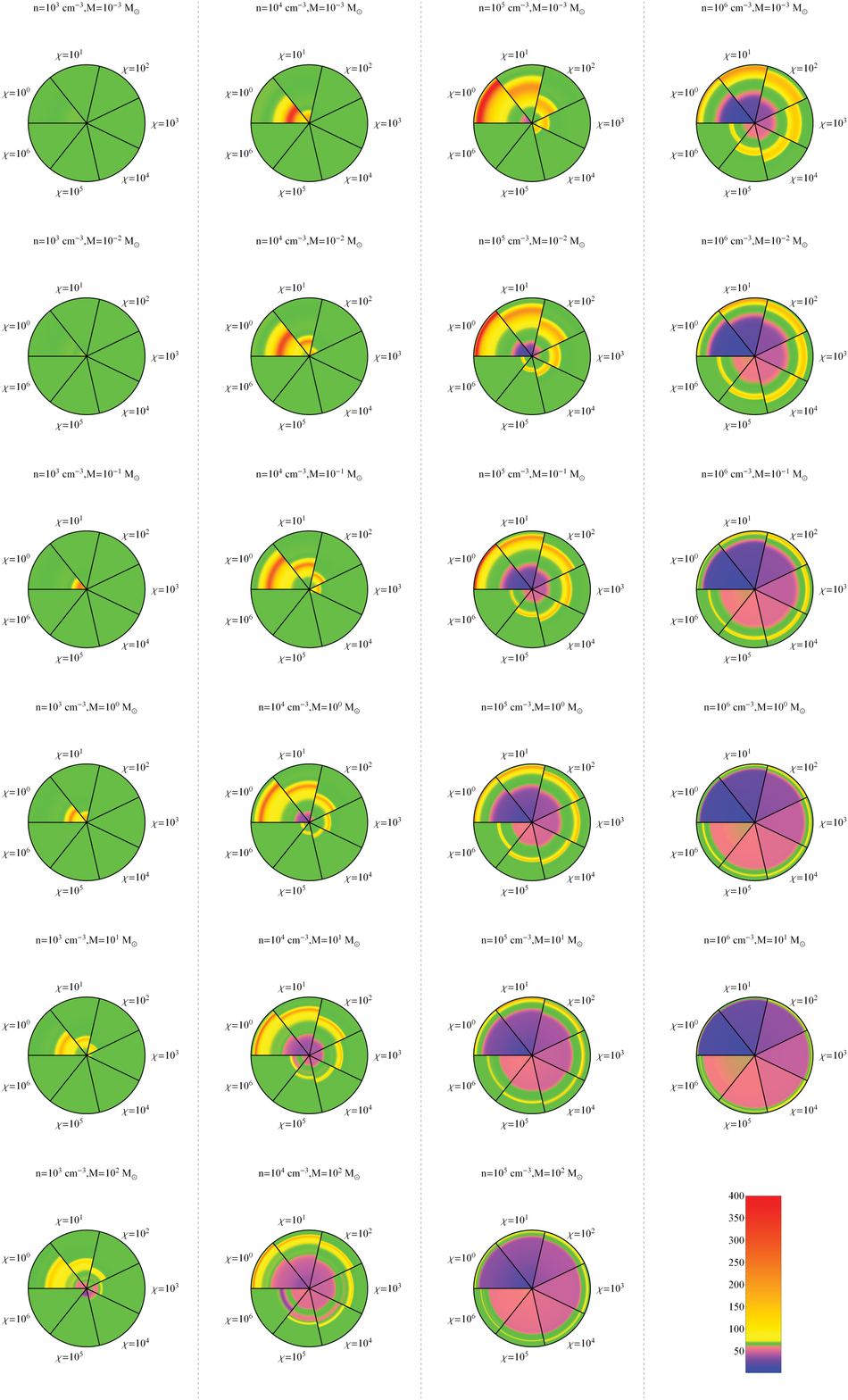}
\caption{Same as Fig.\ref{SP-C+} for \ce{HCO+}.}
\label{SP-HCO+}
\end{figure*}

\begin{figure*}
\centering
\includegraphics[width=16cm]{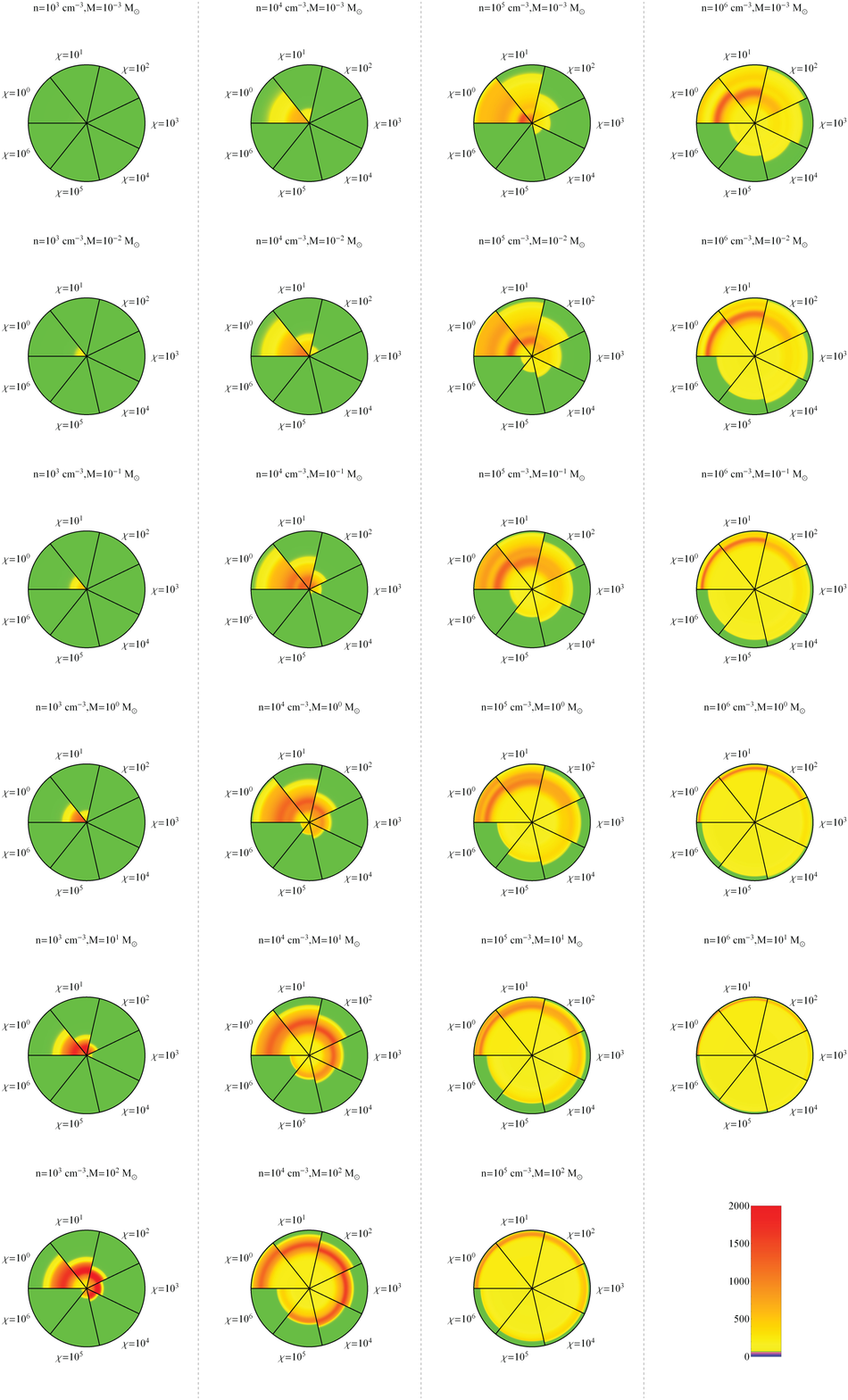}
\caption{Same as Fig.\ref{SP-C+} for \ce{CH}.}
\label{SP-CH}
\end{figure*}

\begin{figure*}
\centering
\includegraphics[width=16cm]{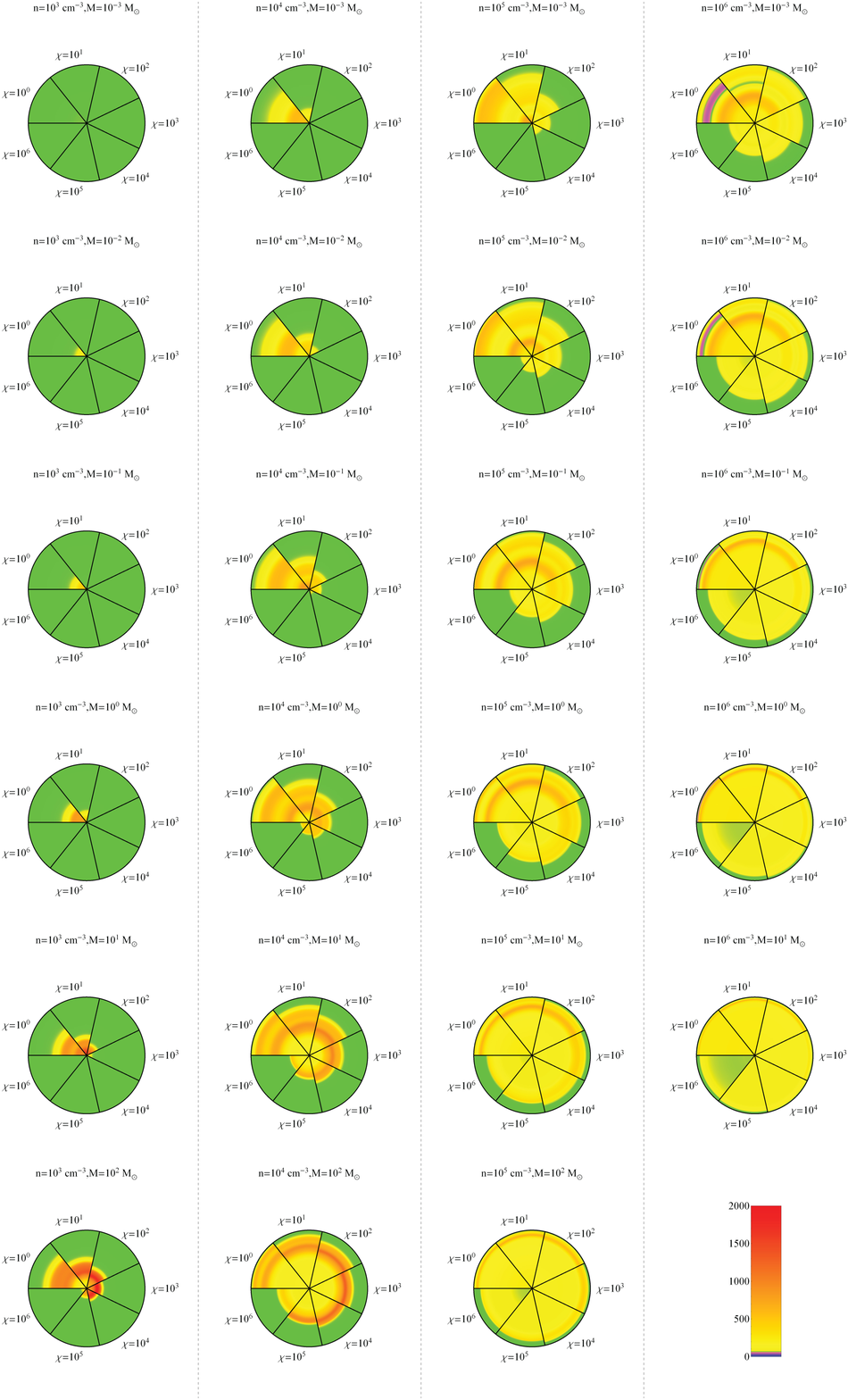}
\caption{Same as Fig.\ref{SP-C+} for \ce{CH+}.}
\label{SP-CH+}
\end{figure*}
\end{appendix}
\end{document}

%% file: journals.tex
\def\aj{AJ}%
\def\actaa{Acta Astron.}%
\def\araa{ARA\&A}%
\def\apj{ApJ}%
\def\apjl{ApJ}%
\def\apjs{ApJS}%
\def\ao{Appl.~Opt.}%
\def\apss{Ap\&SS}%
\def\aap{A\&A}%
\def\aapr{A\&A~Rev.}%
\def\aaps{A\&AS}%
\def\azh{AZh}%
\def\baas{BAAS}%
\def\bac{Bull. astr. Inst. Czechosl.}%
\def\caa{Chinese Astron. Astrophys.}%
\def\cjaa{Chinese J. Astron. Astrophys.}%
\def\icarus{Icarus}%
\def\jcap{J. Cosmology Astropart. Phys.}%
\def\jrasc{JRASC}%
\def\mnras{MNRAS}%
\def\memras{MmRAS}%
\def\na{New A}%
\def\nar{New A Rev.}%
\def\pasa{PASA}%
\def\pra{Phys.~Rev.~A}%
\def\prb{Phys.~Rev.~B}%
\def\prc{Phys.~Rev.~C}%
\def\prd{Phys.~Rev.~D}%
\def\pre{Phys.~Rev.~E}%
\def\prl{Phys.~Rev.~Lett.}%
\def\pasp{PASP}%
\def\pasj{PASJ}%
\def\qjras{QJRAS}%
\def\rmxaa{Rev. Mexicana Astron. Astrofis.}%
\def\skytel{S\&T}%
\def\solphys{Sol.~Phys.}%
\def\sovast{Soviet~Ast.}%
\def\ssr{Space~Sci.~Rev.}%
\def\zap{ZAp}%
\def\nat{Nature}%
\def\iaucirc{IAU~Circ.}%
\def\aplett{Astrophys.~Lett.}%
\def\apspr{Astrophys.~Space~Phys.~Res.}%
\def\bain{Bull.~Astron.~Inst.~Netherlands}%
\def\fcp{Fund.~Cosmic~Phys.}%
\def\gca{Geochim.~Cosmochim.~Acta}%
\def\grl{Geophys.~Res.~Lett.}%
\def\jcp{J.~Chem.~Phys.}%
\def\jgr{J.~Geophys.~Res.}%
\def\jqsrt{J.~Quant.~Spec.~Radiat.~Transf.}%
\def\memsai{Mem.~Soc.~Astron.~Italiana}%
\def\nphysa{Nucl.~Phys.~A}%
\def\physrep{Phys.~Rep.}%
\def\physscr{Phys.~Scr}%
\def\planss{Planet.~Space~Sci.}%
\def\procspie{Proc.~SPIE}%
\let\astap=\aap
\let\apjlett=\apjl
\let\apjsupp=\apjs
\let\applopt=\ao